\documentclass[11pt,a4papee]{article}
\pdfoutput=1
\usepackage{jheppub}
\usepackage{physics}
\usepackage{mathtools}
\usepackage{amssymb}

\title{Fiducial observers and the thermal atmosphere in the black hole quantum throat}

\author{Thomas G. Mertens,}
\author{Thomas Tappeiner,}
\author{Bruno de S. L. Torres}

\affiliation{Department of Physics and Astronomy, Ghent University, \\ Krijgslaan, 281-S9, 9000 Gent, Belgium}

\emailAdd{thomas.mertens@ugent.be}
\emailAdd{thomas.tappeiner@ugent.be}
\emailAdd{bruno.desouzaleaotorres@ugent.be}

\abstract{
We propose a construction of fiducial observers in the throat region of near-extremal black holes within the framework of JT quantum gravity, leading to a notion of local observers in a highly quantum regime of the gravitational field. The construction is based on an earlier proposal for light-ray anchoring to the asymptotic boundary and is uniquely fixed at the semiclassical level by demanding that the notion of time translations for an observer at the asymptotic boundary of JT gravity should be extended into the bulk as the flow of a conformal isometry.  Since conformal isometries are a necessary condition for geometric modular flow, our construction is amenable as a candidate geometric gravitational dressing that may be interpreted via the modular crossed product, potentially connecting our choice of dressing with recent developments on the literature on local observables in quantum gravity. Taking this definition beyond the semiclassical regime, we compute quantum gravitational wormhole contributions to the black hole thermal atmosphere, directly producing a finite thermal entropy and leading to a quantum description of the stretched horizon in this model.}

\begin{document}

\definecolor{darkblue}{rgb}{0.0, 0.0, 0.55}

\maketitle

\section{Introduction}
\label{sec:intro}
Ever since the discovery of the thermal properties of black hole physics \cite{Bardeen:1973gs}, a special role has been played by the so-called fiducial (FIDO) observers. These represent a class of observer worldlines that are accelerating in the exterior of the black hole, and are causally disconnected from the black hole interior. Since this family of observers never crosses the black hole event horizon, they associate a coarse-grained entropy to it, which is the famous Bekenstein-Hawking entropy $S_{BH} = A/4G$. Such observers also generically measure a non-zero temperature, with uniformly accelerated trajectories having their (renormalized)\footnote{By ``renormalized'' here we mean that the magnitude of the radial acceleration is being read off by an asymptotic observer infinitely far away from the black hole. This differs from the local acceleration measured along each observer's worldline by a redshift factor that diverges as one approaches the black hole horizon.} radial acceleration equal to the black hole's surface gravity $\kappa$, which in turn is related to the Hawking temperature by $T_H = \kappa/2\pi$. This class of observers is also important for more general black holes, such as the zero-angular-momentum observers (ZAMO) or Bardeen observers for the Kerr black hole \cite{Bardeen:1970zz}. 

Given the immense impact black hole thermodynamics has had on our understanding of black hole physics, the extension of these ideas beyond classical gravity is vital. However, when doing so, one immediately faces the problem of defining a fiducial observer in a diffeomorphism-invariant way when spacetime is allowed to fluctuate. A priori it seems there are an infinite list of possible extensions, considering various physically independent anchorings or definitions of gravitational dressing \cite{Donnelly:2015hta}. This hampers a clear physical route towards understanding their physics.

In this work, we will advocate for a proposal to address this problem in the context of Jackiw-Teitelboim (JT) quantum gravity \cite{Jackiw:1984je, Teitelboim:1983ux},\footnote{See e.g. \cite{Mertens:2022irh} for a recent review on JT gravity.} which universally describes the near-horizon throat of near-extremal higher-dimensional black holes. Most of our presentation will focus on JT gravity as a standalone theory in its own right, but we will revisit the embedding in the higher-dimensional black hole in the concluding section \ref{sec:concl}.

Our analysis will be motivated by recent developments on operator algebras in quantum gravity~\cite{Leutheusser:2021qhd,Leutheusser:2021frk,WittenCrossedProduct, Chandrasekaran:2022cip, Chandrasekaran:2022eqq}, and rely on the following two observations:
\begin{itemize}
\item As established several years ago \cite{Almheiri:2014cka, Jensen:2016pah, Maldacena:2016upp, Engelsoy:2016xyb}, JT gravity on a manifold with a boundary gives dynamics to the boundary in terms of a dynamical time reparametrization $F(t)$, where $F$ denotes the Poincar\'e reference time on the AdS$_2$ manifold. This is the only (non-topological) gravitational degree of freedom left in the model. The boundary itself is regularized into the worldline parametrized by $(F(t),Z=\epsilon F'(t))$ in Poincar\'e patch coordinates $(F,Z)$.
\item In lower-dimensional gravity models, gravity acts as an ensemble (or statistical) average. This was first articulated in \cite{Saad:2019lba}. A particularly sharp manifestation of this fact is the computation of the low-temperature free energy \cite{Engelhardt:2020qpv,Johnson:2021rsh}, where it was shown that one should consider $\langle \log Z \rangle$ (quenched free energy) instead of $\log\langle Z \rangle$ (annealed free energy), where the bracket denotes the gravitational path integral.\footnote{See \cite{Hernandez-Cuenca:2024icn,Antonini:2025nir} for further recent progress on this distinction.} This means gravity is treated fundamentally different than the other (matter) fields in the theory. Within our setup, we will leverage this to write down matter operators that are functionals of the gravity degrees of freedom. At the end of the computation, these are then to be path-integrated (averaged) over the gravitational field. This distinction is very important on higher topology, but it actually even plays a role at the level of the disk topology in how we phrase our construction.
\end{itemize}
The notion of an observer is intimately tied to that of local \emph{observables}; after all, an observer can be seen as an idealized description of a system that can probe some subset of the degrees of freedom of the physical system of interest. The nature of local observables in gravity is fundamentally relational~\cite{Rovelli:1990ph, Giddings:2005id, Tambornino:2011vg, Donnelly:2016rvo, Goeller:2022rsx}, since diffeomorphism invariance renders the notion of a spacetime point gauge-dependent and thus unphysical. As such, in order to restore some form of locality, one has to define local observables relative to an auxiliary physical system, with the matter degrees of freedom being appropriately dressed to the system that is being used as a reference. 

A natural place to put the reference system is at the asymptotic boundary whenever such a structure is available. One then constructs a notion of diffeomorphism-invariant bulk observables by dressing the operators to the boundary, with a covariant notion of localization in time being achieved by dressing bulk observables with the boundary ADM Hamiltonian, for instance. This general principle can be realized very explicitly in JT gravity, where there are no local bulk gravitational degrees of freedom, and the dynamics can be mapped to observables supported near the asymptotic boundary of spacetime. Since the boundary in this case is also a one-dimensional curve, with the gravitational degree of freedom naturally associated with the dynamics of the boundary trajectory, it is particularly enticing to interpret any geometric dressing in JT gravity as corresponding to some dressing by a ``boundary observer.'' Now, by the timelike tube theorem~\cite{Borchers1961, osti_4665531, Strohmaier:2023hhy, Strohmaier:2023opz}, any observer sharing the same causal patch should be able to account for the same algebra of observables. (Strictly speaking, the timelike tube theorem holds for QFTs on a classical background geometry, without quantum gravity. However, since the bulk geometry in JT gravity is always fixed, we believe the theorem provides useful intuition in our case of interest as well.) The upshot is that for a given such boundary curve parametrized by a time reparametrization $F(t)$, any observer worldline outside of the classical horizon can be taken as this reference observer degree of freedom to construct the same operator algebra. Changing to a different $F(t)$ within a specific gravitational integration space for which the accessible causal region does not change (as happens naturally in the JT Euclidean gravitational path integral), we can define the observer once and for all for the full quantum gravity model---and because of that, there is a sense in which such observer is independent of the gravitational degrees of freedom. \\
This leads us to our main definition of a preferred family of fiducial observers in JT gravity: \\

\begin{figure}[h]
    \centering   \includegraphics[width=0.65\textwidth]{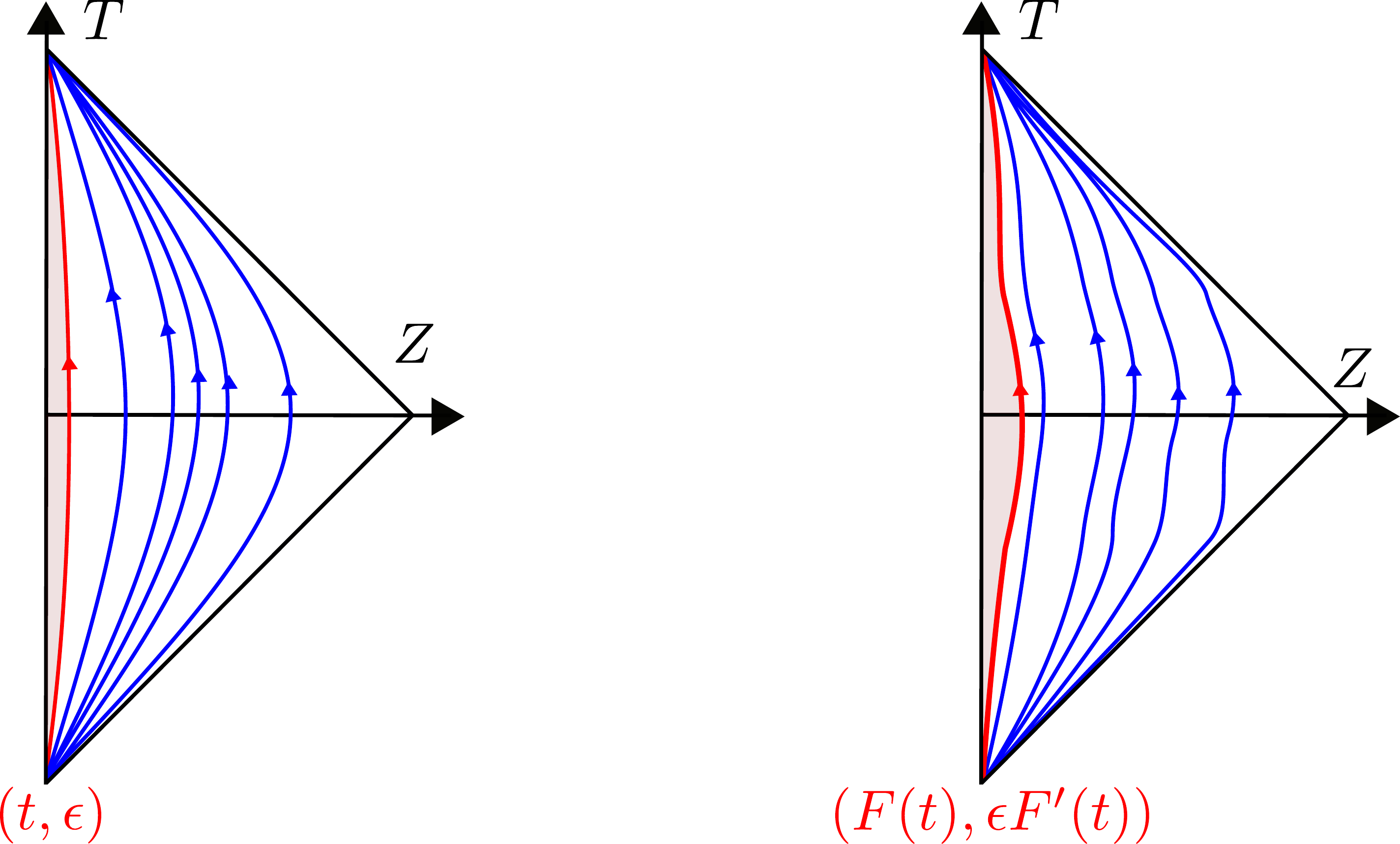}
    \caption{Unique conformal Killing flow (blue) that limits to the Schwarzian wiggly curve (red) and that respects the causal structure of the classical Poincar\'e coordinate region. Left: Semiclassical or saddle flowlines. Right: Illustration of an off-shell flowline family.}
    \label{fig:Flowlines}
\end{figure}

\noindent\fbox{\begin{minipage}{\textwidth}
\emph{Definition}: A family of fiducial (Unruh) observers in JT quantum gravity is defined such that for every off-shell gravity field $F(t)$, their worldlines satisfy the following properties:
\begin{itemize}
\item Their timelike worldlines are a foliation of the exterior of the classical black hole horizon, i.e. they do not intersect and their union is the full exterior of the horizon. This means they are the flowlines of a vector field in AdS$_2$.
\item The vector field whose flowlines correspond to the worldlines of our family of fiducial observers form a \emph{conformal Killing vector field} of AdS$_2$. 
\item The worldlines satisfy the asymptotic boundary condition that they limit to the boundary Schwarzian wiggly curve $(F(t),Z=\epsilon F'(t))$, with the boundary time $t$ matching the conformal Killing time for the bulk conformal Killing vector field. \\
\end{itemize}
\end{minipage}}
\\~\\
We will show below that, for every off-shell profile $F(t)$ of the boundary curve, there is a unique foliation of timelike worldlines that satisfies these properties (see Fig.~\ref{fig:Flowlines} for an illustration). 

Note that we have not defined these observers directly in a canonically quantized quantum gravity framework. Instead, they are defined for each off-shell gravity configuration. Afterwards, one path-integrates over gravity. This corresponds to treating gravity as a statistical (averaging) procedure done at the very end of a calculation, instead of treating it as a fundamental quantum field. This perspective is ubiquitous in lower-dimensional gravity models described as random matrix theories, but it is actually even assumed at the disk level; for instance, the Schwarzian boundary two-point correlator represents the CFT$_1$ answer for a generic reparametrization $F(t)$, that is afterwards path-integrated over all reparametrizations $F(t)$ where configurations are weighted with the exponential of the Schwarzian action.

The key feature in our  proposal, as emphasized in the definition above, is the requirement that boundary time translations can be extended into the bulk as a (quasi)- conformal isometry of AdS$_2$. This is ultimately what allows us to single out a preferred extension of the boundary trajectory in terms of a unique congruence of observer trajectories foliating the black hole exterior. The motivation for this requirement comes from the special role played by conformal isometries in the connection between geometric flow---that is, the one-parameter family of diffeomorphisms generated by flow under a vector field in spacetime---and \emph{modular} flow, which is a rather abstract concept from the theory of operator algebras~\cite{Takesaki:1970aki, takesaki2002theory} that has been central in many recent breakthroughs in quantum field theory and quantum gravity~\cite{ShadiCrossedProduct, kudlerflam2023covariantregulatorentanglemententropy, WittenCrossedProduct, Chandrasekaran:2022cip, Jensen:2023yxy}. We thus view the results presented here as constituting compelling evidence for a gravitational dressing in JT gravity that generalizes the notion of local bulk observers beyond the semiclassical regime, in a way that is consistent with the recent developments on von Neumann algebras in quantum gravity.

\subsection{Outline of the paper}

The paper is organized as follows.
\begin{itemize}
    \item \textbf{Section~\ref{sec:review}} contains a brief review of the relevant technical background for this work, with \textbf{subsection~\ref{sub:JTreview}} providing a short summary of JT gravity and \textbf{subsection~\ref{sub:modularcrossedproduct}} reviewing the notion of modular crossed product. The main purpose of this section is to fix notation and make the paper more conceptually self-contained.
\end{itemize}

Our original contributions are then presented in two main sections:
\begin{itemize}
    \item In \textbf{section \ref{sec:geometricdressing}}, we show that requiring boundary time translations to extend to bulk conformal Killing vectors directly leads to the gravitational dressing proposed in~\cite{Blommaert:2019hjr,Mertens:2019bvy,Blommaert2021}. This strengthens the motivation for the dressing employed in~\cite{Blommaert:2019hjr,Mertens:2019bvy,Blommaert2021}, as it directly implies that this is the only definition of a local bulk frame in JT gravity that may potentially be interpreted in terms of a modular crossed product acting in a geometrically local way in spacetime.
    \item Once equipped with the preferred choice of dressing, in \textbf{section \ref{sec:bhta}} we calculate the bulk entanglement entropy of a CFT minimally coupled to the Schwarzian mode in JT gravity. We perform the calculation in two stages:
    \begin{itemize}
        \item[(a)] A warm-up in \textbf{subsection~\ref{sub:disk}}: calculation with disk topology;
        \item[(b)] The full result in \textbf{subsection~\ref{sub:highertopology}}: non-perturbative calculation including the matrix model corrections to the pair density correlator. 
    \end{itemize}
\end{itemize}
We end in \textbf{section \ref{sec:concl}} with a discussion on several future directions. Some technical steps are relegated to \textbf{appendices \ref{app:appdisk}} and \textbf{\ref{app:appnonpert}}. We end with a novel discussion on the computation of the probability distribution of bulk geodesic lengths (or wormhole size) in \textbf{appendix \ref{app:probdistr}}.

\section{Technical preliminaries}\label{sec:review}
We first collect some prerequisite material of use for the main part of this work.

\subsection{JT gravity}\label{sub:JTreview}
 The action of JT gravity on a manifold $M$ with Lorentzian signature is given by
\begin{equation}
    S_{\text{JT}}[g, \Phi] = \dfrac{1}{16\pi G_N}\int_M d^2x\sqrt{-g}\,\Phi(R+2) + \dfrac{1}{8\pi G_N}\int_{\partial M}dt\sqrt{-h}\,\Phi(K-1) + S_0 \chi(M),
\end{equation}
where $g$ denotes the two-dimensional metric, $R$ is the Ricci scalar, $\Phi$ is the dilaton field, $h$ is the induced metric on the one-dimensional boundary $\partial M$, and $K$ is the extrinsic curvature of $\partial M$. The time $t$ in the second term plays the role of the boundary time, i.e., the time measured by a preferred observer traveling along $\partial M$. The last term on the right-hand side, proportional to the Euler characteristic $\chi(M)$ of the manifold $M$, is a topological invariant which does not play a role at the classical level. It does however become important at the quantum-mechanical level, where one generically has to include effects from higher-order topologies, for instance, when evaluating the gravitational path integral. The prefactor $S_0$ controls the topology fluctuations, with higher topology effects naturally organizing themselves as a perturbative expansion in $e^{-S_0}$.

The dilaton field appears in the bulk action as a Lagrange multiplier, fixing the Ricci scalar to be equal to $R=-2$. In a two-dimensional spacetime, this condition suffices to guarantee that $M$ must correspond to some patch of the AdS$_2$ geometry. The only degree of freedom that remains is the shape of the embedding of $M$ on AdS$_2$, whose dynamics are fixed once we pick boundary conditions for the metric and the dilaton on $\partial M$. The boundary conditions we impose are 

\begin{equation}
    ds^2\big|_{\partial M} =  -\dfrac{dt^2}{\epsilon^2}, \,\,\,\,\, \Phi\big|_{\partial M} = \dfrac{\varphi_b}{\epsilon}.
\end{equation}
The parameter $\epsilon$ serves as a UV cutoff, which moves the location of $\partial M$ inwards relative to the conformal boundary of AdS$_2$. The parameter $\varphi_b$ is the renormalized value of the dilaton, which one can interpret physically as the rate at which the dilaton field diverges close to the boundary.

If we describe the location of $\partial M$ in Poincar\'e coordinates $(T, Z)$ and expand to leading order in $\epsilon$, imposing Poincar\'e boundary conditions for the metric near $\partial M$ implies that the embedding of $\partial M$ in AdS$_2$ is given by a trajectory of the form $(F(t), \epsilon F'(t))$. Thus, the dynamics of JT gravity are encoded in the function $F(t)$ describing the reparametrization from the boundary time $t$ to Poincar\'e time $T$. By then evaluating the Gibbons-Hawking term on this trajectory and taking the limit $\epsilon\to 0$, one can show that the dynamics for the time reparametrization $F(t)$ in JT gravity (omitting the topological term) are described by the so-called Schwarzian action \cite{Jensen:2016pah, Maldacena:2016upp, Engelsoy:2016xyb}
\begin{equation}
    S_{\text{JT}}[g, \Phi] = S_{\text{Schw}}[F] = -C\int dt\,\{F, t\},
\end{equation}
where $C\equiv \frac{\varphi_b}{8\pi G_N}$, and $\{F, t\} \equiv \left(\frac{F''}{F'}\right)' - \frac{1}{2}\left(\frac{F''}{F'}\right)^2$ is the Schwarzian derivative. The coefficient $C$, which has units of length, plays the role of an effective coupling constant of the Schwarzian theory. 

A lot can be learned about the physics of JT gravity by formulating questions of physical interest in Schwarzian language, thanks to the various known results and techniques that allow us to solve the Schwarzian theory exactly~\cite{Mertens:2017mtv, Stanford:2017thb, Blommaert:2018oro}. One famous example is the black hole partition function and density of states in JT gravity at disk level, which can be computed nonperturbatively in $C$ by making use of the fact that the Schwarzian path integral is one-loop exact~\cite{Stanford:2017thb}. In section~\ref{sec:bhta} we will show how to take advantage of this fact to map the computation of the entanglement entropy of matter fields in JT gravity precisely to a definite observable in the Schwarzian theory, using the dressing obtained in section~\ref{sec:geometricdressing}.

\subsection{Primer on the modular crossed product}\label{sub:modularcrossedproduct}

Much of the recent progress in our understanding of local observables in quantum gravity has come from a deeper appreciation of the role of von Neumann algebras in QFT and gravity~\cite{Leutheusser:2021qhd,Leutheusser:2021frk,WittenCrossedProduct, Chandrasekaran:2022cip, Chandrasekaran:2022eqq, Witten:2023qsv, Witten:2023xze, ShadiCrossedProduct, kudlerflam2023covariantregulatorentanglemententropy, Jensen:2023yxy, Akers:2024bel, Kudler-Flam:2024psh, Chen:2024rpx, DeVuyst:2024fxc}. Of particular note for our purposes in this work is the concept of \emph{modular crossed product}, which we now review.\footnote{A formal, mathematically rigorous introduction to algebraic quantum field theory is provided in the classic textbooks~\cite{Takesaki:1970aki, Haag:1996hvx, takesaki2002theory}. Review articles covering the aspects of the subject of most relevance to recent applications in theoretical physics can be found in~\cite{Witten:2018zxz, Sorce:2023fdx}.} The importance of the modular crossed product in our setup will become apparent shortly, due to its role in motivating a preferred form of dressing of bulk observables in JT gravity.

  The basic concept we start with is that of a von Neumann algebra $\mathcal{A}$, which can roughly be thought of as any subalgebra of the set of bounded operators $\mathcal{B}(\mathcal{H})$ acting on some Hilbert space $\mathcal{H}$ that is closed under taking adjoints and is also closed in the weak operator topology.\footnote{For an explicit characterization of all the defining properties of a von Neumann algebra, see e.g.~\cite{Witten:2018zxz}.} The algebra $\mathcal{A}$ will in general possess a nontrivial commutant $\mathcal{A}'$, consisting of all the elements of $\mathcal{B}(\mathcal{H})$ which commute with $\mathcal{A}$. We also need a vector $\ket{\Psi}\in\mathcal{H}$ that is \emph{cyclic} and \emph{separating} for $\mathcal{A}$, where ``cyclic for $\mathcal{A}$'' means that the set $\{a\ket{\Psi}\,\,\big|\,\,a\in\mathcal{A}\}$ is dense in $\mathcal{H}$, and ``separating for $\mathcal{A}$'' means that $\ket{\Psi}$ is cyclic for $\mathcal{A}'$. The notion of a cyclic and separating state is the appropriate algebraic generalization of a maximum-rank entangled state, applicable even in situations where density matrices are not well-defined. This is what happens, for instance, when dealing with the algebra of operators of a QFT restricted to some finite subregion, which is formally described by a von Neumann algebra of type III~\cite{Araki:1964lyc, Driessler:1976ky, Witten:2018zxz, Sorce:2023fdx}. 
  
  Given a von Neumann algebra $\mathcal{A}$ and a cyclic-separating vector $\ket{\Psi}$, one can always define the so-called \emph{modular operator} $\Delta_\Psi$, and subsequently the modular Hamiltonian \mbox{$h_\Psi\coloneqq -\log\Delta_\Psi$}. From $\Delta_\Psi$ one then constructs the notion of \emph{modular flow}, which is the one-parameter group of transformations that acts as\footnote{The way we are introducing the modular operator here is admittedly a bit too abstract, as the details of its definition are not strictly needed for the message of this paper. For a physics-oriented introduction to the basic features of the modular operator and modular flow, see~\cite{Sorce:2023gio}.} 
  
  \begin{equation}\label{eq:modularflow}
      a\mapsto \sigma_s(a) \equiv a_s= \Delta_\Psi^{-is}a\,\Delta^{is}_\Psi, \,\,\,\forall a\in\mathcal{A}.
  \end{equation}
   A key result in modular theory is that modular flow preserves the algebra $\mathcal{A}$: that is, for every $a\in \mathcal{A}$, one also has $\Delta_\Psi^{-is}a\,\Delta^{is}_\Psi\in\mathcal{A}$. In short, modular flow is an \emph{automorphism} of the algebra $\mathcal{A}$, with $\Delta_\Psi^{-is}\mathcal{A} \Delta_\Psi^{is}= \mathcal{A}$. 
   This fact is known in the von Neumann algebra literature as \emph{Tomita's theorem}~\cite{SORCE2024110420}. The parameter $s$ that evolves along modular flow is usually referred to as ``modular time.'' 
  
  Now imagine that, in addition to the Hilbert space $\mathcal{H}$ and the algebra $\mathcal{A}$, we have access to an auxiliary Hilbert space $L^2(\mathbb{R})$ and its associated algebra of observables $\mathcal{B}(L^2(\mathbb{R}))$. One can picture the auxiliary Hilbert space as the space of square-integrable functions of a ``position'' variable $\hat{q}$, and the algebra $\mathcal{B}(L^2(\mathbb{R}))$ as a von Neumann algebra constructed out of bounded functions of $\hat{q}$ and its conjugate momentum $\hat{p}$. The \emph{modular crossed product} is the operation that takes as input the von Neumann algebras $\mathcal{A}$ and $\mathcal{B}(L^2(\mathbb{R}))$, together with the modular operator $\Delta_\Psi$, and returns as output the smallest subalgebra of $\mathcal{A}\otimes \mathcal{B}(L^2(\mathbb{R}))$ that contains the operator
\begin{equation*}
    e^{i\hat{q}t}, \,\,\, t\in\mathbb{R},
\end{equation*}
as well as 
\begin{equation*}
    e^{ih_\Psi\hat{p}}\,a\,e^{-ih_\Psi\hat{p}}, \,\,\,a\in\mathcal{A}.
\end{equation*}
The result of this operation is called the \emph{crossed product algebra}, denoted as $\mathcal{A}\rtimes\mathbb{R}$. We can write this slightly more explicitly as
\begin{equation}\label{eq:dressedquantumtimeMCP}
\mathcal{A}\rtimes\mathbb{R} \coloneqq \left\{e^{ih_\Psi\hat{p}}\,a\,e^{-ih_\Psi\hat{p}}, e^{i\hat{q}t} \,\,\big|\,\, a\in\mathcal{A}, t \in \mathbb{R}\right\}'',
\end{equation}
where here $(\cdot)''$ denotes the double commutant.  

A somewhat intuitive way to understand the crossed product is that we are promoting the modular time (that is, the parameter $s$ that evolves along modular flow) to an operator $\hat{p}$, and adjoining to the original system the Hilbert space where $\hat{p}$ and its canonical conjugate $\hat{q}$ act. The resulting algebra is then generated by quantum-controlled modular evolution of operators in $\mathcal{A}$, together with translations of the ``quantum modular time'' $\hat{p}$ (which are of course generated by its canonical conjugate $\hat{q}$). In this sense, the auxiliary Hilbert space $L^2(\mathbb{R})$ gains the interpretation of the Hilbert space associated to a ``quantum clock,'' and the crossed product algebra~\eqref{eq:dressedquantumtimeMCP} can then be understood as the algebra of observables ``dressed to the time measured by the clock.''

A very important feature of $\mathcal{A}\rtimes\mathbb{R}$ is that it is also the subalgebra of $\mathcal{A}\otimes \mathcal{B}(L^2(\mathbb{R}))$ that commutes with $h_\Psi + \hat{q}$: in other words,
\begin{equation}\label{eq:gaugeinvariantopMCP}
    \mathcal{A}\rtimes\mathbb{R} = \left\{\hat{a}\in \mathcal{A}\otimes \mathcal{B}(L^2(\mathbb{R}))\,\,|\,\,[\hat{a}, h_{\Psi} + \hat{q}] = 0\right\}.
\end{equation}
This means that the crossed product algebra can be equivalently defined as the subset of operators in $\mathcal{A}\otimes \mathcal{B}(L^2(\mathbb{R}))$ that are invariant under the flow generated by $h_\Psi + \hat{q}$. The observation that the definitions~\eqref{eq:dressedquantumtimeMCP} and~\eqref{eq:gaugeinvariantopMCP} are in fact equivalent is known as the commutation theorem for the modular crossed product~\cite{Daele_1978, Klinger:2023auu}.

The modular crossed product has been at the core of many important developments in theoretical high energy physics in recent years. First off, it has been shown to provide a notion of entropy of local regions in quantum field theory that regulates the pervasive UV divergences of entanglement entropy in QFT and exists in the full continuum theory~\cite{ShadiCrossedProduct, kudlerflam2023covariantregulatorentanglemententropy}. This allows for a rigorous treatment of questions related to entanglement entropy in QFT using a regularization method that is fully covariant and does not rely on artefacts such as discretized versions of the theory on a lattice, which often obscure the Lorentz-invariant nature of relativistic QFTs. Secondly, and perhaps most importantly, the crossed product algebra has appeared in the context of quantum gravity as the correct algebra to assign to local subregions in the semiclassical regime~\cite{WittenCrossedProduct, Chandrasekaran:2022cip, Chandrasekaran:2022eqq, Jensen:2023yxy}. This results from a careful analysis of the consequences of imposing diffeomorphism invariance as a gauge symmetry in quantum gravity, which yields important modifications to the nature of local observables even in the semiclassical limit where fluctuations of the background geometry are suppressed. In those settings, the introduction of the auxiliary Hilbert space that appears in the modular crossed product becomes quite natural: it is motivated by the fact that nontrivial observables in diffeomorphism-invariant theories can only be defined in relation to some other physical system, which should itself be equipped with its own internal degrees of freedom that transform under diffeomorphisms for the joint system to display diffeomorphism invariance. 

This central result---namely, that the modular crossed product is also the algebra of diffeomorphism-invariant operators on a subregion in semiclassical gravity---hinges on the assumption that the modular flow generated by $h_\Psi$ can be understood as the action of some diffeomorphism on a fixed geometry.\footnote{To be slightly more precise, one generically has to assume that modular flow is \emph{instantaneously} geometric---i.e., that the modular flow generated by $h_\Psi$ matches the geometric flow by some diffeomorphism in a vicinity of a partial Cauchy slice for the causally complete subregion associated to the algebra $\mathcal{A}$. For more details, see e.g.~\cite{Jensen:2023yxy}.} This is a highly nontrivial assumption, since the modular Hamiltonian for a completely general quantum state is a rather nonlocal object, with no connection \emph{a priori} with any form of geometric flow. There are, however, very important examples in QFT where one can draw a direct connection between modular time and a timelike coordinate in some Lorentzian geometry. These include (i) when $\ket{\Psi}$ is the vacuum state in Minkowski space and $\mathcal{A}$ is the algebra of field operators restricted to one of the Rindler wedges, in which case modular flow reduces to Lorentz boosts; (ii) when $\ket{\Psi}$ is the Hartle-Hawking state in a two-sided Schwarzschild black hole and $\mathcal{A}$ is the algebra of field operators restricted to one of the black hole exterior regions, in which case modular flow corresponds to the flow by the horizon-generating Killing vector field of the exterior geometry; and (iii) when $\ket{\Psi}$ is the Bunch-Davies state in de Sitter space and $\mathcal{A}$ is the algebra of operators in the static patch, in which case modular flow reduces to the flow by the timelike isometry of the static patch. For a more detailed explanation of each of these cases, see e.g.~\cite{Bisognano:1975ih, Bisognano:1976za, Sewell:1982zz, Kay:1988mu}. These are paradigmatic examples that laid the foundation for much of the recent progress on von Neumann algebras in quantum gravity, and guide much of the intuition behind the connection between modular flow and thermal physics. 

Slightly more generally, if one has a family of diffeomorphisms whose flows can be matched to the modular flow of a family of states, then starting with an algebra of field operators in some fixed subregion and making it invariant under this family of diffeomorphisms should directly lead to an algebra of dressed observables which has the structure of modular crossed product. This then begs the question of whether one can define an algebra of local observables in a theory with dynamical gravity that (i) is invariant under a large class of diffeomorphisms, and (ii) contains a dressing that connects to the modular crossed product and is also geometric. 

JT gravity seems to provide an ideal setup to study this question, since the dynamical degree of freedom in quantum JT gravity---i.e., the Schwarzian mode---can be mapped precisely to a timelike parameter in the classical theory (namely, the Poincar\'e time of the boundary curve), which in turn connects to the quantum degree of freedom corresponding to the quantum version of the modular time. In fact, at the semiclassical level, we have some understanding of how this works: for instance, in JT gravity with two asymptotic boundaries, the algebra of operators supported on one of the boundaries (say, the right one) is known to be type II~\cite{Penington:2023dql, Kolchmeyer:2023gwa}; moreover, in the high-temperature limit (which is the semiclassical limit in JT gravity), this algebra encodes, via entanglement wedge reconstruction, the algebra of bulk operators in the right exterior of the two-sided AdS$_2$ black hole, dressed with the modular crossed product by the modular automorphism group of the thermofield double state. We are thus interested in how to take this interpretation further into the bulk more explicitly, in a way that also allows us to go beyond the semiclassical regime.

Our goal in the next section will be to take a first step towards realizing this vision. We will do this by showing that a very elementary property of any geometric modular flow---namely, that it must be implemented by the flow of a conformal isometry of the background spacetime~\cite{SorceGeometricModularFlow}---is already enough to single out a preferred form of mapping from points in the bulk to the Schwarzian mode in JT gravity. Curiously enough, this will lead precisely to the gravitational dressing employed in earlier works in the literature~\cite{Blommaert:2019hjr,Mertens:2019bvy,Blommaert2021}, thus reinforcing the motivation for that choice of dressing and laying the foundation for the calculation that we will perform later in Sec.~\ref{sec:bhta}.

\section{Geometric dressing and fiducial observers}\label{sec:geometricdressing}
Let $U = T+Z$ and $V=T-Z$, where $T$ and $Z$ are the time and radial coordinates of the Poincar\'e patch of AdS$_2$,
\begin{equation}\label{eq:PoincareAdS}
    ds^2 = \dfrac{-dT^2 +dZ^2}{Z^2} = \dfrac{-4dU\,dV}{(U-V)^2}.
\end{equation}
The asymptotic boundary of the Poincar\'e patch sits at $U=V$ (or $Z=0$). 

Our goal is now to set the stage to talk about diffeomorphism-invariant quantities in the Poincar\'{e} patch of AdS. The way we do this, following tradition, is by defining points in the bulk relationally in terms of boundary-intrinsic quantities; in short, we ``dress'' bulk points to the boundary.

The definition of points in the bulk through some anchoring to the boundary can generally be framed as a recipe that assigns
\begin{align}
    U &= \chi(\tau, r), \label{eq:udressing}\\
    V &= \psi(\tau, r), \label{eq:vdressing}
\end{align}
where we take $\tau$ to be a proper time parameter for the boundary trajectory, and $r$ is a second parameter that we will later understand as a proxy for the radial displacement of the bulk point. Formulated this way, a dressing can be simply interpreted as a diffeomorphism connecting an old set of coordinates (in this case, lightcone coordinates in the Poincar\'e patch) to a new pair of coordinates denoted here by $\tau$ and $r$. General diffeomorphism invariance would seem to suggest that the functions $\chi$ and $\psi$ can be completely arbitrary. In order to proceed, however, we will propose a criterion that singles out a preferred form of dressing (or alternatively, a preferred class of frame) that has structurally nice properties and is naturally adapted to describe bulk points from a boundary perspective, as we will now argue. 

The key principle we propose is that the mapping from $(\tau, r)$ to $(U, V)$ must be such that the boundary time $\tau$ extends into the bulk in a way that makes translations in $\tau$ a conformal isometry of AdS$_2$. In other words, the geometric dressing is such that $\partial_\tau$ is a conformal Killing vector of the bulk geometry. The motivation behind this proposal comes from recent developments connecting the construction of diffeomorphism-invariant algebras of observables in quantum gravity and the modular crossed product, as mentioned in section~\ref{sub:modularcrossedproduct}. Conformal Killing vectors are a necessary ingredient in geometric modular flow~\cite{SorceGeometricModularFlow}, so we believe it is natural to demand that they also form part of the construction of a preferred geometric dressing if we want such dressing to bear any chance of being interpretable in connection to the modular crossed product. 

Let $\mathcal{M}$ be a differentiable manifold with metric tensor $g$, and let $\varphi_t: \mathcal{M\to\mathcal{M}}$ be a one-parameter family of diffeomorphisms that generates a vector field $\xi$ via differentiation with respect to the parameter $t$. If such geometrically local flow described by the family of diffeomorphisms $\{\varphi_t\,|\,t\in \mathbb{R}\}$ is unitarily realized in a QFT defined on $\mathcal{M}$, then there must exist a one-parameter family of unitaries $U(t)$ acting on field operators as\footnote{This is a slightly more informal sketch of the proof presented in~\cite{SorceGeometricModularFlow}.}
\begin{equation}\label{eq:fieldmappingdiffeo}
U^\dagger(t) \phi(x)U(t) = \Lambda(\varphi_t x)\,\phi(\varphi_t x),
\end{equation}
where $\varphi_t x$ is the image of the point $x$ under the diffeomorphism $\varphi_t$, and $\Lambda$ is some smooth function on $\mathcal{M}$. 

\begin{figure}[h]
    \centering
    \includegraphics[width=0.25\textwidth]{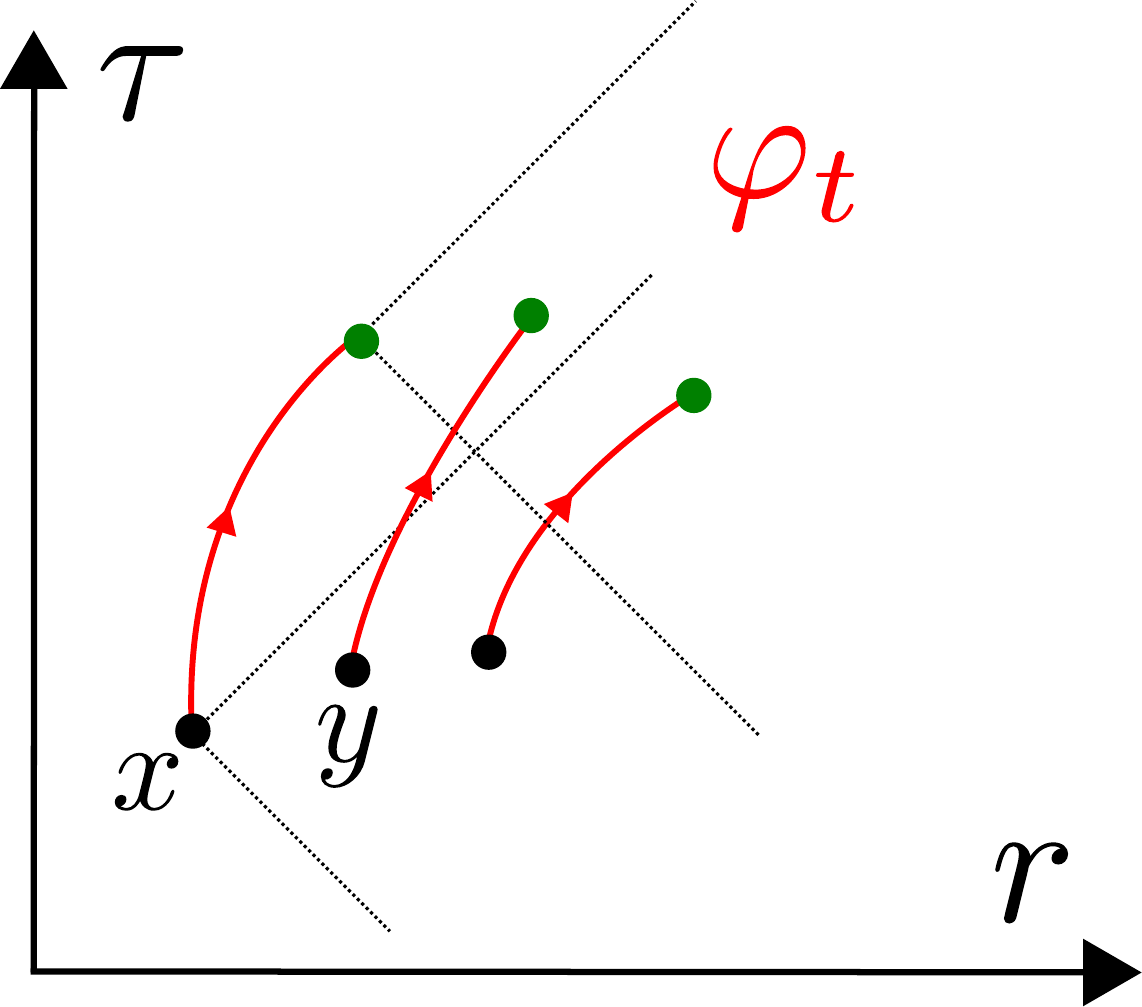}
    \caption{Example of a one-parameter family of diffeomorphisms $\varphi_t$ (red) that maps spacelike separated points (black dots) into spacelike separated points (green dots) in the same coordinate chart.}
    \label{fig:conflow}
\end{figure}

One thing to note at this point is that this family of diffeomorphisms $\{\varphi_t\,|\,t\in \mathbb{R}\}$ maps commuting field operators into commuting field operators: if $[\phi(x),\phi(y)] = 0$, then Eq.~\eqref{eq:fieldmappingdiffeo} immediately leads to $[\phi(\varphi_t x), \phi(\varphi_t y)] = 0$. In a relativistic QFT, any pair of spacelike separated operators must commute (this is just the requirement of microcausality that prevents superluminal signalling), but operators supported on null separated points do not commute. Therefore, the observation above means that whenever $x$ and $y$ are spacelike separated, their images $\varphi_t x$ and $\varphi_t y$ are also spacelike separated (in the same chart as the original coordinates $x$ and $y$); see Fig.~\ref{fig:conflow} for an illustration.

Now, a fundamental theorem in general relativity~\cite{Peleska1984} states that if a diffeomorphism maps spacelike separated points into spacelike separated points, then the transformed metric $\varphi_t^* g_{ab}$ must be related to the old metric $g_{ab}$ by a Weyl (local scale) transformation; in other words, $\varphi_t^\ast g = \Omega^2g$ for some function $\Omega$ (which can depend on the spacetime point and on the parameter $t$). Applying this for all values of $t$, we conclude that the family $\{\varphi_t\,|\,t\in \mathbb{R}\}$ is generated by the flow of a conformal Killing vector field, and each diffeomorphism $\varphi_t$ acts as a conformal transformation on $\mathcal{M}$. 
 
 In slightly more practical terms, starting from the metric~\eqref{eq:PoincareAdS} in lightcone coordinates $(U, V)$, a conformal transformation can be understood as a map to a new coordinate frame $(u, v)$ such that, in terms of the new coordinates, the metric retains the form~\eqref{eq:PoincareAdS} up to an overall prefactor. Now, for a given diffeomorphism $\varphi$ mapping $(u, v)$ to $(U, V)$ via $U = U(u, v)$ and $V=V(u, v)$, we have that
\begin{equation}\label{eq:changelineelement}
    dU\,dV = (\partial_u U\,\partial_vV + \partial_v U\partial_u V)du\,dv + \partial_u U\partial_u Vdu^2 + \partial_vU \partial_v V dv^2.
\end{equation}
This will correspond to a conformal transformation on the metric~\eqref{eq:PoincareAdS} if the right-hand side of the equation above is proportional to $du\,dv$, in which case the coordinate transformation can be written in terms of a pair of chiral functions $U=F(u)$ and $V=G(v)$.\footnote{Technically, a transformation of the form $U=U(v)$ and $V=V(u)$ would also meet this requirement. This additional possibility is superfluous, however, since we are always allowed to swap the roles of $u$ and $v$. Alternatively, if we think not just in terms of a single diffeomorphism but rather a one-parameter group of diffeomorphisms that is continuously connected to the identity (as we should do if the diffeomorphisms are induced by the flow of some vector field), and we take the identity map to be $U=u$ and $V=v$, then we are also immediately led to a solution of the form $U=F(u)$ and $V=G(v)$.} If we then demand that the diffeomorphism preserves the conformal boundary of AdS$_2$---in other words, that $U=V$ is mapped to $u=v$ in the new coordinates---then the two functions $F$ and $G$ are constrained to be equal, and we obtain the most general solution
\begin{equation}
U = F(u), \qquad V=F(v),
\end{equation}
for a single reparametrization $F(.)$. We thus conclude that the group of conformal Killing vector fields that are relevant for our construction is just Diff$^+(\mathbb{R})$, the group of all differentiable and invertible time reparametrizations. This has a subgroup PSL$(2,\mathbb{R})$ which are the genuine Killing vectors of the bulk AdS$_2$ geometry.

Now that we have established what to expect from the most general form of conformal transformation that is tangent to the AdS boundary, it is instructive to see more explicitly how to connect the new frame $(u, v)$ to the dressing parameters $(\tau, r)$ that we introduced earlier. In general, given a dressing of the form ~\eqref{eq:udressing} and~\eqref{eq:vdressing}, the flow by $\tau$ for fixed $r$ generates the vector field
\begin{equation}\label{eq:velocityvectorfield}
    \dot{\gamma} = \left(\dfrac{\partial \chi}{\partial \tau}\right)_r\partial_U+\left(\dfrac{\partial \psi}{\partial \tau}\right)_r \partial_V,
\end{equation}
where $\left(\frac{\partial(\cdot)}{\partial \tau}\right)_r$ denotes a partial derivative with respect to $\tau$ while keeping $r$ fixed. (This is to distinguish from $\partial_U$ and $\partial_V$, which are taken while holding $V$ and $U$ fixed, respectively.)

On the other hand, a general conformal Killing vector $\xi$ on $d$-dimensional a spacetime with metric $g$ satisfies the equation
\begin{equation}
    \nabla_{(a} \xi_{b)} = \dfrac{1}{d}(\nabla_c\xi^c)\,g_{ab}.
\end{equation}
For the metric~\eqref{eq:PoincareAdS}, this admits a general solution given by
\begin{equation}\label{eq:generalCKV}
    \xi = f(U)\partial_U + g(V)\partial_V,
\end{equation}
where $f$ and $g$ are two functions which we assume to be differentiable but are otherwise arbitrary. Constraining $\dot{\gamma}$ in Eq.~\eqref{eq:velocityvectorfield} to be of the form~\eqref{eq:generalCKV} then leads to
\begin{align}
    \left(\dfrac{\partial \chi}{\partial \tau}\right)_r = f(\chi), \label{eq:constraint1}\\
    \left(\dfrac{\partial \psi}{\partial \tau}\right)_r = g(\psi). \label{eq:constraint2}
\end{align}
One can directly integrate Eqs.~\eqref{eq:constraint1} and~\eqref{eq:constraint2} to obtain
\begin{align}
    h(\chi(\tau, r)) \equiv \int^{\chi(\tau, r)}\dfrac{dq}{f(q)}= \tau + \alpha(r), \label{eq:handfintegral1} \\
    p(\psi(\tau, r)) \equiv \int^{\psi(\tau, r)}\dfrac{dq}{g(q)}= \tau + \beta(r), \label{eq:handfintegral2}
\end{align}
where $\alpha$ and $\beta$ are arbitrary functions of the radial parameter $r$. This gives us the general form
\begin{align}
    \psi(\tau, r) = F(\tau + \alpha(r)),\\
    \chi(\tau, r) = G(\tau + \beta(r)),
\end{align}
where $F$ is the inverse of $h$, $G$ is the inverse of $p$, and $p$ and $q$ are primitives of $1/f$ and $1/g$ respectively, as made implicit in Eqs.~\eqref{eq:handfintegral1} and~\eqref{eq:handfintegral2}. 

It is natural to demand that the radial coordinate be such that $r=0$ is at the asymptotic boundary of the Poincar\'e patch of AdS$_2$. Since the asymptotic boundary corresponds to $U=V$, this constrains us to have
\begin{equation}
    F(\tau+\alpha(0)) = G(\tau + \beta(0)),
\end{equation}
or, by shifting $\tau \mapsto \tau - \beta(0)$,
\begin{equation}
    G(\tau) = F(\tau+\alpha(0)-\beta(0)).
\end{equation}
This means that the function $G$ is actually fully determined by $F$. Going back to the general expression of the dressing, we have thus far
\begin{align}
    U=\chi(\tau, r) &= F(\tau +\alpha(r)),\\
    V=\psi(\tau, r) &= F(\tau + \beta(r) + \alpha(0)),
\end{align}
where we redefined $\beta(r)-\beta(0)\mapsto \beta(r)$ and we now only consider functions $\beta$ with $\beta(0)=0$. By further shifting the origin of the $\tau$ coordinate by an amount $\alpha(0)$, we can also restrict to functions $\alpha(r)$ satisfying $\alpha(0)=0$. We can further constrain the functions $\alpha(r)$ and $\beta(r)$ by requiring that the vector field $d/dr$, obtained by varying $r$ while keeping $\tau$ fixed, be orthogonal to $d/d\tau$. This will yield $\alpha'(r)+\beta'(r)=0$, and therefore $\alpha(r)+\beta(r) = C$ for some constant $C$. Using the fact that $\alpha(0)=\beta(0)=0$, we conclude that we must have $C=0$, and thus we arrive at
\begin{align}
    U = F(\tau+\alpha(r)), \\
    V = F(\tau -\alpha(r)).
\end{align}
At this point, we can simply rename $\alpha(r)\equiv z$, and use $z$ as the new radial coordinate. This only leaves one nontrivial degree of freedom in the dressing, which is characterized by the function $F(x)$, as we had concluded earlier.

By setting $z=0$ and recalling $T = (U+V)/2$, we see that the relation between the Poincar\'{e} time coordinate $T$ and the boundary time $\tau$ is directly given by $T=F(\tau)$. Therefore, when applied to points at the asymptotic boundary of the Poincar\'e patch of AdS, $F$ can be interpreted as the reparametrization from boundary time to Poincar\'e time, as is commonly used when describing boundary dynamics in JT gravity. Our construction then amounts to extending this boundary time reparametrization into the bulk by defining a radial coordinate $z$ and using the null coordinates $u=\tau+z$ and $v=\tau-z$ as the new bulk frame, which is connected to the old $(U, V)$ frame by $U=F(u)$, $V=F(v)$, as we had previously established. 

This result precisely reconstructs the radar definition of a bulk frame based on boundary data in JT gravity proposed  in~\cite{Blommaert:2019hjr,Mertens:2019bvy,Blommaert2021}. 
Its geometrical interpretation is very simple. $\tau-z$ and $\tau+z$ correspond to the two null coordinates of a point with temporal coordinate $\tau$ and radial coordinate $z$ in two-dimensional Minkowski space. The construction above can then be pictured as anchoring a bulk point $p$ to the two points of the boundary that are reached by future- and past-pointing null geodesics emanating from $p$. If $\tau_1$ is the boundary time at which the incoming null geodesic leaves the boundary and $\tau_2$ is the boundary time at which the outgoing null geodesic hits the boundary, then the dressing above amounts to a definition of the radial coordinate as $z = (\tau_2-\tau_1)/2$. We can therefore summarize this construction by saying that, up to reparameterizations by arbitrary functions of $\tau_2-\tau_1$, this definition of a bulk radial coordinate is the only one that leads to boundary time translations being extended into the bulk as conformal isometries of the AdS$_2$ geometry.

\begin{figure}[h]
    \centering
    \includegraphics[width=0.65\textwidth]{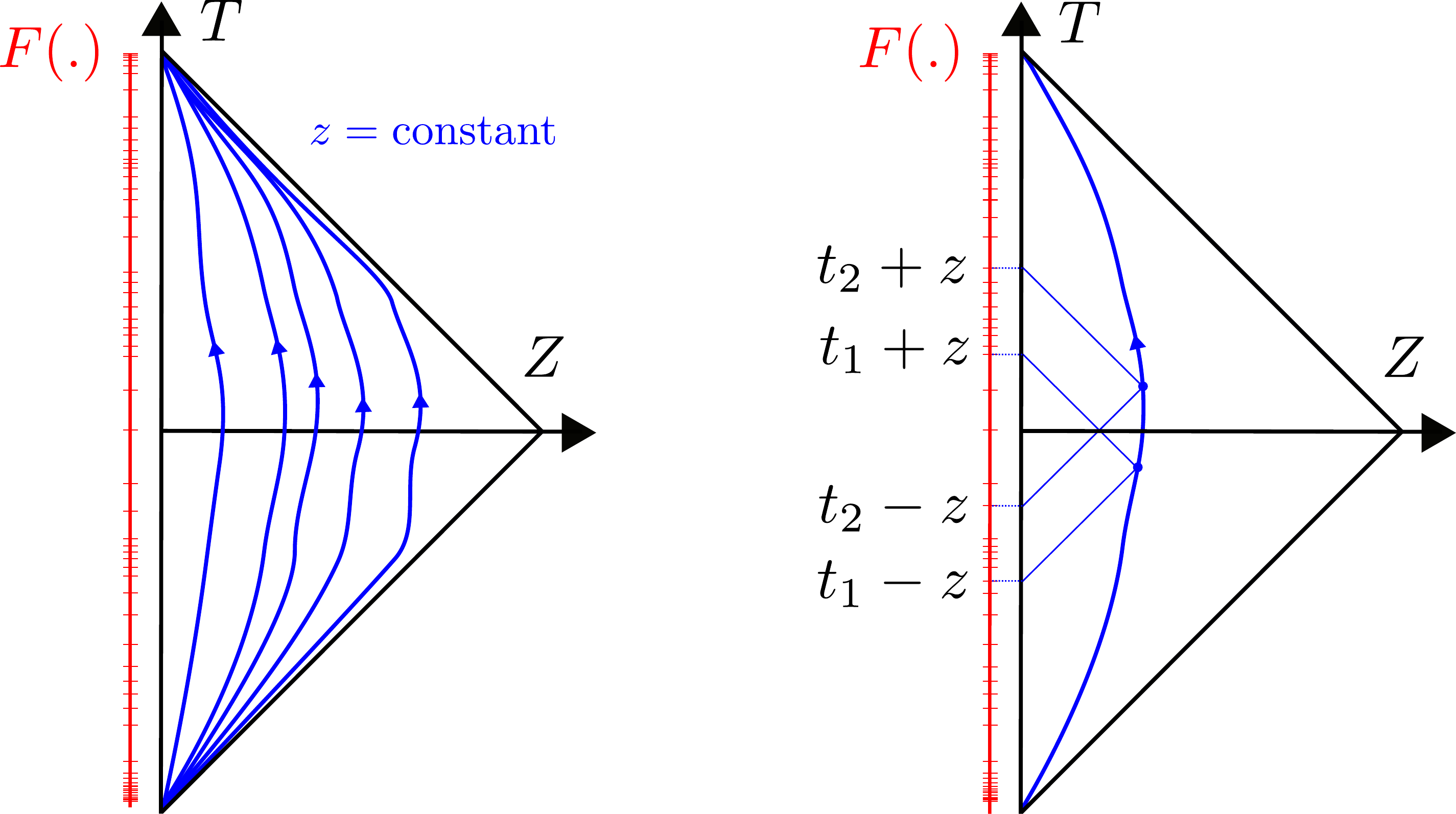}
    \caption{For a given off-shell time reparametrization $F(.)$ (represented here in red with a clock-ticking pattern), there is a unique conformal Killing vector field that limits to this time translation near the boundary. Left: The resulting flowlines can be viewed as ``wiggled'' versions of the semiclassical boost flowlines. Right: Each of these worldlines can be equivalently defined using the Einstein antenna construction \cite{Blommaert:2019hjr,Mertens:2019bvy, Blommaert2021} for a fixed $z$.}
    \label{fig:ConfKilling}
\end{figure}

This establishes our proposal for a fiducial observer in JT gravity for each off-shell configuration of the Schwarzian mode $F(t)$: one simply declares that the observer's location is labeled by the parameter $z$, which is a boundary-intrinsic quantity that is independent from the embedding of the boundary curve into AdS$_2$, and then the trajectory of the observer in the bulk is determined in lightcone coordinates $U$ and $V$ by $U=F(t+z)$ and $V=F(t-z)$, with $z$ fixed and $t$ ranging from $-\infty$ to $+\infty$. The resulting construction of the conformal Killing vector field and bulk frame is illustrated in Fig.~\ref{fig:ConfKilling}.
The wiggled flowlines limit to the Schwarzian wiggly curve close to $z \approx 0$, and limit to the black hole horizons as $z\to +\infty$. 

As a final remark, we note that in order for a given vector field $\xi$ to implement geometric modular flow, it is also necessary that $\xi$ be timelike and future-oriented~\cite{SorceGeometricModularFlow}. The timelike condition for a vector field of the form~\eqref{eq:velocityvectorfield} with $\chi(\tau, z) = F(\tau +z)$ and $\psi(\tau, z)=F(\tau-z)$ yields $F'(\tau+z)F'(\tau-z)>0$, and future-directedness implies $F'(\tau+z)>0$. Both conditions are thus automatically satisfied if we take the function $F(x)$ to be such that $F'(x)>0$ everywhere in its domain. Of course, the condition $F'\neq0$ is guaranteed if $F$ is a time reparametrization, and the condition that it is positive simply means that we are taking orientation-preserving diffeomorphisms on the real line.

\section{Application: black hole thermal atmosphere}
\label{sec:bhta}
As a physically interesting application of our construction of the class of fiducial operators in JT quantum gravity, we look at the computation of the entropy of the thermal atmosphere of quantum fields surrounding the black hole horizon.\footnote{Other applications of these gravitationally dressed observables were considered in \cite{Blommaert:2019hjr,Mertens:2019bvy, Blommaert2021, Blommaert:2020seb, DeVuyst:2022bua}. In \cite{Blommaert:2019hjr} a quantum-corrected metric was considered, as a measure for how fiducial observers measure distances close to the black hole horizon. Also bulk matter correlation functions were computed. Wormhole and non-perturbative corrections to these were studied in \cite{Blommaert:2020seb}. In \cite{Mertens:2019bvy} quantum gravity corrected occupation numbers in the thermal atmosphere were computed, which were later augmented by wormhole and non-perturbative corrections in \cite{Blommaert2021}. An application to the entanglement island prescription was presented in \cite{DeVuyst:2022bua}.} This calculation is famously divergent in QFT in curved spacetime approaches, and is usually regularized with the help of a ``brick wall'' cut-off at Planckian distance from the event horizon \cite{tHooft:1984kcu,Susskind:1994sm}. We explore in this section how the fiducial observer family in JT quantum gravity introduced in the previous sections improves on this calculation.

\subsection{At the disk level}\label{sub:disk}
We are ultimately interested in the effect of quantum corrections on the near-horizon behavior of the entanglement entropy of thermal matter surrounding the black hole. A first approximation of this quantity  is to compute the relevant entanglement entropy on the disk topology. This approach was explored in \cite{Mertens:2019bvy} where it was shown that the disk topology is not sufficient to change the leading behavior compared to the classical result. Since this work aims to generalize the approach used to derive this result, let us briefly review this computation while fixing conventions. 

One major hurdle towards computing entanglement entropies in quantum gravitational theories is the specification of the codimension-$2$ entangling surface in an intrinsic manner.  This problem reduces to finding a diffeomorphism-invariant definition of bulk points. This is precisely what we addressed in section~\ref{sec:geometricdressing}, where we concluded that a physically motivated way to pick a dressing to solve this puzzle is to specify a bulk point in terms of its null geodesic distances to a number of reference points on the boundary, implementing the famous Einstein radar definition of bulk frames. From a boundary observer's point of view, this is understood as reflecting lightrays off of a fictitious mirror placed at the bulk point and labeling the point by half the boundary time it takes to receive the reflected signal. 

In JT gravity this implementation takes a particularly simple form \cite{Blommaert:2019hjr}, since the JT path integral can be fully described in terms of the boundary reparametrization mode $F(t)$ of AdS$_2$ in Poincar\'e coordinates. Given such a reparametrization, the corresponding bulk metric is constructed as 
\begin{equation}\label{eq:AdSSchwarziandressing}
    ds^2 = \frac{F'(u) F'(v)}{(F(u) - F(v))^2} (- dt^2+dz^2),
\end{equation}
where the observer labels each bulk point by the two boundary clock times $t_1 \equiv v = t -z$ and $t_2 \equiv u = t + z$  at which he sends and receives the signal (after reflection).
The quantum gravitational path integral of scalar, diff-invariant, local operators $\mathcal{O}_h(u,v)$ is then realized by integrating the insertion $\mathcal{O}_h(f(u),f(v))$ over all inequivalent reparametrizations $F(\tau) \equiv \tan \frac{\pi}{\beta} f(\tau)$ weighted by the Schwarzian action as:
\begin{equation}
    \langle{\mathcal{O}_h}\rangle_\beta = \int_\mathcal{_\mathcal{M}} [\mathcal{D}f]\mathcal{O}[f] e^{C\int_0^\beta d\tau \left\{\tan \frac{\pi}{\beta}f,\tau\right\}}.
\end{equation}
We are interested in applying this formalism to the calculation of the entanglement entropy of matter minimally coupled to JT gravity. This means we consider a theory of JT gravity with matter where the interaction is mediated solely through the metric, and the action is given by
\begin{equation}
    S[\psi_{\text{mat}}, g, \Phi] = S_{\text{JT}}[g, \Phi] + S_{\text{mat}}[\psi_{\text{mat}}, g],
\end{equation}
where $\psi_{\text{mat}}$ denotes the collection of matter fields, and $S_{\text{mat}}$ is the matter action. 

Our goal now is to compute the entanglement entropy of matter in the Poincar\'e vacuum, with the location of the entangling surface on the Cauchy slice being specified via the radar construction. For concreteness, we will assume that the matter theory is described by a two-dimensional CFT.
In this setup, and with a fixed geometry specified via the reparametrization mode $F(t)$, the entanglement entropy of matter can be computed as
\begin{equation}
\label{eq:opint}
    S_\text{ent} = \frac{c}{12} \ln \frac{(F(u) - F(v))^2 }{\delta^2 F'(u)F'(v)},
\end{equation}
where $c$ is the central charge of the CFT, and $\delta$ is a UV cutoff that smears the endpoints of the partial Cauchy slice by a distance $\delta$ as measured by the observer in the coordinate frame $(u,v)$ \cite{Holzhey:1994we}. The quantum gravitational matter entanglement entropy is then computed by evaluating the Schwarzian path integral with the expression in Eq.~\eqref{eq:opint} being added as an operator insertion.  We will regularize the entanglement entropy by removing this $\delta$ contribution, which can be interpreted to stem from the infinite number of degrees of freedom per unit volume. 

A first value for this entropy is derived from the thermal saddle $F(t) = \tanh \frac{\pi}{\beta} t$, leading to
\begin{equation}
    S_{\text{ent}} + \frac{c}{6}\delta = \frac{c}{6}\ln \frac{\beta}{\pi} \sinh \frac{\pi}{\beta}(2z) ,
\end{equation}
which scales as
\begin{equation}
    S_{\text{ent}} + \frac{c}{6}\delta \, \simeq \,\frac{c}{6}\frac{2 \pi }{\beta} z
\end{equation}
as the entangling surface is moved towards the black hole horizon at $z \to + \infty$ \cite{Mertens:2019bvy}. This is exactly the volume divergence mentioned above, which is normally cured by introducing a brick wall---i.e., a cutoff close to the horizon. In what follows, we will show that this brick wall regularization of the volume divergence is not necessary in JT quantum gravity; instead, the inclusion of wormholes in the gravitational path integral removes this divergence automatically. 

As a first step towards this, let us see how the inclusion of gravitational effects on the disk topology affects this result. It turns out that the gravitational path integral with this operator insertion is of a particularly nice form, allowing for an exact and explicit computation of this quantity far beyond the semiclassical regime.\footnote{This observation was first made in~\cite{Yang:2018gdb} in the context of a calculation of the Einstein-Rosen bridge length.} At a technical level, using the identity $\ln(x) = \lim_{h \to 0} \partial_h x^h$, we can rewrite
\begin{equation}\label{eq:matterentropyexpval}
    \expval{S_\text{ent}}_\beta +  \frac{c}{6} \ln \delta = -\frac{c}{12} \left\langle \lim_{h\to 0} \partial_h \left(\frac{F'(u)F'(v)}{(F(u) - F(v))^2 }\right)^h\right\rangle_\beta, 
\end{equation}
which one can identify as the derivative of the (analytically continued) boundary two-point function in Schwarzian QM with respect to the conformal weight $h$ such that 
\begin{align}
\label{eq:Strick}
    \expval{S_\text{ent}}_\beta + \frac{c}{6}\ln{\delta} &= -\frac{c}{12} \lim_{h \to 0}\partial_h\expval{\mathcal{O}_h(u,v)}_{\beta}.
\end{align}
Inserting the well-known bilocal correlator on the disk \cite{Mertens:2017mtv} in terms of the Schwarzian density of states $d\mu(k) =dk k\sinh(2\pi k)$ and the Schwarzian partition function $Z(\beta) = (2\pi C /\beta)^{3/2} e^{2\pi^2C/\beta}$ then leads to the explicit double integral expression
\begin{equation}
\label{eq:entdisk}
     S_\text{ent} + \frac{c}{6}\ln{\delta} = - \frac{c}{6\pi^2 Z(\beta)}\int d\mu(k_1)d\mu(k_2)  e^{2iz \frac{k_1^2}{2C}} e^{- \left(  \beta + 2iz \right) \frac{k^2_2}{2C}} \frac{\partial  }{\partial h }\left[\frac{\Gamma\left(  h \pm i k_1 \pm i k_2 \right) }{(2C)^{2h} \Gamma\left(  2h \right)} \right]_{h = 0}.
\end{equation}
While a general analytic solution to this integral is unknown, progress can be made in the large (macroscopic) black hole regime $\beta \ll C$ in the limit where one approaches the horizon, $C \ll z$. In this limit, the leading-order contribution to the entanglement entropy can be computed via a saddle point argument. We review some technical details of this computation in Appendix \ref{app:appdisk}. The result to leading order in the $\beta \ll C \ll z$ regime is
\begin{equation}
\expval{S_{\text{ent}}}_\beta -\frac{c}{6}\ln \delta \, \simeq \, \frac{c}{6}\frac{2 \pi}{\beta}z.
\end{equation}
One thus finds that the linear divergence of the matter entanglement entropy as the entangling surface is moved towards the horizon persists far beyond the semiclassical regime for large black holes. The quantity $\lim_{z\to +\infty }z$ is directly interpreted as the total volume of space close to the black hole horizon in tortoise coordinates.

\subsection{Beyond the disk}\label{sub:highertopology}
The result presented at the end of the previous section contains all quantum corrections to the Schwarzian path integral with fixed (disk) topology. However, the full JT gravity path integral also contains wormholes, which contribute to the correlators. As discovered in the seminal work \cite{Saad:2019lba}, the wormhole contributions can be treated in a genus expansion using perturbation theory in the parameter $e^{S_0 \chi}$, with $\chi$ the Euler characteristic of the surface, and $S_0$ the parameter multiplying the Einstein-Hilbert action (or the extremal black hole entropy when treating JT as coming from a higher-dimensional black hole). This genus expansion is in turn an asymptotic series, which needs to be non-perturbatively completed. This was done in \cite{Saad:2019lba} in terms of a specific double-scaled hermitian matrix integral.

\begin{figure}[h]
    \centering   \includegraphics[width=0.27\textwidth]{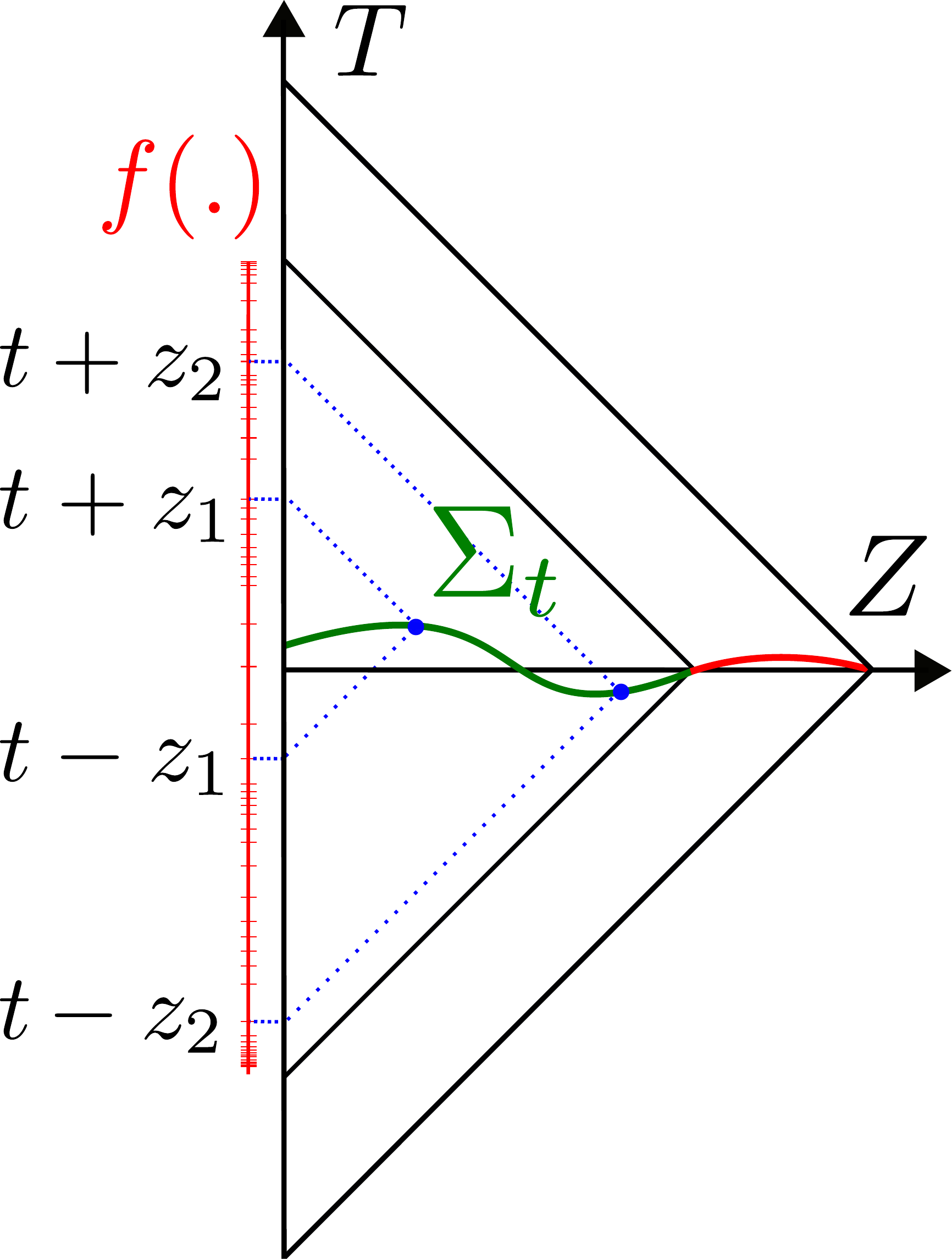}
    \caption{As the $z$-coordinate gets larger ($z_2 > z_1$), the anchoring points on the boundary are further apart, probing the IR strong quantum regime of the model. Ultimately as $z\to +\infty$, we are computing the matter entanglement entropy between the green and red segments of the Cauchy slice $\Sigma_t$ in the Poincar\'e vacuum, here drawn at time $t$ as specified by the anchoring.}
    \label{fig:EntEnt}
\end{figure}

At exponentially late times (in $S_0$), contributions from nontrivial topologies can and do dominate the path integral. This leads to a drastically different behavior of certain quantities in JT gravity, and is directly related e.g. to the emergence of the ramp and plateau regions in the spectral two-point function \cite{Saad:2018bqo,Saad:2019lba}. From a physical perspective, a similar behavior is to be expected for the matter entanglement entropy, since the radar construction of the bulk frame implies that moving the entangling surface exponentially close to the black hole horizon corresponds to an exponential separation in boundary time of the anchoring points. This is illustrated in Fig.~\ref{fig:EntEnt}.

From a more pragmatic standpoint, the geometric formulation of the matter entanglement entropy \eqref{eq:entdisk} as a derivative of the two-point function is quadratic in the spectral density $\rho_0(E) = e^{S_0}2C \sinh(2 \pi\sqrt{2CE}) \equiv e^{S_0}2C \sinh{2\pi k}$, and as such, it should be sensitive to the higher topology corrections of the spectral pair density correlator.  Assuming this geometric structure to persist beyond the disk thus automatically reduces this question to the calculation of two-point functions in the matrix model description of JT.  This problem was addressed in \cite{Saad:2019pqd}, where it was shown (building on \cite{Blommaert:2019hjr} and \cite{Saad:2018bqo}) that the inclusion of higher topology in JT gravity can be reproduced from a matrix model description via the inclusion of the exact spectral $n$-point density correlator in the Laplace transformation, while the matrix elements are chosen to match their disk values. 

\begin{figure}[h]
    \centering
    \includegraphics[width=0.65\textwidth]{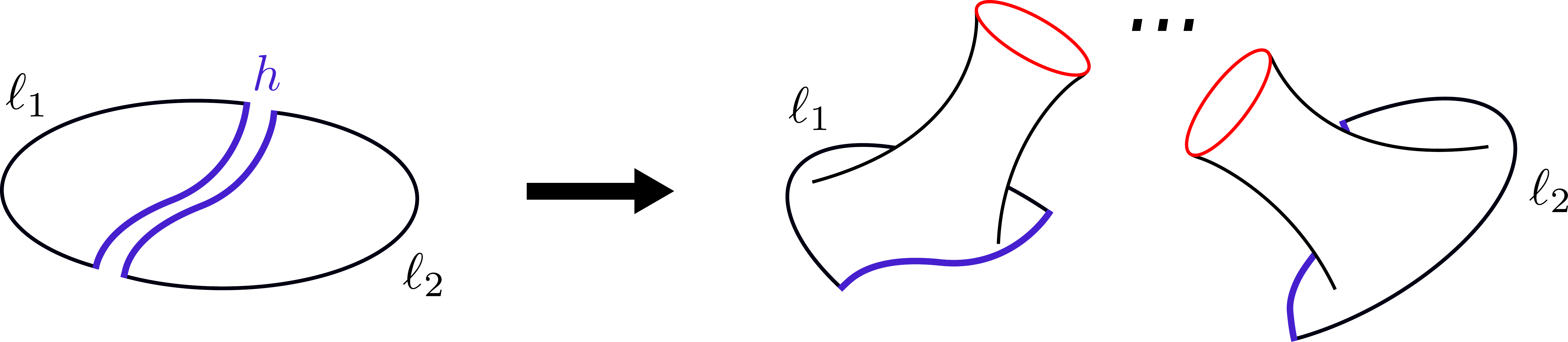}
    \caption{The disk surface is cut along the bilocal line into two disks. Each part then is decorated with wormholes, including connections between both pieces. These are the same wormhole contributions as those of the spectral form factor.}
    \label{fig:wormholes1}
\end{figure}

\begin{figure}[h]
    \centering
    \includegraphics[width=0.7\textwidth]{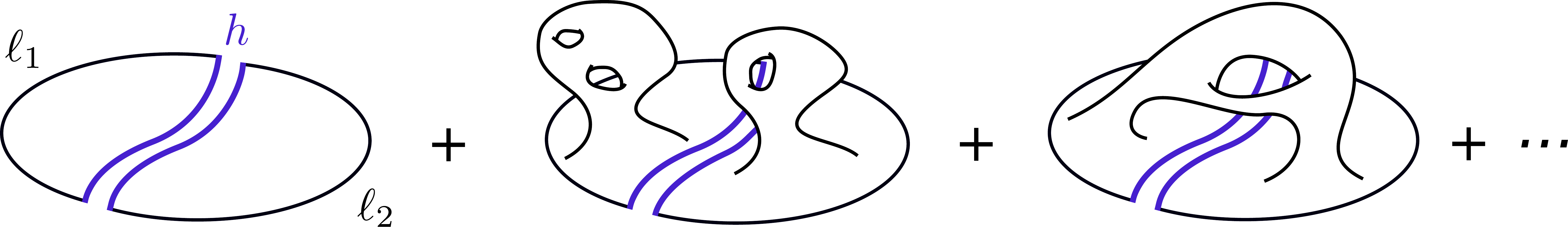}
    \caption{Coupling the gravitational chaotic system to a matter sector leads to a summation of diagrams where wormholes are included connecting the different sectors of the cut disk.}
    \label{fig:wormholes}
\end{figure}

More explicitly, the recipe of \cite{Saad:2019pqd} to replace $\rho_0 \ldots \rho_0 \to \langle \rho \ldots \rho\rangle_{\text{RMT}} $ was argued for there by an analysis based on ETH of the matter-gravity coupled system.\footnote{For more than two boundary operators, one also has to account for different ``classes'' of configurations as described generally in \cite{Blommaert:2020seb} and for the OTOC specifically in \cite{Stanford:2021bhl}.} Here we will describe it directly in terms of ``operator'' insertions in the underlying matrix integral. For the computation of $\langle \mathcal{O}_h\rangle$, the prescription can be viewed directly as a matrix integral insertion of\footnote{The precise JT matrix integral is a double-scaled matrix integral \cite{Saad:2019lba}, for which the specification of the matrix potential $V(M)$ is implicit in the choice of leading (disk) density of states $\rho_0(E)$. We will here use the notation as if it is a finite-dimensional matrix.}
\begin{equation}
\langle \mathcal{O}_h\rangle =\int [dM] \, \left(\sum_{i,j}e^{-\ell_1 \lambda_i}e^{-\lambda_j \ell_2} \frac{\Gamma(h \pm i\sqrt{\lambda_i} \pm i \sqrt{\lambda_j})}{\Gamma(2h)}\right) e^{- \text{Tr}V(M)},
\label{eq:operatormatrixmodel}
\end{equation}
where we denoted $\text{spectrum}(M) = \{\lambda_i, i=1 \hdots \text{dim}M\}$. This can be viewed as a sampling of the disk bilocal correlator at energy eigenvalues $\lambda_i$ that are now distributed according to the JT random matrix weight factor $e^{- \text{Tr}V(M)}$. This matrix operator insertion however is not writable immediately solely in terms of the matrix $M$. However, using the integral identity
\begin{equation}
\frac{\Gamma(h \pm i\sqrt{\lambda_i} \pm i \sqrt{\lambda_j})}{\Gamma(2h)} = \int_{-\infty}^{+\infty}d\phi \,e^{2h\phi} K_{2i \sqrt{\lambda_i}}(e^\phi)K_{2i \sqrt{\lambda_j}}(e^\phi),
\end{equation}
we can rewrite this as
\begin{equation}
\label{eq:wormholes}
{\color{darkblue}\int_{-\infty}^{+\infty}d\phi \, e^{2h\phi}}
\int [dM] \, \left(\text{Tr}\left[ e^{-\ell_1 M }K_{2i \sqrt{M}}(e^\phi)\right]\text{Tr}\left[ e^{-\ell_2 M }K_{2i \sqrt{M}}(e^\phi)\right]\right) e^{- \text{Tr}V(M)}.
\end{equation}
The quantity in black here is interpretable geometrically as a two-boundary correlator just like the spectral form factor (Fig.~\ref{fig:wormholes1}).
The Bessel-K insertions in the matrix integral are functions of the matrix $M$, which are defined as the formal series expansion in $\sqrt{M}$:
\begin{equation}
K_{2i \sqrt{M}}(e^\phi) = K_{0}(e^\phi) + 2i \partial_x K_x(e^\phi)\vert_{x=0} \sqrt{M} + \hdots
\end{equation}
The matrix integral over $M$ leads to the known topological expansion in wormholes of Fig.~\ref{fig:wormholes}.
Finally, the two disk-shaped segments are glued together in the full expression \eqref{eq:wormholes} due to the operator insertion with the $\phi$-integral.\footnote{The variable $\phi$ is related to the physical (renormalized) geodesic length $\ell$ as $\phi=-\ell/2 + \text{constant}$.} Higher topology is inserted in exactly the same way as for the spectral form factor simply because it is a double-trace matrix integral insertion. In more detail, we write the operator insertion as
\begin{equation}
\text{Tr}\left[ e^{-\ell M }K_{2i \sqrt{M}}(e^\phi)\right] = \sum_{i}e^{-\ell \lambda_i }K_{2i \sqrt{\lambda_i}}(e^\phi) = \int dE \, e^{-\ell E} \left(\sum_{i}\delta(E-\lambda_i)\right)K_{2i \sqrt{\lambda_i}}(e^\phi),
\end{equation}
leading to the $E_1,E_2$ integral of the quantity
\begin{equation}
\label{eq:wormholes2}
\int [dM] \, \rho(E_1)\rho(E_2) e^{- \text{Tr}V(M)} \equiv \expval{\rho(E_1)\rho(E_2)}_{\text{RMT}}, \qquad \rho(E) \equiv \sum_{i}\delta(E-\lambda_i),
\end{equation}
which is just the pair density correlator of the JT matrix ensemble. 

\begin{figure}[htpb]
    \centering
    \includegraphics[width=\textwidth]
    {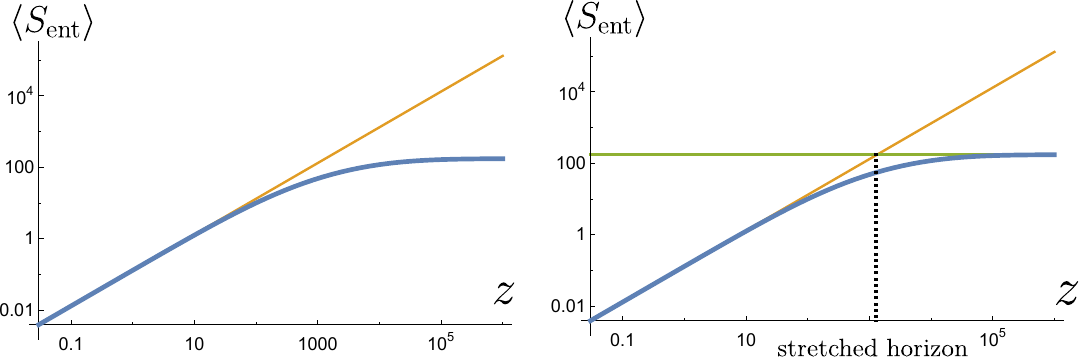}
        \caption{Left: Thermal atmosphere entropy (blue) $\langle S_{\text{ent}}\rangle$ as a function of the radial coordinate label $z$ (brick wall cut-off) for $\beta = 10$ and $C = 1$. The disk approximation (including the Schwarzian path integral) is depicted in orange. Right: The asymptotic finite value of the exact answer (green), and the effective definition of the stretched horizon at the radial location where one deviates from the linear growth.}
    \label{fig:numeric}
\end{figure}

Motivated by this, and following \cite{Saad:2019pqd, Iliesiu:2021ari}, we posit that the entanglement entropy should be extended to arbitrary topology and into the non-perturbative regime via the introduction of the exact matrix model spectral density correlator $\expval{\rho(k_1)\rho(k_2)}$ instead of the disconnected disk contribution $\rho_0(k_1)\rho_0(k_2)$. Explicitly 
\begin{equation}\label{eq:ententropyRMT}
    \expval{S_\text{ent}} + \frac{c}{6}\ln{\delta} = -\frac{c}{12} \frac{e^{-2S_0}}{C^2 Z(\beta)} \int dk_1 dk_2 \frac{k_1k_2 \expval{\rho(k_1) \rho (k_2)}}{\left(k_1^2 - k_2^2\right) \sinh\left(  \pi \left(  k_1 \pm k_2 \right)  \right) } e^{\frac{iz}{C} \left(  k_1^2 - k_2^2 \right)  -\frac{\beta}{2C} k_2^2},
\end{equation}
where 
\begin{equation}
\expval{\rho(E_1)\rho(E_2)}_{\text{RMT}} \approx \rho_0(E_1)\rho_0(E_2)  - \frac{\sin^2\left[  \pi \rho_0 (E_2) \left(  E_1 -E_2 \right)  \right] }{\pi^2 \left(  E_1 - E_2 \right)^2 } + \rho_0(E_1)\delta(E_1 - E_2), 
\end{equation}
for small energy separations which, since we are interested in the large $z$ limit, coincides with the dominant region of integration. A technical analysis of an analytic continuation (in the anchoring points of the two-point function) of exactly this integral was performed in the context of the geodesic wormhole length (or holographic complexity) in \cite{Iliesiu:2021ari}. 
We can thus import much of the technical analysis presented in~\cite{Iliesiu:2021ari} and mostly focus on the interpretation of the results and assumptions of this calculation in the following. A review of the technical details of this computation in our setting can be found in Appendix \ref{app:appnonpert}, where it is shown that in the $\beta \ll C\ll z$ limit the integral reduces to leading order to
\begin{align}
    \left\langle S_\text{ent} \right\rangle  + \frac{c}{6}\ln\left(  \delta \right)                                               &= \frac{c}{12} \frac{ e^{-2S_0}}{Z(\beta)}\frac{\pi}{3C^3} \int_{s_*}^{\infty} ds \frac{s^2 e^{-\frac{\beta}{2C} s^2} \rho_0(s)^3}{\sinh(2 \pi s)} \left[\frac{z}{\pi \rho_0(s)} - 1 \right]^3,
    \label{eq:nonpert}
\end{align}
with $s_* = \rho_0^{-1}\left(\frac{z}{\pi}\right)$.
In this form, the remaining integral can easily be solved numerically and the results of this are shown in Fig.~\ref{fig:numeric}. However, before we discuss these results, let us perform a sanity check on the above result and fix an ambiguity introduced by the calculation.

Firstly we may notice that we have neglected $z$-independent terms both in the calculation on disk topology and beyond. As mentioned in Appendix \ref{app:appnonpert}, such terms can be interpreted as a choice of calibration of the boundary observer's measurement apparatus. A natural choice is to define the (regularized) entanglement entropy to vanish as the entangling surface reaches the boundary i.e. both anchor points coincide. As such, we shift the entanglement entropy by its value in the $z \to 0$ limit. In this small $z$ limit,\footnote{Here small $z$ is the regime $\beta \ll C \ll z \ll Ce^{S_0}$.} we can also perform a sanity check on the above result matching the linear $z$ coefficient to the disk amplitude in the regime where topological contributions are suppressed. Expanding the integral in small $z$ to linear order directly reduces the problem to solving Gaussian integrals with non-trivial bounds. Explicitly one finds 
\begin{equation}
    \left\langle S_\text{ent} \right\rangle  + \frac{c}{6}\ln\left(  \delta \right) = C_0 + zC_1,
\end{equation}
with 
\begin{align*}
    C_0 &= - \frac{c}{36} e^{S_0}\left[\left(1 + \frac{16 C \pi^2}{\beta}\right)e^{6C \pi^2 /\beta} - e^{-2C \pi^2 / \beta} \right],\\
    C_1 &= \frac{c}{12} \left[ \frac{2}{C \pi} \sqrt{\frac{2 \pi C}{\beta}} e^{-2C \pi^2 /\beta} + \frac{1}{C \pi}\left(1 + \frac{4 C \pi^2}{\beta}\right) \text{Erf}\left(\sqrt{\frac{2 \pi^2 C}{\beta}}\right)\right].
\end{align*}
To match with the disk answer, we shift by $C_0$ and further approximate in the classical black hole limit $\beta \ll C$ to find
\begin{equation}
    \left\langle S_\text{ent} \right\rangle  + \frac{c}{6}\ln\left(  \delta \right) - C_0 \approx \frac{c}{6}\frac{2\pi}{\beta}z,
\end{equation}
which is indeed the desired match. 
Turning to the full (shifted) result shown in Fig.~\ref{fig:numeric} (left), we find that at exponentially late times the behavior is indeed drastically different from the disk answer. Most importantly, the entanglement saturates to $-C_0$ as the entangling surface reaches the horizon. This fully removes the need for the introduction of a brick wall cutoff ubiquitous with this type of calculation. 

There are several interesting limits one can consider:
\begin{itemize}
\item
Large (macroscopic) black holes correspond to the regime $C \gg \beta$, for which we obtain
\begin{equation}
\label{eq:largebh}
S \approx ce^{S_0}\frac{4\pi^2}{9}\frac{C}{\beta}e^{6\pi^2\frac{C}{\beta}}.
\end{equation}
This is a very large entropy, much larger than the bare area term $S_{BH} = 4\pi^2 \frac{C}{\beta}$. In the spirit of the early literature \cite{Susskind:1994sm,Solodukhin:2011gn}, these terms lead to a renormalization of Newton's constant as
\begin{equation}
S_{BH} = 4\pi^2 \frac{C}{\beta} + ce^{S_0}\frac{4\pi^2}{9}\frac{C}{\beta}e^{6\pi^2\frac{C}{\beta}} \equiv 4\pi^2 \frac{C_{\text{eff}}}{\beta},
\end{equation}
where the renormalized gravitational constant is defined through $C_{\text{eff}} \sim 1/G_{N,\text{eff}}$.
The fact that the matter contribution is very large might seem strange at first sight in the macroscopic limit. However, the thermal atmosphere gets entropy contributions all the way into the Planckian regime, and it is hence impossible to obtain a matter entropy that is both finite and semiclassical; the expression \eqref{eq:largebh} is non-perturbative and hence is not even series expandable in the small $G_N \sim 1/C$ limit. This exemplifies the fact that even large black holes have strong quantum gravity effects at the horizon according to fiducial observers.
\item
Small black holes on the other hand satisfy $C \ll \beta$. Extrapolating the above results back to small black holes, we obtain $S \approx \frac{c}{36}24\pi^2\frac{C}{\beta}e^{S_0}$. This vanishes in the leading approximation, matching the fact that there is no black hole left at $\beta \to \infty$. Note that this is an extrapolation of the semiclassical black hole regime towards small black holes, where we might not trust these results anymore. Nonetheless, when only considering the physics of the atmosphere (as we do here), this is still an interesting limit to take. (In the same limit the leading $z$ contribution also vanishes as it scales with $\sqrt\frac{2 \pi C}{\beta}$).
\end{itemize}

\begin{figure}[h]
    \centering   \includegraphics[width=0.4\textwidth]{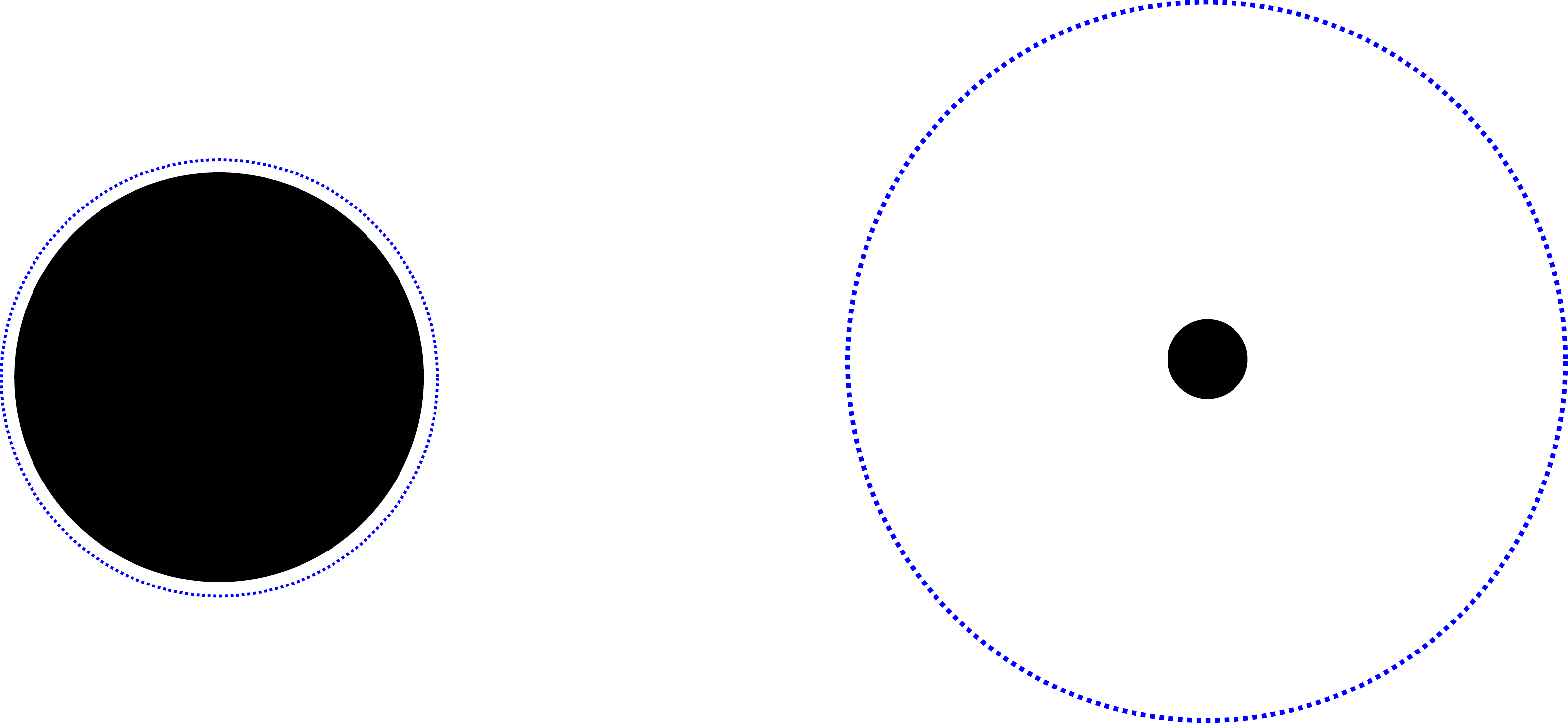}
    \caption{Cartoon of the stretched horizon of a black hole at proper length $\ell_{\text{sh}}$, defined as the demarcation surface where the growth in entanglement entropy stops and where non-perturbative quantum gravity effects become important. For smaller black holes, the stretched horizon moves outwards.}
    \label{fig:SHcartoon}
\end{figure}

A crude approximation of the different regimes is given by the intersection of $-C_0$ and $C_1$ which, yields
\begin{equation}
    \frac{z e^{-S_0}}{C} = \frac{4 \pi}{3} e^{6 C \pi^2 /\beta}.
\end{equation}
We thus find that the plateau region is pushed, exponentially quickly, to large $z$ in the large black hole limit. 
The proper distance $\ell$ to the classical black hole horizon is related to the $z$-coordinate as\footnote{We remark that as $z$ gets bigger, the geometry locally is very deformed, and hence the local distance changes as operationally defined in \cite{Blommaert:2019hjr}. The line element $ds$ in a quantum average $\langle ds \rangle$ is larger than the classical length. This means the operationally measured distance would actually be slightly larger than the value we give here. However, the region where it differs from the classical case is tiny compared to the full answer so that we can safely neglect this distinction. For a Planck-size black hole however, the distinction can be important, so equation \eqref{eq:sh} should be viewed as a \emph{lower} bound if defining this distance operationally.}
\begin{equation}
\ell \approx  \frac{\beta}{\pi} e^{-\frac{2\pi}{\beta}z}.
\end{equation}
Plugging in the above value of $z$, we obtain
\begin{equation}
\label{eq:sh}
\ell_{\text{sh}} \approx \frac{\beta}{\pi} e^{-\frac{2\pi}{\beta}Ce^{S_0} \frac{4\pi}{3} e^{\frac{6\pi^2}{\beta}C}},
\end{equation}
which is doubly non-perturbative in $G_N$ or in both  $1/C$ and $1/S_0$. We can view this length $\ell_{\text{sh}}$ as the location of a stretched horizon or membrane, where QFT in curved spacetime results are replaced by UV complete answers, here modeled via the topological completion of the model, see Fig.~\ref{fig:numeric} (right). As the temperature decreases, and the black hole shrinks, $\ell_{\text{sh}}$ increases unboundedly, and the stretched horizon moves outwards and away from the event horizon. This effect only becomes appreciable at doubly exponentially small (in $S_0$) temperatures. We illustrate this in Fig.~\ref{fig:SHcartoon}. Even though the stretched horizon moves outwards, the total entropy contained in the matter gas around the black hole decreases monotonically.

Such behavior is qualitatively the same as observed in the past in string theoretic models of black holes, where the stringy thermal atmosphere expands \cite{Kutasov:2005rr,Giveon:2005mi} as the correspondence point is reached where the black hole itself becomes string size \cite{Susskind:1993ws,Horowitz:1996nw,Chen:2021emg,Chen:2021dsw,Urbach:2022xzw,Agia:2023skp}.\footnote{
In AdS spacetimes, this corresponds to the case of the small black hole (with negative heat capacity). The large black hole branch is immaterial for the string-black hole transition.
} 
This fits with the general theme that doubly non-perturbative effects in lower-dimensional gravity model features from the underlying microscopic discrete system, i.e. where one expects a UV-complete model such as string theory to deviate from an effective model such as disk gravity plus fields. Computing entanglement entropy in perturbative string theory has a long history \cite{Dabholkar:1994ai,Lowe:1994ah,He:2014gva,Mertens:2016tqv,Balasubramanian:2018axm,Witten:2018xfj,Mazenc:2019ety,Dabholkar:2022mxo,Dabholkar:2023ows}, but it is generally believed that the perturbative corrections (in $g_s$) are UV-finite order by order (as fitting for a perturbatively UV-finite theory).

As a final remark, we considered here the expectation value of the matter entanglement entropy, but one might also be interested in the full probability distribution of the length variable from which the entanglement entropy derives. We present some formulas and plots in Appendix \ref{app:probdistr} to that effect.

\section{Concluding remarks}\label{sec:concl}

In this work, we have provided a natural definition of fiducial observers in JT quantum gravity. We showed that this definition coincides with an earlier antenna construction \cite{Blommaert:2019hjr,Mertens:2019bvy, Blommaert2021}, with a strengthened motivation thanks to the promising connection to the recent body of work in operator algebras and modular flows. In the second part of this work, we have applied our construction to study the matter entanglement entropy of quantum fields across the black hole event horizon. This is the entropy of the thermal Unruh gas, which is famously divergent due to the near-horizon region \cite{tHooft:1984kcu}. We showed that non-perturbative gravity effects cure this behavior and lead to a finite thermal atmosphere entropy, reaching an entanglement plateau, in parallel to the complexity plateau in earlier work \cite{Iliesiu:2021ari}. \\

We end with several more speculative discussions and name some further research directions that we largely leave to future work.\\

\emph{Thermality of states with geometric modular flow and gravitationally dressed observables} \\
It is a well-known result in modular theory that
\begin{equation}\label{eq:KMSmodularflow}
    \bra{\Psi}a_s b\ket{\Psi} = \bra{\Psi}ba_{s+i}\ket{\Psi},
\end{equation}
where $a, b$ are elements of a von Neumann algebra $\mathcal{A}$, $\ket{\Psi}$ is any cyclic-separating vector for $\mathcal{A}$, and $a_s$ denotes the modular flow of $a$ as defined in~\eqref{eq:modularflow}. This shows that correlation functions of operators evolved via modular flow satisfy the \emph{KMS condition}, and therefore, every cyclic-separating state for an algebra $\mathcal{A}$ is ``thermal'' with respect to modular flow.\footnote{The KMS condition is the most general basis for a definition of thermality in cases where the state of the system does not admit a description in terms of a density matrix. For more details, see e.g.~\cite{Kubo:1957mj, Martin:1959jp, Sorce:2023gio}.}  The inverse temperature in units of modular time can be immediately recognized from Eq.~\eqref{eq:KMSmodularflow} as $\beta=1$.

It is natural to ask in what circumstances the thermality noted above is manifest to physical observers probing the system locally. For a given observer following a fixed trajectory in spacetime, the notion of time evolution that is relevant for thermality is not the one given by modular flow, but the one given by the geometric flow along the observer's own worldline. This means that the observer will not read off a thermal signature from the system unless the KMS condition is satisfied for correlation functions probed along the observer's trajectory and written as a function of the oberver's \emph{proper time} $\tau$. In a general state $\ket{\Psi}$, there is no relation between the modular time $s$ and the proper time $\tau$ of any timelike trajectory, because modular flow itself is in general nonlocal. When modular flow is geometric, however, the situation is more promising: in this case, evolution in modular time $s$ corresponds to the flow of a timelike vector field $d/ds \equiv \xi$ in spacetime, and the proper time along any integral curve of this vector field is then related to modular time by
\begin{equation}\label{eq:propertime}
    \left(\dfrac{d\tau}{ds}\right)^2 = -\xi^a\xi_a.
\end{equation}
If the norm of the vector field $\xi$ is constant along its integral curves, we can immediately integrate Eq.~\eqref{eq:propertime} to write $\tau = ||\xi||s+\text{const.}$, where $||\xi||=\sqrt{-\xi^a\xi_a}$. In this case, it is clear that the KMS condition in modular time~\eqref{eq:KMSmodularflow} directly implies the KMS condition in proper time, with the inverse temperature given by $\beta=||\xi||$ in units of proper time. 

As we noted in Sec.~\ref{sec:geometricdressing}, demanding that modular flow be geometric implies that its associated vector field $\xi$ must be a conformal Killing vector field. If, on top of that, we also require that the norm of $\xi$ is constant along its flow lines, we conclude that $\xi$ must be not just a conformal Killing vector, but a true Killing vector~\cite{Caminiti:2025hjq}. Therefore, in situations where $\xi$ is not an exact isometry but only a conformal isometry---as will be the case for most of the off-shell Schwarzian profiles $F(t)$ used to construct the dressing in this paper---any state whose modular flow is implemented by $\xi$ does not satisfy the KMS condition from the point of view of an observer following its integral curves. 

Having said that, although one cannot expect to extract a local temperature for each individual off-shell clock $F(t)$, an averaged notion of thermality is also available in our model. This can be seen, for instance, by observing that the gravitationally dressed two-point function of a bulk CFT coupled to JT gravity in the thermofield double state (as constructed explicitly, e.g., in~\cite{Blommaert:2019hjr}) satisfies the KMS condition along any trajectory at a fixed value of the bulk coordinate $z$, at inverse temperature $\beta$ in units of boundary time $t$. This can then be rescaled by the local redshift factor for each value of $z$, $\sqrt{g_{00}}$ where $g_{00}$ is extracted from Eq.~\eqref{eq:AdSSchwarziandressing}. This redshift factor is independent of $t$ after the Schwarzian path integral is performed, reflecting the fact that the ``averaged'' norm of the conformal Killing vector fields associated to each off-shell Schwarzian profile $F(t)$ is time-independent. It would be interesting to understand if this effective local notion of thermality for fiducial observers, even after quantum-gravity effects are taken into account, could lead to new insights to the near-horizon physics of quantum black holes.
\\

\emph{Crossed product algebra and states with geometric modular flow in JT gravity} \\
As we have emphasized in the main text, the form of dressing that we used to map bulk quantities to boundary 
observables of the JT gravity theory in Schwarzian language is fully fixed by the requirement that boundary 
time translations 
be extended into the bulk as the flow of a conformal isometry of AdS$_2$. We motivated this requirement 
by appealing to the tight connection between conformal isometries and geometric modular flow: in essence, the fact 
that every geometric modular flow must be implemented by a conformal isometry implies that, in the case 
of AdS$_2$, every geometric modular flow that asymptotes to the notion of time translation of the boundary trajectory
in JT gravity must be of the form given here, for some reparametrization function $F(t)$. However, we have not established the converse 
statement---namely, that every conformal isometry defined by extending the configuration of the Schwarzian 
mode $F(t)$ into the bulk corresponds to modular flow of some state in JT gravity. 
Because of this, we cannot decisively claim at present that the dressed observables constructed with 
this form of dressing can be univocally interpreted as observables in a crossed product algebra. Relatedly, 
the entanglement entropy that we are computing is not explicitly the von Neumann entropy of the full crossed product algebra; rather, it is simply the matter entanglement 
entropy of a CFT that is minimally coupled to gravity, with the gravitational path integral being performed at the very end as a statistical ensemble.

Despite still being conjectural at its current stage, we believe that
the interpretation of the radar definition of gravitational dressing as a realization of the algebra of gravitational observables dressed by the modular crossed product is rather promising. 
It would be very interesting to understand this connection in more depth in the future. A few ideas to pursue when trying to make progress in this direction inspired by recent literature are the following:

\begin{itemize}
\item In~\cite{Gao:2024gky}, a set of partially entangled thermal states (PETS) in JT gravity is constructed by inserting heavy matter operators in the preparation of the thermofield double state. It is then shown that the modular flow of those states acting on the algebra of operators restricted to one of the boundaries is geometric in the semiclassical limit. It would be interesting to see whether that set of states can be connected (at least in the semiclassical limit) to a definite profile for the Schwarzian mode, which we can then use to translate to the dressing used here and see whether it does indeed match with the geometric modular flow obtained there.
\item Similarly, in~\cite{Caminiti:2025hjq}, an explicit construction of states with geometric modular flow is presented for CFTs in two-dimensional Minkowski spacetime. A natural question to ask, motivated by our setup in this paper, is how to transport their strategy from flat spacetime to AdS$_2$, and compare the dressing by the modular crossed product of that family of states with the geometric dressing with the antenna construction that we have used in our work.
\end{itemize}

\emph{Other families of observers}\\ 
Our work focused solely on the accelerating or fiducial family of observers. Of course other observer families are interesting. In particular, the infalling worldlines are of considerable relevance. Clearly, one of the three constraints of our definition in section \ref{sec:intro} would have to be relaxed. For recent analyses along these lines, we refer to \cite{Jafferis:2020ora,deBoer:2022zps} where a probe black hole was used as an anchor. Other anchorings in JT gravity were studied in \cite{Nitti:2024iyj}. \\

\begin{figure}[h]
    \centering
    \includegraphics[width=0.55\textwidth]{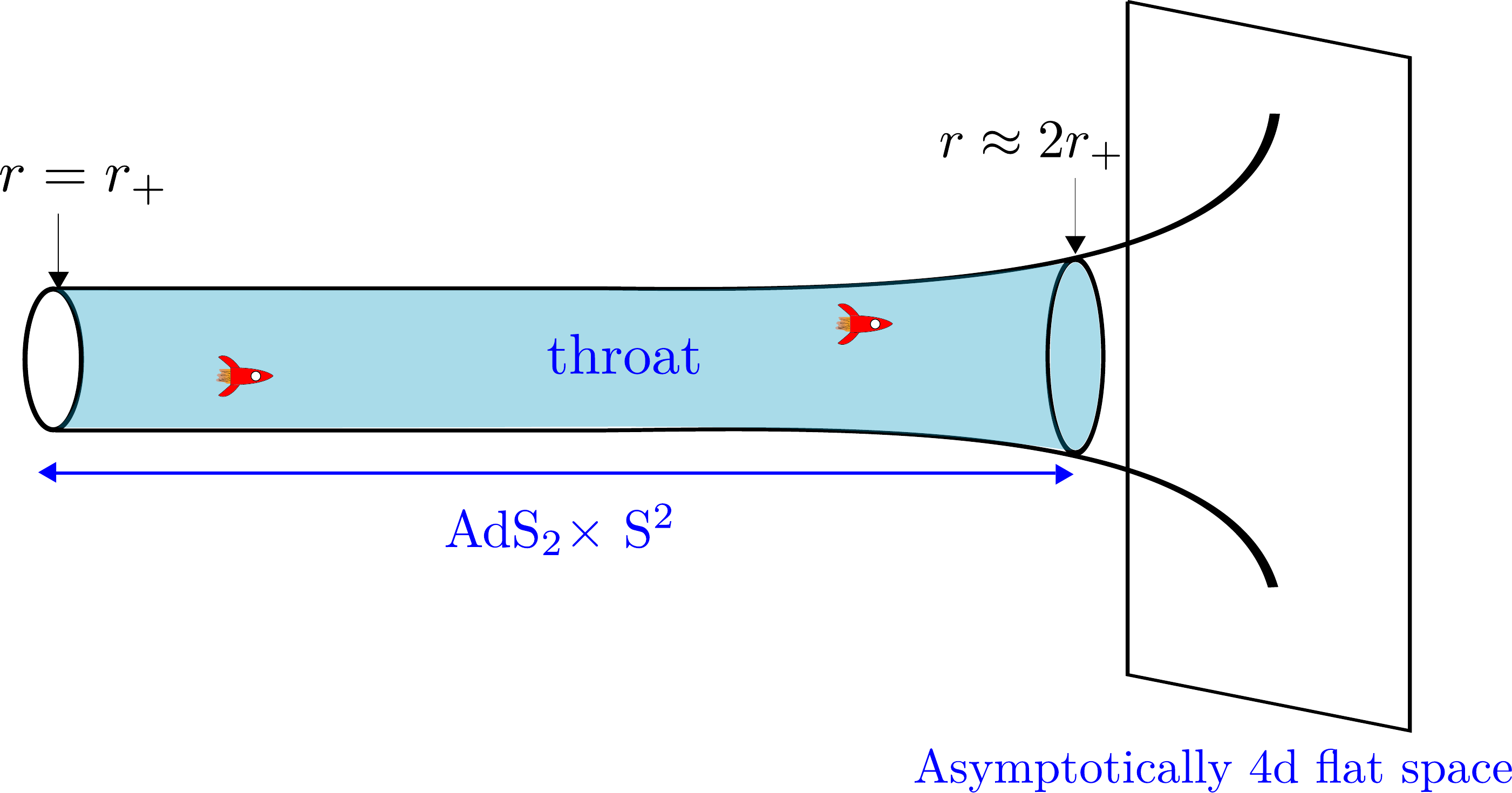}
    \caption{Near-horizon spatial geometry of a 4d Reissner-Nordstr\"om black hole with outer horizon radius $r_+$. Fiducial/accelerating observers in the long throat (depicted here by stationary spaceships) have been effectively defined in the quantum JT model in this work.}
    \label{fig:NEBH}
\end{figure}

 \emph{Fiducial observers in the higher-dimensional black hole throat} \\
Within the context of a higher-dimensional near-extremal black hole, there is no sharp boundary where the Schwarzian lives. Instead, it is to be viewed as an effective highly quantum degree of freedom that needs to be path-integrated over. For any fixed such off-shell choice $F(t)$, we have a unique construction of a bulk conformal Killing vector field that defines accelerating worldline trajectories in the quantum throat region (Fig.~\ref{fig:NEBH}).
Quantum throat physics has attracted much attention recently \cite{Brown:2024ajk,Lin:2025wof,Maulik:2025hax,Emparan:2025sao,Biggs:2025nzs,Emparan:2025qqf, Betzios:2025sct}. 
In particular, Hawking evaporation corrected by the quantum black hole throat region was discussed in some of these works. The corrections that are discussed there to represent quantum gravity in the near-horizon throat manifestly match with those as constructed by our fiducial observers, which we view as further evidence of the importance of our observer construction. The main new ingredient within the higher-dimensional setup is the inclusion of the classical greybody factor representing the coupling of the throat to the asymptotic region. 
In this work however, we constructed local observables deep in the throat region, so we do not have to worry about coupling back to the asymptotic region for the higher-dimensional black hole. 
One important discussion to be had is whether the RMT corrections in JT gravity discussed in subsection \ref{sub:highertopology} are also directly dominant for the very near-horizon region of the higher dimensional black hole. Our perspective on this is that the ``quenched'' matrix model we utilized to model the wormhole contributions, ``minimally'' couples a matter sector to a quantum chaotic JT black hole background system, and as such should have a high degree of universality along the lines of \cite{Cotler:2016fpe}. Effectively, the physics of the entanglement plateau is just spectral level repulsion of a quantum chaotic black hole. \\

\begin{figure}[h]
    \centering
    \includegraphics[width=0.3\textwidth]{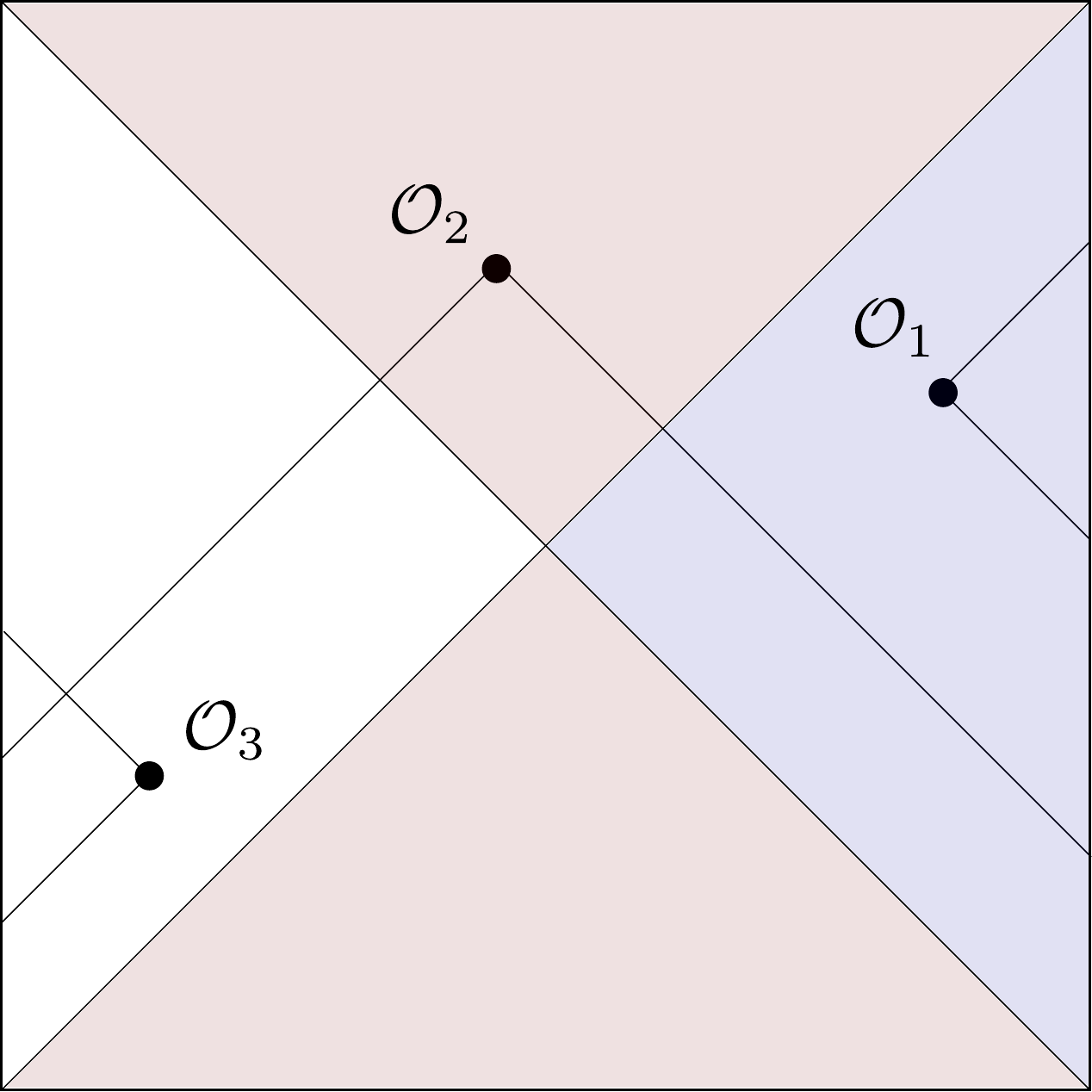}
    \caption{Our gravitational dressing extended to the other wedges. A bulk point in the black (and white) hole interior (red regions) has to be anchored to both asymptotic boundaries.}
    \label{fig:Interior}
\end{figure}

\emph{The black hole interior} \\
From a geometrical point of view, the gravitational dressing we utilized can be naturally extended in the other three wedges of the eternal two-sided black hole. To do this, we analytically continue the dressing by anchoring one of the null lines to the second holographic boundary upon crossing an event horizon (Fig.~\ref{fig:Interior}). This naturally matches with causality considerations in the black hole interior.
The interior of the black hole is then locally constructed by taking data from \emph{both} asymptotic regions, which is very natural from causality arguments. The resulting bulk frame constructed in the black hole interior is somewhat peculiar, with the usual time $\leftrightarrow$ space swap. Of particular interest would be a cross-horizon correlation function, such as $\langle \mathcal{O}_1\mathcal{O}_2 \rangle$ and $\langle \mathcal{O}_1\mathcal{O}_3 \rangle$ in the notation of Fig.~\ref{fig:Interior}. We leave this for the future. \\

\emph{Quenched vs annealed disorder as gravity} \\
Lower-dimensional gravity models famously display features similar to those of disordered spin-glass systems  exhibiting features of quenched disorder \cite{Castellani:2005fkd}.
For annealed disordered systems, the disorder average (here the gravitational path integral) is on the same footing as the other matter sectors, i.e.:
\begin{equation}
\beta F_{\text{annealed}} = - \log \int [\mathcal{D}g_{\mu\nu}] [\mathcal{D}\phi] \, e^{-S[g_{\mu\nu},\phi]}.
\end{equation}
In the quenched disorder set-up on the other hand, the metric field varies slowly compared to the other matter fields:
\begin{equation}
\beta F_{\text{quenched}} = - \int [\mathcal{D}g_{\mu\nu}] \log \left(\int[\mathcal{D}\phi] \, e^{-S[g_{\mu\nu},\phi]}\right).
\end{equation}
This can be intuitively argued as follows: in lower-dimensional gravity models, quantum effects are strong in the IR, and hence at long timescales. This means quantum fluctuations in the gravity sector vary slowly compared to the other fields, and this sector can be treated as a quenched disorder average. Below we will comment on how this feature shows up in our framework. First note however, that his feature should be present in supergravity models as well, even those that contain gauge fields in the gravitational multiplet. As a first step in the $\mathcal{N}=1$ JT supergravity case, the disk late time wormhole length was computed in \cite{Fan:2021wsb} and was linearly growing as well. Analytically continuing that calculation then gives the same linearly increasing (and hence ultimately divergent) entropy of the black hole atmosphere. The non-perturbative wormhole contributions could then be read from \cite{Stanford:2019vob}. The wormhole length for $\mathcal{N}=2$ JT supergravity was discussed in \cite{Lin:2022zxd}.  \\

\emph{Gravity as a quenched statistical ensemble}  \\
In this work we explicitly showed that the IR divergences of matter entanglement entropy are cured by the matrix model completion of the gravitational sector of JT gravity. On the other hand, other pathological features, especially the short distance divergences well known from flat space QFT calculations, still persist in this setup. This is far from surprising and rather an expected feature of the setup we chose. In what we call a single matrix model completion, completing the model by replacing the spectral correlator $\rho_0 \ldots \rho_0 \to \langle \rho \ldots \rho\rangle_{\text{RMT}} $ by its matrix model expression following \cite{Saad:2019pqd} can be viewed as microscopically completing the gravitational sector of the coupled theory while treating the matter sector as a standard QFT on curved spacetimes. However, since the short distance singularities are expected to arise from the short distance modes of the matter sector, resolving this singularity instead requires a microscopic description of the full matter $\times$ gravity system. In this hypothetical system on the other hand, the notion of matter entanglement entropy may not be well-defined since a full theory may be expected to induce strong mixing of the degrees of freedom leading to a loss of separation between the entropy attributed to the black hole system and that of its thermal atmosphere.\footnote{Of course the full von Neumann entropy of the total system should still be well-defined.} Instead one can view our calculation as probing how the thermal atmosphere is influenced by the chaotic nature of the quantum black hole. We view this physical \textbf{choice} of treating matter differently, at least conceptually, as something akin to quenched rather than an annealed disorder in the system and summarize this approach via the simple mantra: \textit{ Simplify first, extend later.} 

\begin{figure}[h]
    \centering
    \includegraphics[width=0.55\textwidth]{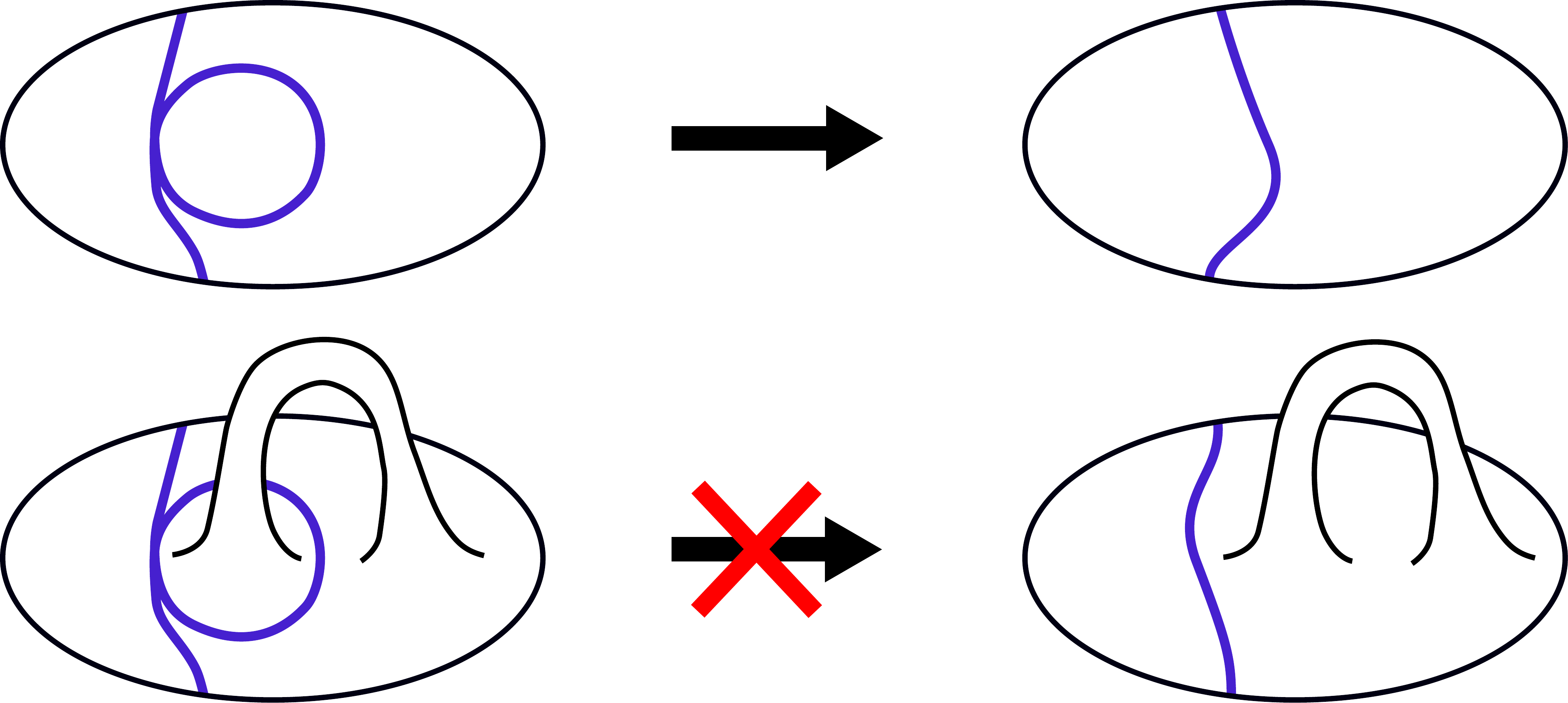}
    \caption{Self-intersections can always be trivialized on the disk. This is no longer true on higher topology due to topologically supported loops.}
    \label{fig:selfintersect}
\end{figure}

Let us exemplify some of the consequences of this choice and contrast this approach to different options like the two-matrix model pioneered in \cite{Jafferis:2022wez,Jafferis:2022uhu}. As a first example, take the two-point function on the disk (see Fig. \ref{fig:selfintersect} for an example). 
A priori, one may choose the path integral to include or exclude self-intersecting configurations of the matter lines. It turns out that at disk level these prescriptions are identical, since every such self-intersection can be uncrossed without changing the two-point function \cite{Blommaert:2018oro}. 
Including topological fluctuations in the gravitational path integral, however, leads to configurations topologically supporting the self-intersection. As such, the order of operations matters, since it leads to a different space of allowed configurations in the path integral. As explained around \eqref{eq:operatormatrixmodel}, the single matrix model leads to the exclusion of these contributions, in line with the approach of simplifying first. A second example is discussed in Appendix \ref{app:probdistr} (around eq. \eqref{eq:Prmt}) where the entire probability distribution of geodesic lengths (or bulk matter entanglement entropy) is discussed in a similar manner.

One benefit of our writing of the effective ``quenched'' gravity matrix model in \eqref{eq:wormholes} explicitly in terms of operator insertions is that it is directly generalizable to other models. The only ingredient we need to write down \eqref{eq:wormholes}, besides the density of states, is the two-boundary gravitational wavefunction. This object has been extensively studied in a variety of models the past few years and is known in the JT supergravity models \cite{Fan:2021wsb,Lin:2022zxd,Belaey:2024dde}, Liouville gravity and minimal string models \cite{Mertens:2020hbs,Fan:2021bwt}, and sine dilaton gravity and DSSYK models \cite{Berkooz:2018jqr,Berkooz:2022mfk,Lin:2022rbf,Blommaert:2023opb,Belaey:2025ijg}. One can then immediately write down the corresponding quenched coupled matter-gravity matrix integral. 

\begin{figure}[h]
    \centering
    \includegraphics[width=0.7\textwidth]{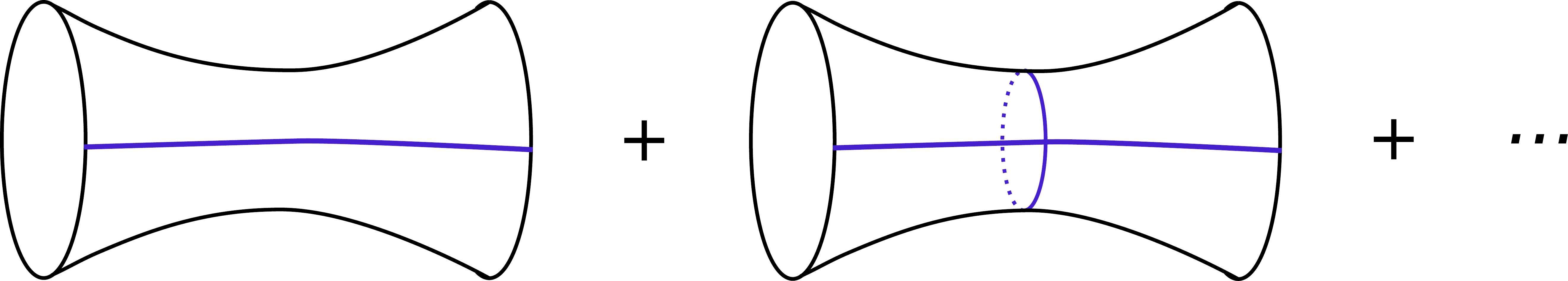}
    \caption{Leading contributions to the two-point function on the trumpet geometry in the two-matrix model \cite{Jafferis:2022wez,Jafferis:2022uhu}. The quenched disorder system specified by a single matrix model corresponds to the exclusion of all bulk loop contributions. }
    \label{fig:trumpet}
\end{figure}

Finally, it is worth briefly mentioning how our approach compares to the two-matrix model approach described in \cite{Jafferis:2022wez,Jafferis:2022uhu}, where both the mater sector and the gravity sector are treated via a matrix model. Geometrically, this treatment differs from our approach by the inclusion of matter vacuum loops wrapping the non-trivial topologies. Even though the matter sector is a priori free, the gravitational average leads to an effective coupling between the closed bulk lines and other matter line insertions, as shown in Fig.~\ref{fig:trumpet}.
These matter then naturally lead to the familiar tachyon divergences akin to bosonic string theory. Technically the transition from this setup to the quenched setup utilized in out work can be realized by a different order of operations within the path integral. Explicitly the quenched answer is reached by schematically computing 
\begin{align}
    \langle \mathcal{OO}\rangle \sim \int \left[ \mathcal{D}g_{\mu\nu}\right] e^{-S_{\text{grav}}[g_{\mu\nu}]}Z_m^{-1}[g_{\mu\nu}]\int [\mathcal{D}\phi] \mathcal{O}\mathcal{O}e^{-S_m[\phi;g_{\mu\nu}]},
\end{align}
where the normalization by the fixed background matter partition function, or in other words the thermalization of the matter operators in the fixed geometry subspace, ensures the removal of the tachyonic contributions.  \\

\emph{Fiducial observers and universal Teichm\"uller space} \\
In this work we argued for the identification of the space of fiducial observers with the space of finite flows induced by conformal Killing vector fields with certain boundary conditions on the hyperbolic disk, modulo the space of true Killing vector fields. It is understood that for Riemann surfaces this space has a natural description in terms of Teichm\"uller space of the corresponding Riemann surface \cite{hubbard:hal-01297628}.
In the case of JT gravity one may wonder if such an identification is valid after the inclusion of the asymptotic boundary  conditions. It turns out that this is indeed the case within the framework of universal Teichm\"uller space $T(1)$ \cite{bers_automorphic_1965}. Explicitly, on the disk, we considered the space $\text{Diff}^+(S^1)/\text{PSL}(2,\mathbb{R})$ which was shown to exactly comprise the smooth subset of universal Teichm\"uller space, which equivalently corresponds to the space of deformations of the hyperbolic disk with smooth boundary curves \cite{Nag:1990dj,nag1992tangentspaceuniversalteichmuller}. It would be interesting to investigate this connection in order to extend the definition of fiducial observers to geometries beyond the disk or to other models such as supersymmetric JT gravity. Indeed the generalization to different geometries seems promising as small modification allow for the inclusion of the double trumpet in the framework of universal Teichm\"uller space \cite{Choi:2023syx}.

\section*{Acknowledgments}
We thank Andreas Blommaert for early discussions on some of the ideas of the second part of this work. We also thank Luca Ciambelli, Josh Kirklin, Jacopo Papalini, Jonathan Sorce, Gabriel Wong, and Qi-Feng Wu for discussions. TT is supported by a Research Foundation - Flanders (FWO) doctoral fellowship. TM and BT acknowledge financial support from the European Research Council (grant BHHQG-101040024). Funded by the European Union. Views and opinions expressed are however those of the author(s) only and do not necessarily reflect those of the European Union or the European Research Council. Neither the European Union nor the granting authority can be held responsible for them.

\appendix
\section{The large black hole limit on the disk}
\label{app:appdisk}
In this appendix, we will provide some details on the derivation of the matter entanglement entropy on the disk---i.e., the solution to 
\begin{equation}
\label{eq:diskent}
      S_\text{ent} + \frac{c}{6}\ln{\delta} = - \frac{c}{6\pi^2 Z(\beta)}\int d\mu(k_1)d\mu(k_2)  e^{2iz \frac{k_1^2}{2C}} e^{- \left(  \beta + 2iz \right) \frac{k^2_2}{2C}} \frac{\partial  }{\partial h }\left[\frac{\Gamma\left(  h \pm i k_1 \pm i k_2 \right) }{(2C)^{2h} \Gamma\left(  2h \right)} \right]_{h = 0},
\end{equation}
in the limit $\beta \ll C \ll z$. 
We begin by presenting two equivalent computations of the derivative of the matrix element,
\begin{equation}
    \frac{\partial  }{\partial h }\left[\frac{\Gamma\left(  h \pm i k_1 \pm i k_2 \right) }{(2C)^{2h} \Gamma\left(  2h \right)} \right]_{h = 0},
\end{equation}
both of which will be used later on. 

Since $k_1$ and $k_2$ are real, the derivative can be taken directly in the case $k_1 \neq k_2$ by using the pole structure of the digamma function to evaluate the limit $l \to 0$. In that case, one finds that the only non-vanishing contribution to the limit is given by the term 
\begin{equation}
     \frac{\partial  }{\partial h }\left[\frac{\Gamma\left(  h \pm i k_1 \pm i k_2 \right) }{(2C)^{2h} \Gamma\left(  2h \right)} \right]_{h = 0} = - 2\lim_{h \to 0} (2C)^{-2h}\Gamma(h \pm ik_1 \pm ik_2) \frac{\Gamma'(2h)}{\Gamma(2h)^2},
\end{equation}
since for all other terms the numerator is regular as $h \to 0$. This limit can then be directly evaluated to yield
\begin{equation}
    \frac{\partial  }{\partial h }\left[\frac{\Gamma\left(  h \pm i k_1 \pm i k_2 \right) }{(2C)^{2h} \Gamma\left(  2h \right)} \right]_{h = 0} = 2\Gamma\left(  \pm i (k_1 \pm k_2)\right) = \frac{2\pi^2}{(k_1^2 - k_2^2)\sinh(\pi(k_1  \pm k_2))},
\end{equation}
where we use the identity $    \Gamma\left(  \pm ix \right) = \frac{\pi}{x \sinh(\pi x)}$.
If $k_1 \to k_2$ a more detailed analysis is needed, since the ordering of limits is non-trivial. However, this subtlety turns out to not be important for our results: in this limit, the integrand of \eqref{eq:diskent} becomes fully independent of the bulk point, and therefore does not produce corrections to the leading $z$ behavior of the entropy. Physically, the contribution of such terms can be viewed as defining the boundary value of the entanglement entropy---i.e., constant contributions which are purely a choice of calibration of the measurement apparatus from a boundary observer's perspective \cite{Iliesiu:2021ari}.  

The second method that can be used to compute the derivative of the matrix element starts by using the integral identity 
\begin{equation}
    \frac{\Gamma\left(  h \pm i\left(  k_1-k_2 \right)  \right) }{(2C)^{2h}\Gamma\left(  2h \right) } = 2 \int_{-\infty}^\infty dy \frac{e^{2i\left(  k_1-k_2 \right)y} }{\left(  4C \cosh y \right)^{2h}} = 2 \int_{-\infty}^\infty dy \frac{\cos\left(  2y\left(  k_1 -k_2 \right) \right) }{(4C \cosh y)^{2h}},
\end{equation}
as well as the identities $\ln x = \lim_{h \to 0} \frac{d}{dh} x^h$ and $\lim_{x \to 0} \frac{\Gamma\left(  h \pm ix \right) }{\Gamma(2h)} = 2\pi \delta(x)$, to show that 
\begin{equation}
     \frac{\partial  }{\partial h }\left[\frac{\Gamma\left(  h \pm i k_1 \pm i k_2 \right) }{(2C)^{2h} \Gamma\left(  2h \right)} \right]_{h = 0} = -4\int_{-\infty}^{\infty} dy \frac{\pi \ln(4C \cosh y)}{(k_1 +k_2) \sinh(\pi (k_1+ k_2))}e^{2i\left(  k_1 - k_2 \right)y},
\end{equation}
up to the $k_1 = k_2$ contribution 
\begin{equation}
    \left. 2 \pi \delta\left(  k_1-k_2 \right) \frac{\partial}{\partial h} \Gamma\left(  h \pm i \left(  k_1 + k_2 \right)  \right)\right\vert_{h = 0},
\end{equation}
that we will again not be interested in.
The entanglement entropy thus takes the form 
\begin{equation}
     \frac{c}{12}\frac{2}{\pi^2}\frac{4 \pi}{Z} \int_{-\infty}^{\infty} dy \int_0^{\infty} d\mu(k_1) d \mu(k_2)\frac{ \ln\left( 4C \cosh y\right) }{(k_1+k_2) \sinh \left(  \pi \left(  k_1 + k_2  \right)  \right)} e^{\frac{2iz}{2C}\left(  k_1^2 - k_2^2 \right) - \beta \frac{k_2^2}{2C} - 2i \left(  k_1 - k_2 \right)y},
\end{equation}
where again $d\mu(k) = dk\, k \sinh 2 \pi k$. In the limit we are interested in, the $k_1,k_2$ integrals are well-approximated by their saddle point contributions. Note that, in the large black hole limit, the term $e^{-\beta k_2^2/2C}$ is subleading; as such, both integrals can be solved on the same saddlepoint. We thus have an integral of the form 
\begin{equation}
    \int_0^\infty dk f(k) e^{2i\left(\frac{z}{2C} k^2 - ky\right)},
 \end{equation}
 where $f(k)$ denotes all terms constant in $z,y$. Substituting $k = k_* + \sqrt\frac{2C}{z} \tilde{k}, k_* = \frac{Cy}{z}$ reduces the integral to 
 \begin{equation}
       \int_0^\infty dk f(k) e^{2i(\frac{z}{2C} k^2 - ky)} \approx \sqrt{\frac{2C}{z}}  e^{-\frac{iy^2C}{z}}  f(k_*) \int_{-k_* \sqrt{\frac{z}{2C}}}^{\infty} d \tilde{k} e^{2i \tilde{k}^2},
 \end{equation}
 where we used $f(k) = f(k_*) + \mathcal{O}(z^{-1/2})$. The solution of this integral now depends on the behavior of the lower bound. This scales as $y/\sqrt{Cz}$ and as such, as we are integrating over all values of $y$, there are different cases that need to be considered. Firstly, if $y$ is of order one compared to $Cz$, the lower bound tends to $0$. However here $f(k_*) \to f(0) = 0$, meaning that this contribution is subleading. A similar argument works for $y \ll z$ since $k \geq 0$. Thus at least $0 < y$ and $y \sim \Omega(z/C)$ where the lower bound tends to $-\infty$ and $f(k_*)$ is nontrivial. As such, we can write 
 \begin{equation}
     \int_0^\infty dk_1 dk_2 f(k_1,k_2) e^{\frac{2iz}{2C}\left(  k_1^2 - k_2^2 \right) - 2i \left(  k_1 - k_2 \right)y} \approx f(k_*,k_*) \frac{2C}{z} \frac{\pi}{2},
 \end{equation}
 with the condition that $y$ is restricted to be at least of order $z/C$. Thus,
 \begin{equation}
     S_\text{ent}  + \frac{c}{6} \ln \delta \approx \frac{c}{12}\frac{4}{Z(\beta)}\frac{C}{z}\int_a^\infty dy k_* \sinh \left(  2 \pi k_* \right) \ln(4C \cosh y) e^{-\beta \frac{k_*^2}{2C}}.
\end{equation}
Now, using $y \sim \mathcal{O}(z/C)$, one approximates $\ln(4C \cosh y) \simeq y + \ln(2C) \simeq y$, after which the integral reduces to
\begin{equation}
    S_\text{ent} + \frac{c}{6}\ln \delta \simeq 
    -C \frac{c}{12} \frac{4}{Z(\beta)} \frac{z}{C} \frac{\partial}{\partial \beta} \left[ e^{\frac{2C\pi^2}{\beta}}\int_{u(a) > 0}^\infty du e^{-\left(  \sqrt{\frac{\beta}{2C}} u  - \sqrt{\frac{2C \pi^2}{\beta}}  \right)^2 } -   e^{-\left(  \sqrt{\frac{\beta}{2C}} u +\sqrt{\frac{2C \pi^2}{\beta}}  \right)^2 }\right] 
\end{equation}
where $u = Cy/z$ and $a$ denotes the lower bound. The entanglement entropy thus reduces to a difference of gaussian integrals. In the limit $\beta \ll C$ the peak at $u = 2C\pi /\beta$ is dominant, for which the lower bound can be approximated to be $-\infty$ such that 
\begin{align}
    S_\text{ent} + \frac{c}{6}\ln \delta &\approx
    -C \frac{c}{12} \frac{4}{Z(\beta)} \frac{z}{C} \frac{\partial}{\partial \beta} \left[ e^{\frac{2C\pi^2}{\beta}}\int_{u(a) > 0}^\infty du e^{-\left(  \sqrt{\frac{\beta}{2C}} u  - \sqrt{\frac{2C \pi^2}{\beta}}  \right)^2 } -   e^{-\left(  \sqrt{\frac{\beta}{2C}} u +\sqrt{\frac{2C \pi^2}{\beta}}  \right)^2 }\right] \\
    &  \approx -C \frac{c}{12} \frac{4}{Z(\beta)} \frac{z}{C}\frac{\partial}{\partial \beta } \left[  \sqrt{\frac{2 C \pi}{\beta}} e^{\frac{2 C \pi^2}{\beta}}  \right] \approx \frac{c}{6}\frac{2 \pi z}{\beta}.
\end{align}
\section{Solving the integral~\eqref{eq:ententropyRMT}}
\label{app:appnonpert}
In this section we show how to compute the entanglement entropy beyond the disk. More specifically, we give some technical details on going from the integral \eqref{eq:ententropyRMT} to the expression~\eqref{eq:nonpert} close to the horizon and in the large black hole limit, i.e., $\beta \ll C\ll z$. We emphasize once again that this calculation can be viewed as an analytic continuation of the wormhole length calculation performed in~\cite{Iliesiu:2021ari}; as such, the material below is essentially a review of the calculation outlined there. 

The staring point is the integral~\eqref{eq:ententropyRMT}
\begin{equation}
    \expval{S_\text{ent}} + \frac{c}{6}\ln{\delta} = -\frac{c}{12} \frac{e^{-2S_0}}{C^2 Z(\beta)} \int dk_1 dk_2 \frac{k_1k_2 \expval{\rho(k_1) \rho (k_2)}}{\left(k_1^2 - k_2^2\right) \sinh\left(  \pi \left(  k_1 \pm k_2 \right)  \right) } e^{\frac{iz}{C} \left(  k_1^2 - k_2^2 \right)  -\frac{\beta}{2C} k_2^2},
\end{equation}
where $\expval{\rho(k_1)\rho(k_2)}$ denotes the spectral two-point function, which is given at small energy separations $\abs{E_1 - E_2} = \abs{ k_1^2/2C - k_2^2/2C} \ll 1$ by \cite{Saad:2019lba}
\begin{equation}
\expval{\rho(E_1)\rho(E_2)} = \rho_0(E_1)\rho_0(E_2) + \rho_0(E_1)\delta(E_1 - E_2) - \frac{\sin^2\left[  \pi \rho_0 (E_2) \left(  E_1 -E_2 \right)  \right] }{\pi^2 \left(  E_1 - E_2 \right)^2 }.  
\end{equation}
As a first simplification, we note that we can drop the contact term in the spectral two-point function immediately. This is because such terms are to be viewed as calibration choices performed by the boundary observer and therefore do not carry any important physical interpretation, as discussed before. Introducing the average and sum variables $s = k_1 + k_2$, $\omega = k_1 - k_2$ in the remaining integral then leads to the explicit expression
\begin{align}
      \expval{S_\text{ent}} + \frac{c}{6}\ln{\delta} = -\frac{c}{12} \frac{e^{-2S_0}}{C^2 Z(\beta)} &\int_0^{\infty} \int_{-2s}^{2s} ds d\omega \frac{s^2 - \frac{\omega^2}{4}}{2 \omega s} \frac{\rho_0(s + \frac{\omega}{2}) \rho_0(s - \frac{\omega}{2})}{\sinh\left( 2 \pi s  \right) \sinh(\pi \omega) }  e^{\frac{2iz}{C}\omega s -\frac{\beta}{2C}\left(  s -\frac{\omega}{2} \right)^2 } \\
   & \times \left[  1 - \left(\frac{C}{\pi \omega s }\right)^2 \frac{\sin^2\left(\frac{\pi}{C} \omega s\rho_0(s - \frac{\omega}{2})\right) }{\rho_0(s + \frac{\omega}{2}) \rho_0(s - \frac{\omega}{2})}\right].
\end{align}
We are now tasked to simplify and solve this integral in the large $z$ limit. This can be done by analysing the leading contributions of the integrand. The exponential $e^{2iz\omega s/C}$ implies that the integral is dominated by $\omega s \sim \mathcal{O}(1/z)$. Moreover, one can check that $\omega \sim \mathcal{O}(s)$ automatically by the integral bounds and that the suppression of the integrand is minimized if $\omega z \sim \mathcal{O}(s)$. This naturally suggests rescaling $w = z\omega$ and expanding the integrand in $w/z$ . One then finds that, to leading order in $z$, the integral takes the form 
\begin{align}
    \expval{S_\text{ent}} + \frac{c}{6}\ln{\delta} = - \frac{c}{12} \frac{ e^{-2S_0}}{C^2Z(\beta)} &\int_0^{\infty} \int_{-2sz}^{2sz} dsdw \frac{s}{2w} \frac{\rho_0(s)^2}{\sinh\left( 2 \pi s  \right) \sinh(\pi \frac{w}{z}) }  e^{\frac{2i}{C}w s -\frac{\beta}{2C} s^2} \\
    & \times \left[  1 - \left(\frac{Cz}{\pi w s \rho_0(s) }\right)^2 \sin^2\left(\frac{\pi ws\rho_0(s)}{Cz}\right)\right]. 
\end{align}
Noticing that the integral will now be dominated by $\omega \sim s$ and the integrand vanishes in the $s \to 0$ regime allows for the approximation $2zs \to \infty$ and $\sinh(\pi w/z) \approx \pi w/z$, leading to 
\begin{align}
   \expval{S_\text{ent}} + \frac{c}{6}\ln{\delta} = - \frac{c}{12} \frac{ e^{-2S_0}}{C^2 Z(\beta)}\frac{z}{2\pi} &\int_0^{\infty} \int_{-\infty}^{\infty} dsdw \frac{s}{w^2} \frac{\rho_0(s)^2}{\sinh\left( 2 \pi s  \right)}  e^{\frac{2i}{C}w s -\frac{\beta}{2C} s^2} \\                         
   & \times \left[  1 - \left(\frac{Cz}{\pi w s \rho_0(s) }\right)^2 \sin^2\left(\frac{\pi ws\rho_0(s)}{Cz}\right)\right]. 
\end{align}
Remarkably, this is the last necessary simplification in $w$; the integral can now be solved exactly via residues, since the inclusion of the sine kernel leads to regularity of the integrand at $w =0$. This allows for the deformation of the $w$-integration contour on the upper half plane to avoid $w = 0$. Let us focus on the $w$ integral for now. After rewriting $\sin^2(x)$, this takes the form 
\begin{equation}
   \int_\mathcal{C} dw\left[\frac{e^{\frac{2i}{C}ws}}{w^2} + \frac{1}{2}\left(\frac{Cz}{\pi s\rho_0(s)w^4}\right)^2\left(e^{\frac{2i}{C}w s \left(  1 + \pi\rho_0(s) /z \right)} +   e^{\frac{2i}{C}w s \left(  1 - \pi\rho_0(s) /z \right)}-e^{\frac{2i}{C}w s}\right)\right].
\end{equation}
Since $s,z,\rho_0\geq 0$ the contour can be closed in the upper half plane for the first, second, and last summands, where the respective integrands are holomorphic since the singularity at $w =0$ is avoided. The solution to the remaining integral, 
\begin{equation}
  \frac{1}{2}\left(\frac{Cz}{\pi s\rho_0(s)}\right)^2 \int_\mathcal{C} dw \frac{1}{w^4}e^{\frac{2i}{C}w s \left(  1 - \pi\rho_0(s) /z \right)},
\end{equation}
will depend on the value of $s$. Since $\rho_0(s)$ is monotonous, we can define $s_* \equiv \rho^{-1}_0(\frac{z}{\pi})$ such that 
\begin{equation}
    1 - \frac{ \pi \rho_0(s)}{z} \geq 0 \Leftrightarrow s \leq s_*.
\end{equation}
In this regime, the same argument as above holds. If $s > s_*$ the contour can be closed in the lower half plane, where we find 
\begin{align}
    \int_{\mathcal{C}} dw \frac{1}{w^4} e^{\frac{2i}{C}w s \left(  1 - \pi\rho_0(s) /z \right)} &= -2 \pi i \Res\left[   \frac{1}{w^4} e^{\frac{2i}{C}w s \left(  1 - \pi\rho_0(s) /z \right)}\right]_{w = 0}  \\
    & =- \frac{8 \pi}{3 C^3} s^3 \left[  1 - \frac{\pi \rho_0(s)}{z} \right]^3.
\end{align}
Plugging this back into the original expression, we arrive at a simple integral expression for the entanglement entropy:
\begin{align}
    \left\langle S_\text{end} \right\rangle  - \frac{c}{6}\ln\left(  \delta \right)    &=\frac{c}{12} \frac{ e^{-2S_0}}{Z(\beta)}\frac{z^3}{3C^3\pi^2} \int_{s_*}^{\infty} ds \frac{s^2 e^{-\frac{\beta}{2C} s^2}}{\sinh(2 \pi s)} \left[  1 - \frac{\pi \rho_0(s)}{z} \right]^3  \\                      
    &= \frac{c}{12} \frac{ e^{-2S_0}}{Z(\beta)}\frac{\pi}{3C^3} \int_{s_*}^{\infty} ds \frac{s^2 e^{-\frac{\beta}{2C} s^2} \rho_0(s)^3}{\sinh(2 \pi s)} \left[  \frac{z}{\pi \rho_0(s)} - 1 \right]^3.
\end{align}
This is precisely Eq.~\eqref{eq:nonpert} in the main text.
\section{Probability distribution of geodesic length (or matter entanglement entropy)}
\label{app:probdistr}
\begin{figure}[h]
\centering
\includegraphics[width=0.2\textwidth]{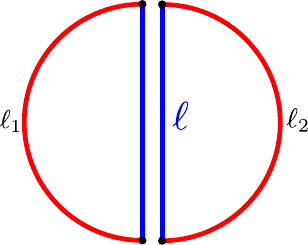}
\caption{The Schwarzian path integral with the constraint that the geodesic length equals $\ell$ between two given boundary times $\tau_1$ and $\tau_2$, leads to a product of two WdW wavefunctions.}
\label{figMagneSurface}
\end{figure}

One can implement a fixed geodesic length constraint within the Schwarzian path integral by using a delta-function as
\begin{align}
\int &[\mathcal{D}F] \, \delta\left(\ell + \log \frac{\dot{F}_1\dot{F}_2}{(F_1-F_2)^2}\right) e^{-S_{\text{Schw}}}, \qquad S_{\text{Schw}} = -C\int_0^\beta d\tau \left\{\tan \frac{\pi}{\beta}f,\tau\right\},
\end{align}
where $-\log \frac{\dot{F}_1\dot{F}_2}{(F_1-F_2)^2}$ is the renormalized geodesic length in AdS$_2$ between two boundary endpoints at Euclidean times $\tau_1$ and $\tau_2$, and $F(\tau) \equiv \tan \frac{\pi}{\beta}f(\tau)$, and is set to $\ell$ in the above path integral. This length $\ell$ is related to the matter entanglement entropy by $S_\text{ent} = \frac{c}{12}\ell$, upon suitable Wick rotation of the two times $\tau_1$ and $\tau_2$ in this Euclidean path integral, as explained in the main text in subsection \ref{sub:disk}.
This path integral can be readily evaluated as
\begin{align}
& \int_{-i\infty}^{+i \infty} dt \, e^{\ell t}\int [\mathcal{D}F] \, \left(\frac{\dot{F}_1\dot{F}_2}{(F_1-F_2)^2}\right)^{t} e^{-S_{\text{Schw}}} \\
&= \int d\mu(k_i) \, e^{-\ell_i k_i^2} \int_{-i\infty}^{+i \infty} dt \, e^{\ell t} \frac{\Gamma(t\pm i k_1 \pm i k_2)}{\Gamma(2t)} \\
\label{eq:finfixl}
&= \prod_i \int d\mu(k_i) \, e^{-\ell_i k_i^2} K_{2ik_i}(2e^{-\ell/2}),
\end{align}
where we made use of the identity
\begin{equation}
K_\mu(z)K_\nu(z) = \frac{1}{8\pi i} \int_{-i\infty}^{+i\infty} \frac{\Gamma(t\pm \mu/2 \pm \nu/2)}{\Gamma(2t)}(z/2)^{-2t}dt.
\end{equation}
The final result \eqref{eq:finfixl} is just the product of two (half-disk) WdW wavefunctions, and can be diagrammatically represented as in Fig.~\ref{figMagneSurface}.

\begin{figure}[h]
\centering
\includegraphics[width=0.5\textwidth]{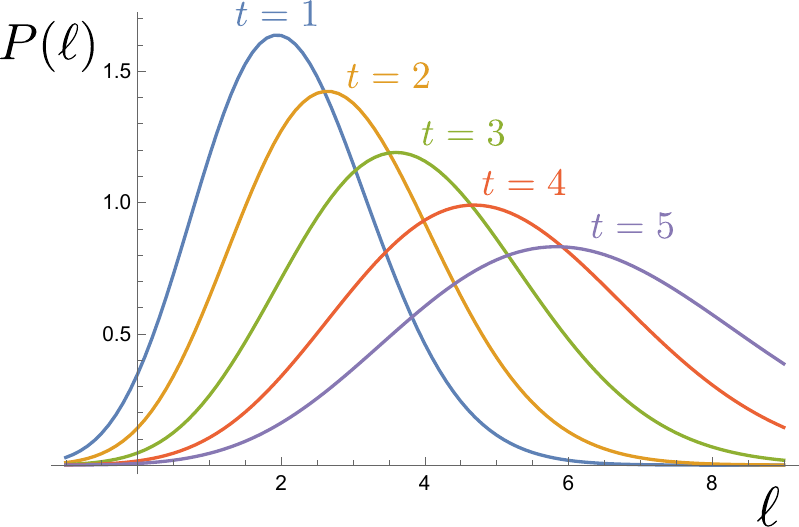}
\caption{Geodesic length probability distributions \eqref{eq:PL} on the disk topology for increasing Lorentzian time $t$. The peaks shift upwards linearly in $t$, and the distribution broadens. Units chosen where $C=1/2$, $\beta=2\pi$.}
\label{figprobdis}
\end{figure}

This answer can be interpreted directly as the probability distribution of the variable $\ell$ in (Schwarzian) quantum gravity, since the latter is given by the ratio of path integrals of those that respect the constraint versus those that do not necessarily do so:
\begin{align}
\label{eq:PL}
P[\ell] \equiv \frac{\int [\mathcal{D}F] \,\delta\left(\ell + \log \frac{\dot{F}_1\dot{F}_2}{(F_1-F_2)^2}\right) e^{-S_{\text{Schw}}}}{\int [\mathcal{D}F] \,e^{-S_{\text{Schw}}}},
\end{align}
normalized as $\int_{-\infty}^{+\infty} d\ell \, P[\ell] = 1$. The moments of this distribution can be computed equivalently by differentiating the boundary two-point correlator w.r.t. $h$ as in equation \eqref{eq:Strick}:
\begin{equation}
\langle \ell^n \rangle \equiv \int_{-\infty}^{+\infty} d\ell\, \ell^n P[\ell] = \frac{d^n}{dh^n}\langle \mathcal{O}_h(\ell_1,\ell_2)\rangle \vert_{h=0},
\end{equation}
which are readily observed by \eqref{eq:finfixl} to be manifestly the same computation. For the geodesic wormhole length, we illustrate this probability distribution in Fig.~\ref{figprobdis}.
This $P[\ell]$ has been obtained before in section 7.3 of \cite{Iliesiu:2024cnh}, where it was phrased as a center of mass collision energy (or Casimir).
The distribution for the thermal atmosphere matter entropy is qualitatively similar, though more challenging to plot due to the lack of exponential suppression of one of the integrands.\footnote{One also has to be careful in that case to take the real part of the expression in the end or, alternatively, to average the two time-orderings in the real-time two-point function.} Hence also the probability distribution of the matter entanglement entropy becomes broader as one moves the entangling surface at larger locations $z$ (closer to the black hole horizon).

Including non-perturbative corrections, we are then led to the probability distribution:
\begin{align}
\label{eq:Prmt}
P[\ell] &\sim \int dE_1 dE_2 \,\langle\rho(E_1)\rho(E_2)\rangle_{\text{RMT}} \, e^{-\ell_1 E_1}e^{-\ell_2 E_2} K_{2i\sqrt{E_1}}(2e^{-\ell/2}) K_{2i\sqrt{E_2}}(2e^{-\ell/2}) \\
&=\int [dM] \, \left(\text{Tr}\left[ e^{-\ell_1 M }K_{2i \sqrt{M}}(2e^{-\ell/2})\right]\text{Tr}\left[ e^{-\ell_2 M }K_{2i \sqrt{M}}(2e^{-\ell/2})\right]\right) e^{- \text{Tr}V(M)},
\end{align}
which is just a sampling of the energy eigenvalues according to the JT RMT ensemble as earlier. For $\langle \ell \rangle$, this matches with the approach detailed in section 5 of \cite{Iliesiu:2024cnh}. If we want to proceed analytically, just as before, we can approximate the pair density correlator as
\begin{equation}
\label{eq:rmtpdc}
\expval{\rho(E_1)\rho(E_2)}_{\text{RMT}} \approx \rho_0(E_1)\rho_0(E_2)  - \frac{\sin^2\left[  \pi \rho_0 (E_2) \left(  E_1 -E_2 \right)  \right] }{\pi^2 \left(  E_1 - E_2 \right)^2 } + \rho_0(E_1)\delta(E_1 - E_2).
\end{equation}
The final contact term leads as before to a $t$-independent additive  contribution:
\begin{equation}
\int dE \rho_0(E) e^{-\beta E} (K_{2i\sqrt{E}}(2e^{-\ell/2}))^2 \,\, \overset{?}{\subset} \,\,P[\ell].
\end{equation}
However, the LHS can be shown to scale linearly in $\ell$ for large $\ell$. Just as earlier, this contact term is just giving a divergent contribution to the calculation that pushes us towards the large $\ell$ regime. Since it is independent of the actual variable of interest ($t$ or $z$), we again drop it here. 
The qualitative behavior of the remaining terms in \eqref{eq:rmtpdc} for $P[\ell]$ is very different at late times compared to the disk answer due to the sine kernel contribution. Indeed, in terms of the energy separation variable $\omega \equiv E_1-E_2$, at very late times we have $\omega \approx 0$, and the full RMT result exhibits quadratic level repulsion:
\begin{equation}\expval{\rho(E_1)\rho(E_2)}_{\text{RMT}} \approx \frac{\pi^2}{3}\omega^2 \rho_0(E)^4, \qquad \omega \ll e^{-S_0}/C.
\end{equation}
This means the $\omega$-integral in \eqref{eq:Prmt} gets its leading contribution from
\begin{equation}
\sim \int d\omega \, \omega^2 e^{it\omega}f(E), \qquad f(E) \approx \rho_0(E)^4 (K_{2i\sqrt{E}}(2e^{-\ell/2}))^2,
\end{equation}
where we used that the BesselK-function at hand varies slowly as $\omega \to 0$.\footnote{This is the same late-time vanishing that the spectral form factor has. In that case, the contact term in \eqref{eq:rmtpdc} gives the late-time plateau offset.} This resulting answer goes to zero faster at late times than the disk answer, which goes as $\sim \int d\omega \, e^{it\omega}\tilde{f}(E)$ which does not have the $\omega^2$-suppression. This indicates the entire probability distribution is almost everywhere further suppressed 
and, as we discussed above, with an average that reaches a plateau value at large times $t$ (or large radial coordinate $z$).

\begin{figure}[h]
\centering
\includegraphics[width=0.2\textwidth]{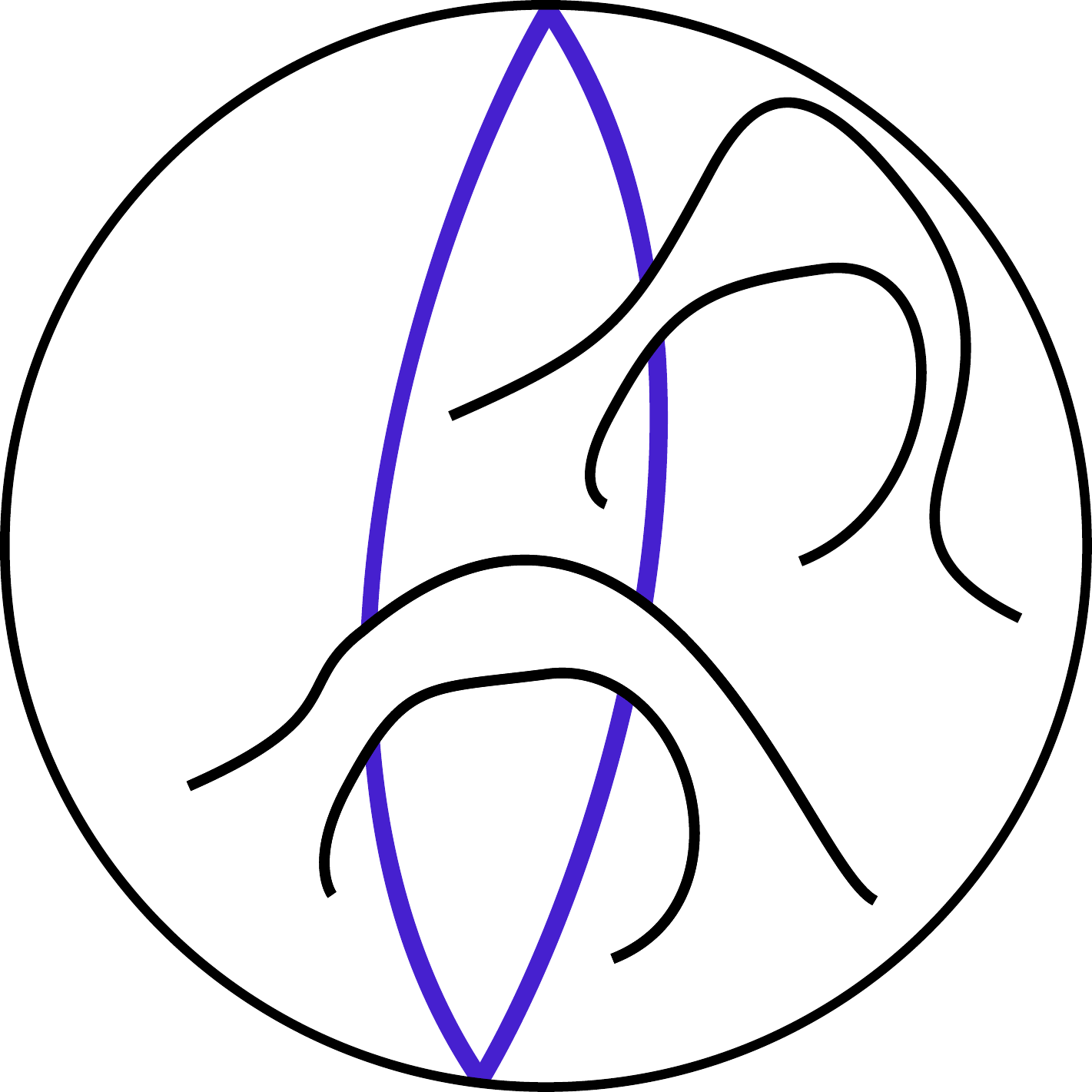}
\caption{Various wormhole contributions for the calculation of the variance $\langle \ell^2\rangle$. }
\label{figwormvarian}
\end{figure}

However, there are choices one makes in how one precisely defines the non-perturbative completion of the model. These ambiguities were also commented on in \cite{Iliesiu:2021ari,Iliesiu:2024cnh}; here we used a different definition than that studied in \cite{Iliesiu:2021ari}, based in our case on the ``quenched'' RMT ensemble. As an example of these choices, if one considers say the second moment, one can consider including or excluding various wormhole contributions, as shown in Fig.~\ref{figwormvarian}.
The non-perturbative expression \eqref{eq:Prmt} does not contain the wormholes that connect the ``inner'' region with the outer regions; only those directly connecting outer regions. This corresponds to first merging the bilocal lines on the disk (which is equivalent at the disk level), and only afterwards adding higher topology. This is in the spirit of the ``quenched'' matter-gravity matrix model utilized in this work.
On a technical level, including the ``inner'' wormhole contributions would lead to a dependence of the probability distribution on higher spectral correlators of the matrix model, as is also the case (but due to different reasons) in the computation of the variance in  \cite{Iliesiu:2021ari}.  

\bibliographystyle{ourbst}
\bibliography{bibliography.bib}

@article{Blommaert:2018oro,
    author = "Blommaert, Andreas and Mertens, Thomas G. and Verschelde, Henri",
    title = "{The Schwarzian Theory - A Wilson Line Perspective}",
    eprint = "1806.07765",
    archivePrefix = "arXiv",
    primaryClass = "hep-th",
    doi = "10.1007/JHEP12(2018)022",
    journal = "JHEP",
    volume = "12",
    pages = "022",
    year = "2018"
}

@article{Choi:2023syx,
    author = "Choi, Sangmin and Larsen, Finn",
    title = "{AdS$_{2}$ holography and effective QFT}",
    eprint = "2302.13917",
    archivePrefix = "arXiv",
    primaryClass = "hep-th",
    reportNumber = "LCTP-22-10, CPHT-RR035.052022",
    doi = "10.1007/JHEP11(2023)151",
    journal = "JHEP",
    volume = "11",
    pages = "151",
    year = "2023"
}

@article{Jafferis:2022uhu,
    author = "Jafferis, Daniel Louis and Kolchmeyer, David K. and Mukhametzhanov, Baur and Sonner, Julian",
    title = "{Matrix Models for Eigenstate Thermalization}",
    eprint = "2209.02130",
    archivePrefix = "arXiv",
    primaryClass = "hep-th",
    doi = "10.1103/PhysRevX.13.031033",
    journal = "Phys. Rev. X",
    volume = "13",
    number = "3",
    pages = "031033",
    year = "2023"
}

@article{Nag:1990dj,
    author = "Nag, S. and Verjovsky, A.",
    title = "{Diff S1 and the Teichmuller spaces}",
    doi = "10.1007/BF02099878",
    journal = "Commun. Math. Phys.",
    volume = "130",
    pages = "123--138",
    year = "1990"
}

@misc{nag1992tangentspaceuniversalteichmuller,
      title={On the Tangent Space to the Universal Teichmuller Space}, 
      author={Subhashis Nag},
      year={1992},
      eprint={alg-geom/9205007},
      archivePrefix={arXiv},
      primaryClass={alg-geom},
      url={https://arxiv.org/abs/alg-geom/9205007}, 
}

@book{hubbard:hal-01297628,
  TITLE = {{Teichm{\"u}ller Theory and Applications to Geometry, Topology, and Dynamics}},
  AUTHOR = {Hubbard, John H.},
  URL = {https://hal.science/hal-01297628},
  PUBLISHER = {{Matrix Editions}},
  SERIES = {Surface Homeomorphisms and Rational Functions},
  VOLUME = {2},
  PAGES = {262 pages, 108 color illustrations.},
  YEAR = {2016},
  MONTH = Apr,
  HAL_ID = {hal-01297628},
  HAL_VERSION = {v1},
}

@inproceedings{bers_automorphic_1965,
	address = {Berlin, Heidelberg},
	title = {Automorphic {Forms} and {General} {Teichmüller} {Spaces}},
	isbn = {978-3-642-48016-4},
	booktitle = {Proceedings of the {Conference} on {Complex} {Analysis}},
	publisher = {Springer Berlin Heidelberg},
	author = {Bers, L.},
	editor = {Aeppli, Alfred and Calabi, Eugenio and Röhrl, Helmut},
	year = {1965},
	pages = {109--113},
}

@article{Saad:2019lba,
    author = "Saad, Phil and Shenker, Stephen H. and Stanford, Douglas",
    title = "{JT gravity as a matrix integral}",
    eprint = "1903.11115",
    archivePrefix = "arXiv",
    primaryClass = "hep-th",
    month = "3",
    year = "2019"
}

@article{Saad:2018bqo,
    author = "Saad, Phil and Shenker, Stephen H. and Stanford, Douglas",
    title = "{A semiclassical ramp in SYK and in gravity}",
    eprint = "1806.06840",
    archivePrefix = "arXiv",
    primaryClass = "hep-th",
    month = "6",
    year = "2018"
}

@article{Saad:2019pqd,
    author = "Saad, Phil",
    title = "{Late Time Correlation Functions, Baby Universes, and ETH in JT Gravity}",
    eprint = "1910.10311",
    archivePrefix = "arXiv",
    primaryClass = "hep-th",
    month = "10",
    year = "2019"
}

@article{Jensen:2016pah,
    author = "Jensen, Kristan",
    title = "{Chaos in AdS$_2$ Holography}",
    eprint = "1605.06098",
    archivePrefix = "arXiv",
    primaryClass = "hep-th",
    doi = "10.1103/PhysRevLett.117.111601",
    journal = "Phys. Rev. Lett.",
    volume = "117",
    number = "11",
    pages = "111601",
    year = "2016"
}

@article{Bardeen:1973gs,
    author = "Bardeen, James M. and Carter, B. and Hawking, S. W.",
    title = "{The Four laws of black hole mechanics}",
    doi = "10.1007/BF01645742",
    journal = "Commun. Math. Phys.",
    volume = "31",
    pages = "161--170",
    year = "1973"
}

@article{Bardeen:1970zz,
    author = "Bardeen, James M.",
    title = "{Kerr Metric Black Holes}",
    doi = "10.1038/226064a0",
    journal = "Nature",
    volume = "226",
    pages = "64--65",
    year = "1970"
}

@article{Driessler:1976ky,
    author = "Driessler, W.",
    title = "{On the Type of Local Algebras in Quantum Field Theory}",
    reportNumber = "Print-76-0833 (BIELEFELD)",
    doi = "10.1007/BF01609853",
    journal = "Commun. Math. Phys.",
    volume = "53",
    pages = "295",
    year = "1977"
}

@article{Sorce:2023fdx,
    author = "Sorce, Jonathan",
    title = "{Notes on the type classification of von Neumann algebras}",
    eprint = "2302.01958",
    archivePrefix = "arXiv",
    primaryClass = "hep-th",
    reportNumber = "MIT-CTP/5527",
    doi = "10.1142/S0129055X24300024",
    journal = "Rev. Math. Phys.",
    volume = "36",
    number = "02",
    pages = "2430002",
    year = "2024"
}

@article{Kubo:1957mj,
    author = "Kubo, Ryogo",
    title = "{Statistical mechanical theory of irreversible processes. 1. General theory and simple applications in magnetic and conduction problems}",
    doi = "10.1143/JPSJ.12.570",
    journal = "J. Phys. Soc. Jap.",
    volume = "12",
    pages = "570--586",
    year = "1957"
}

@article{Martin:1959jp,
    author = "Martin, Paul C. and Schwinger, Julian S.",
    editor = "Milton, K. A.",
    title = "{Theory of many particle systems. 1.}",
    doi = "10.1103/PhysRev.115.1342",
    journal = "Phys. Rev.",
    volume = "115",
    pages = "1342--1373",
    year = "1959"
}

@article{Araki:1964lyc,
    author = "Araki, Huzihiro",
    title = "{Type of von Neumann Algebra Associated with Free Field}",
    doi = "10.1143/ptp.32.956",
    journal = "Prog. Theor. Phys.",
    volume = "32",
    number = "6",
    pages = "956--965",
    year = "1964"
}

@article{Sorce:2023gio,
    author = "Sorce, Jonathan",
    title = "{An intuitive construction of modular flow}",
    eprint = "2309.16766",
    archivePrefix = "arXiv",
    primaryClass = "hep-th",
    reportNumber = "MIT-CTP/5622",
    doi = "10.1007/JHEP12(2023)079",
    journal = "JHEP",
    volume = "12",
    pages = "079",
    year = "2023"
}

@article{SorceGeometricModularFlow,
    author = "Sorce, Jonathan",
    title = "{Analyticity and the Unruh effect: a study of local modular flow}",
    eprint = "2403.18937",
    archivePrefix = "arXiv",
    primaryClass = "hep-th",
    reportNumber = "MIT-CTP/5699",
    doi = "10.1007/JHEP09(2024)040",
    journal = "JHEP",
    volume = "24",
    pages = "040",
    year = "2024"
}

@article{Iliesiu:2021ari,
    author = "Iliesiu, Luca V. and Mezei, M\'ark and S\'arosi, G\'abor",
    title = "{The volume of the black hole interior at late times}",
    eprint = "2107.06286",
    archivePrefix = "arXiv",
    primaryClass = "hep-th",
    reportNumber = "CERN-TH-2021-102",
    doi = "10.1007/JHEP07(2022)073",
    journal = "JHEP",
    volume = "07",
    pages = "073",
    year = "2022"
}

@article{Hernandez-Cuenca:2024icn,
    author = "Hern{\'a}ndez-Cuenca, Sergio",
    title = "{Entropy and spectrum of near-extremal black holes: semiclassical brane solutions to non-perturbative problems}",
    eprint = "2407.20321",
    archivePrefix = "arXiv",
    primaryClass = "hep-th",
    reportNumber = "MIT-CTP/5741",
    doi = "10.1007/JHEP05(2025)020",
    journal = "JHEP",
    volume = "05",
    pages = "020",
    year = "2025"
}

@article{Antonini:2025nir,
    author = "Antonini, Stefano and Iliesiu, Luca V. and Rath, Pratik and Tran, Patrick",
    title = "{A Black Hole Airy Tail}",
    eprint = "2507.10657",
    archivePrefix = "arXiv",
    primaryClass = "hep-th",
    month = "7",
    year = "2025"
}

@article{Mertens:2019bvy,
    author = "Mertens, Thomas G.",
    title = "{Towards Black Hole Evaporation in Jackiw-Teitelboim Gravity}",
    eprint = "1903.10485",
    archivePrefix = "arXiv",
    primaryClass = "hep-th",
    doi = "10.1007/JHEP07(2019)097",
    journal = "JHEP",
    volume = "07",
    pages = "097",
    year = "2019"
}

@article{Blommaert:2019hjr,
    author = "Blommaert, Andreas and Mertens, Thomas G. and Verschelde, Henri",
    title = "{Clocks and Rods in Jackiw-Teitelboim Quantum Gravity}",
    eprint = "1902.11194",
    archivePrefix = "arXiv",
    primaryClass = "hep-th",
    doi = "10.1007/JHEP09(2019)060",
    journal = "JHEP",
    volume = "09",
    pages = "060",
    year = "2019"
}

@article{Blommaert2021,
    author = "Blommaert, Andreas and Mertens, Thomas G. and Verschelde, Henri",
    title = "{Unruh detectors and quantum chaos in JT gravity}",
    eprint = "2005.13058",
    archivePrefix = "arXiv",
    primaryClass = "hep-th",
    doi = "10.1007/JHEP03(2021)086",
    journal = "JHEP",
    volume = "03",
    pages = "086",
    year = "2021"
}

@article{ShadiCrossedProduct,
    author = "Ali Ahmad, Shadi and Jefferson, Ro",
    title = "{Crossed product algebras and generalized entropy for subregions}",
    eprint = "2306.07323",
    archivePrefix = "arXiv",
    primaryClass = "hep-th",
    doi = "10.21468/SciPostPhysCore.7.2.020",
    journal = "SciPost Phys. Core",
    volume = "7",
    pages = "020",
    year = "2024"
}

@article{kudlerflam2023covariantregulatorentanglemententropy,
    author = "Kudler-Flam, Jonah and Leutheusser, Samuel and Rahman, Adel A. and Satishchandran, Gautam and Speranza, Antony J.",
    title = "{Covariant regulator for entanglement entropy: Proofs of the Bekenstein bound and the quantum null energy condition}",
    eprint = "2312.07646",
    archivePrefix = "arXiv",
    primaryClass = "hep-th",
    doi = "10.1103/PhysRevD.111.105001",
    journal = "Phys. Rev. D",
    volume = "111",
    number = "10",
    pages = "105001",
    year = "2025"
}

@article{WittenCrossedProduct,
    author = "Witten, Edward",
    title = "{Gravity and the crossed product}",
    eprint = "2112.12828",
    archivePrefix = "arXiv",
    primaryClass = "hep-th",
    doi = "10.1007/JHEP10(2022)008",
    journal = "JHEP",
    volume = "10",
    pages = "008",
    year = "2022"
}

@article{tHooft:1984kcu,
    author = "'t Hooft, Gerard",
    title = "{On the Quantum Structure of a Black Hole}",
    reportNumber = "Print-84-0924 (UTRECHT)",
    doi = "10.1016/0550-3213(85)90418-3",
    journal = "Nucl. Phys. B",
    volume = "256",
    pages = "727--745",
    year = "1985"
}

@article{Stanford:2017thb,
    author = "Stanford, Douglas and Witten, Edward",
    title = "{Fermionic Localization of the Schwarzian Theory}",
    eprint = "1703.04612",
    archivePrefix = "arXiv",
    primaryClass = "hep-th",
    doi = "10.1007/JHEP10(2017)008",
    journal = "JHEP",
    volume = "10",
    pages = "008",
    year = "2017"
}

@article{Kolchmeyer:2023gwa,
    author = "Kolchmeyer, David K.",
    title = "{von Neumann algebras in JT gravity}",
    eprint = "2303.04701",
    archivePrefix = "arXiv",
    primaryClass = "hep-th",
    doi = "10.1007/JHEP06(2023)067",
    journal = "JHEP",
    volume = "06",
    pages = "067",
    year = "2023"
}

@article{Gao:2024gky,
    author = "Gao, Ping",
    title = "{Modular flow in JT gravity and entanglement wedge reconstruction}",
    eprint = "2402.18655",
    archivePrefix = "arXiv",
    primaryClass = "hep-th",
    doi = "10.1007/JHEP06(2024)151",
    journal = "JHEP",
    volume = "06",
    pages = "151",
    year = "2024"
}

@article{DeVuyst:2024fxc,
    author = "De Vuyst, Julian and Eccles, Stefan and Hoehn, Philipp A. and Kirklin, Josh",
    title = "{Crossed products and quantum reference frames: on the observer-dependence of gravitational entropy}",
    eprint = "2412.15502",
    archivePrefix = "arXiv",
    primaryClass = "hep-th",
    doi = "10.1007/JHEP07(2025)063",
    journal = "JHEP",
    volume = "07",
    pages = "063",
    year = "2025"
}

@article{Betzios:2025sct,
    author = "Betzios, Panos and Papadoulaki, Olga and Zhou, Yanjun",
    title = "{Near-extremal quantum cross-section for charged fields and superradiance}",
    eprint = "2507.13896",
    archivePrefix = "arXiv",
    primaryClass = "hep-th",
    month = "7",
    year = "2025"
}

@article{Penington:2023dql,
    author = "Penington, Geoff and Witten, Edward",
    title = "{Algebras and States in JT Gravity}",
    eprint = "2301.07257",
    archivePrefix = "arXiv",
    primaryClass = "hep-th",
    month = "1",
    year = "2023"
}

@Article{Peleska1984,
author={Peleska, Jan},
title={A characterization for isometries and conformal mappings of pseudo-Riemannian manifolds},
journal={aequationes mathematicae},
year={1984},
month={Mar},
day={01},
volume={27},
number={1},
pages={20-31},
issn={1420-8903},
doi={10.1007/BF02192656},
url={https://doi.org/10.1007/BF02192656}
}

@book{Takesaki:1970aki,
    author = "Takesaki, M.",
    title = "{Tomita's Theory of Modular Hilbert Algebras and its Applications}",
    doi = "10.1007/bfb0065832",
    publisher = "Springer-Verlag",
    series = "Lecture Notes in Mathematics",
    year = "1970"
}

@book{takesaki2002theory,
  title={Theory of Operator Algebras II},
  author={Takesaki, M.},
  isbn={9783540429142},
  lccn={2003542762},
  series={Encyclopaedia of Mathematical Sciences},
  year={2002},
  publisher={Springer Berlin Heidelberg}
}

@article{Chandrasekaran:2022eqq,
    author = "Chandrasekaran, Venkatesa and Penington, Geoff and Witten, Edward",
    title = "{Large N algebras and generalized entropy}",
    eprint = "2209.10454",
    archivePrefix = "arXiv",
    primaryClass = "hep-th",
    doi = "10.1007/JHEP04(2023)009",
    journal = "JHEP",
    volume = "04",
    pages = "009",
    year = "2023"
}

@article{Chandrasekaran:2022cip,
    author = "Chandrasekaran, Venkatesa and Longo, Roberto and Penington, Geoff and Witten, Edward",
    title = "{An algebra of observables for de Sitter space}",
    eprint = "2206.10780",
    archivePrefix = "arXiv",
    primaryClass = "hep-th",
    doi = "10.1007/JHEP02(2023)082",
    journal = "JHEP",
    volume = "02",
    pages = "082",
    year = "2023"
}

@article{Jensen:2023yxy,
    author = "Jensen, Kristan and Sorce, Jonathan and Speranza, Antony J.",
    title = "{Generalized entropy for general subregions in quantum gravity}",
    eprint = "2306.01837",
    archivePrefix = "arXiv",
    primaryClass = "hep-th",
    doi = "10.1007/JHEP12(2023)020",
    journal = "JHEP",
    volume = "12",
    pages = "020",
    year = "2023"
}

@article{Nitti:2024iyj,
    author = "Nitti, Francesco and Piazza, Federico and Taskov, Alexander",
    title = "{Relativity of the event: examples in JT gravity and linearized GR}",
    eprint = "2402.01847",
    archivePrefix = "arXiv",
    primaryClass = "hep-th",
    doi = "10.1007/JHEP10(2024)092",
    journal = "JHEP",
    volume = "10",
    pages = "092",
    year = "2024"
}

@article{Brown:2024ajk,
    author = "Brown, Adam R. and Iliesiu, Luca V. and Penington, Geoff and Usatyuk, Mykhaylo",
    title = "{The evaporation of charged black holes}",
    eprint = "2411.03447",
    archivePrefix = "arXiv",
    primaryClass = "hep-th",
    month = "11",
    year = "2024"
}

@article{Lin:2025wof,
    author = "Lin, Guanda and Iliesiu, Luca V. and Usatyuk, Mykhaylo",
    title = "{The evaporation of black holes in supergravity}",
    eprint = "2504.21077",
    archivePrefix = "arXiv",
    primaryClass = "hep-th",
    month = "4",
    year = "2025"
}

@article{Maulik:2025hax,
    author = "Maulik, Sabyasachi and Meng, Xin and Pando Zayas, Leopoldo A.",
    title = "{Quantum-Corrected Hawking Radiation from Near-Extremal Kerr-Newman Black Holes}",
    eprint = "2501.08252",
    archivePrefix = "arXiv",
    primaryClass = "hep-th",
    reportNumber = "LCTP-24-22",
    month = "1",
    year = "2025"
}

@article{Emparan:2025sao,
    author = "Emparan, Roberto",
    title = "{Quantum cross-section of near-extremal black holes}",
    eprint = "2501.17470",
    archivePrefix = "arXiv",
    primaryClass = "hep-th",
    doi = "10.1007/JHEP04(2025)122",
    journal = "JHEP",
    volume = "04",
    pages = "122",
    year = "2025"
}

@article{Biggs:2025nzs,
    author = "Biggs, Anna",
    title = "{Following the state of an evaporating charged black hole into the quantum gravity regime}",
    eprint = "2503.02051",
    archivePrefix = "arXiv",
    primaryClass = "hep-th",
    month = "3",
    year = "2025"
}

@article{Castellani:2005fkd,
    author = "Castellani, Tommaso and Cavagna, Andrea",
    title = "{Spin-glass theory for pedestrians}",
    doi = "10.1088/1742-5468/2005/05/P05012",
    journal = "J. Phys. A",
    volume = "2005",
    number = "05",
    pages = "P05012",
    year = "2005"
}

@article{Susskind:1994sm,
    author = "Susskind, Leonard and Uglum, John",
    title = "{Black hole entropy in canonical quantum gravity and superstring theory}",
    eprint = "hep-th/9401070",
    archivePrefix = "arXiv",
    reportNumber = "SU-ITP-94-1",
    doi = "10.1103/PhysRevD.50.2700",
    journal = "Phys. Rev. D",
    volume = "50",
    pages = "2700--2711",
    year = "1994"
}

@article{DeVuyst:2022bua,
    author = "De Vuyst, Julian and Mertens, Thomas G.",
    title = "{Operational islands and black hole dissipation in JT gravity}",
    eprint = "2207.03351",
    archivePrefix = "arXiv",
    primaryClass = "hep-th",
    doi = "10.1007/JHEP01(2023)027",
    journal = "JHEP",
    volume = "01",
    pages = "027",
    year = "2023"
}

@article{Leutheusser:2021qhd,
    author = "Leutheusser, Samuel and Liu, Hong",
    title = "{Causal connectability between quantum systems and the black hole interior in holographic duality}",
    eprint = "2110.05497",
    archivePrefix = "arXiv",
    primaryClass = "hep-th",
    reportNumber = "MIT-CTP/5335",
    doi = "10.1103/PhysRevD.108.086019",
    journal = "Phys. Rev. D",
    volume = "108",
    number = "8",
    pages = "086019",
    year = "2023"
}

@article{Leutheusser:2021frk,
    author = "Leutheusser, Samuel Aaron Wehlau and Liu, Hong",
    title = "{Emergent Times in Holographic Duality}",
    eprint = "2112.12156",
    archivePrefix = "arXiv",
    primaryClass = "hep-th",
    reportNumber = "MIT-CTP/5382",
    doi = "10.1103/PhysRevD.108.086020",
    journal = "Phys. Rev. D",
    volume = "108",
    number = "8",
    pages = "086020",
    year = "2023"
}

@article{Donnelly:2015hta,
    author = "Donnelly, William and Giddings, Steven B.",
    title = "{Diffeomorphism-invariant observables and their nonlocal algebra}",
    eprint = "1507.07921",
    archivePrefix = "arXiv",
    primaryClass = "hep-th",
    reportNumber = "NSF-KITP-15-133",
    doi = "10.1103/PhysRevD.93.024030",
    journal = "Phys. Rev. D",
    volume = "93",
    number = "2",
    pages = "024030",
    year = "2016",
    note = "[Erratum: Phys.Rev.D 94, 029903 (2016)]"
}

@article{Stanford:2021bhl,
    author = "Stanford, Douglas and Yang, Zhenbin and Yao, Shunyu",
    title = "{Subleading Weingartens}",
    eprint = "2107.10252",
    archivePrefix = "arXiv",
    primaryClass = "hep-th",
    doi = "10.1007/JHEP02(2022)200",
    journal = "JHEP",
    volume = "02",
    pages = "200",
    year = "2022"
}

@article{Blommaert:2020seb,
    author = "Blommaert, Andreas",
    title = "{Dissecting the ensemble in JT gravity}",
    eprint = "2006.13971",
    archivePrefix = "arXiv",
    primaryClass = "hep-th",
    doi = "10.1007/JHEP09(2022)075",
    journal = "JHEP",
    volume = "09",
    pages = "075",
    year = "2022"
}

@article{Jafferis:2020ora,
    author = "Jafferis, Daniel Louis and Lamprou, Lampros",
    title = "{Inside the hologram: reconstructing the bulk observer\textquoteright{}s experience}",
    eprint = "2009.04476",
    archivePrefix = "arXiv",
    primaryClass = "hep-th",
    doi = "10.1007/JHEP03(2022)084",
    journal = "JHEP",
    volume = "03",
    pages = "084",
    year = "2022"
}

@article{deBoer:2022zps,
    author = "de Boer, Jan and Jafferis, Daniel Louis and Lamprou, Lampros",
    title = "{On black hole interior reconstruction, singularities and the emergence of time}",
    eprint = "2211.16512",
    archivePrefix = "arXiv",
    primaryClass = "hep-th",
    month = "11",
    year = "2022"
}

@article{Belaey:2024dde,
    author = "Belaey, Andreas and Mariani, Francesca and Mertens, Thomas G.",
    title = "{Gravitational wavefunctions in JT supergravity}",
    eprint = "2405.09289",
    archivePrefix = "arXiv",
    primaryClass = "hep-th",
    doi = "10.1007/JHEP10(2024)037",
    journal = "JHEP",
    volume = "10",
    pages = "037",
    year = "2024"
}

@article{Yang:2018gdb,
    author = "Yang, Zhenbin",
    title = "{The Quantum Gravity Dynamics of Near Extremal Black Holes}",
    eprint = "1809.08647",
    archivePrefix = "arXiv",
    primaryClass = "hep-th",
    doi = "10.1007/JHEP05(2019)205",
    journal = "JHEP",
    volume = "05",
    pages = "205",
    year = "2019"
}

@article{Maldacena:2016upp,
    author = "Maldacena, Juan and Stanford, Douglas and Yang, Zhenbin",
    title = "{Conformal symmetry and its breaking in two dimensional Nearly Anti-de-Sitter space}",
    eprint = "1606.01857",
    archivePrefix = "arXiv",
    primaryClass = "hep-th",
    doi = "10.1093/ptep/ptw124",
    journal = "PTEP",
    volume = "2016",
    number = "12",
    pages = "12C104",
    year = "2016"
}

@article{Mertens:2022irh,
    author = "Mertens, Thomas G. and Turiaci, Gustavo J.",
    title = "{Solvable models of quantum black holes: a review on Jackiw{\textendash}Teitelboim gravity}",
    eprint = "2210.10846",
    archivePrefix = "arXiv",
    primaryClass = "hep-th",
    doi = "10.1007/s41114-023-00046-1",
    journal = "Living Rev. Rel.",
    volume = "26",
    number = "1",
    pages = "4",
    year = "2023"
}

@article{Bisognano:1975ih,
    author = "Bisognano, J. J and Wichmann, E. H.",
    title = "{On the Duality Condition for a Hermitian Scalar Field}",
    doi = "10.1063/1.522605",
    journal = "J. Math. Phys.",
    volume = "16",
    pages = "985--1007",
    year = "1975"
}

@article{Sewell:1982zz,
    author = "Sewell, Geoffrey L.",
    title = "{Quantum fields on manifolds: PCT and gravitationally induced thermal states}",
    doi = "10.1016/0003-4916(82)90285-8",
    journal = "Annals Phys.",
    volume = "141",
    pages = "201--224",
    year = "1982"
}

@article{Kay:1988mu,
    author = "Kay, Bernard S. and Wald, Robert M.",
    title = "{Theorems on the Uniqueness and Thermal Properties of Stationary, Nonsingular, Quasifree States on Space-Times with a Bifurcate Killing Horizon}",
    reportNumber = "PRINT-88-0840 (CHICAGO)",
    doi = "10.1016/0370-1573(91)90015-E",
    journal = "Phys. Rept.",
    volume = "207",
    pages = "49--136",
    year = "1991"
}

@article{Bisognano:1976za,
    author = "Bisognano, J. J and Wichmann, E. H.",
    title = "{On the Duality Condition for Quantum Fields}",
    doi = "10.1063/1.522898",
    journal = "J. Math. Phys.",
    volume = "17",
    pages = "303--321",
    year = "1976"
}

@article{Witten:2018zxz,
    author = "Witten, Edward",
    title = "{APS Medal for Exceptional Achievement in Research: Invited article on entanglement properties of quantum field theory}",
    eprint = "1803.04993",
    archivePrefix = "arXiv",
    primaryClass = "hep-th",
    doi = "10.1103/RevModPhys.90.045003",
    journal = "Rev. Mod. Phys.",
    volume = "90",
    number = "4",
    pages = "045003",
    year = "2018"
}

@article{Jackiw:1984je,
      author         = "Jackiw, R.",
      title          = "{Lower Dimensional Gravity}",
      booktitle      = "{1984 Meeting of the Division of Particles and Fields of
                        the APS Santa Fe, New Mexico, October 31-November 3,
                        1984}",
      journal        = "Nucl. Phys.",
      volume         = "B252",
      year           = "1985",
      pages          = "343-356",
      doi            = "10.1016/0550-3213(85)90448-1",
      reportNumber   = "MIT-CTP-1203",
      SLACcitation   = "%%CITATION = NUPHA,B252,343;%%"
}

@article{Teitelboim:1983ux,
      author         = "Teitelboim, C.",
      title          = "{Gravitation and Hamiltonian Structure in Two Space-Time
                        Dimensions}",
      journal        = "Phys. Lett.",
      volume         = "126B",
      year           = "1983",
      pages          = "41-45",
      doi            = "10.1016/0370-2693(83)90012-6",
      SLACcitation   = "%%CITATION = PHLTA,126B,41;%%"
}

@article{Almheiri:2014cka,
      author         = "Almheiri, Ahmed and Polchinski, Joseph",
      title          = "{Models of AdS$_{2}$ backreaction and holography}",
      journal        = "JHEP",
      volume         = "11",
      year           = "2015",
      pages          = "014",
      doi            = "10.1007/JHEP11(2015)014",
      eprint         = "1402.6334",
      archivePrefix  = "arXiv",
      primaryClass   = "hep-th",
      SLACcitation   = "%%CITATION = ARXIV:1402.6334;%%"
}

@article{Witten:2023qsv,
    author = "Witten, Edward",
    title = "{Algebras, regions, and observers.}",
    eprint = "2303.02837",
    archivePrefix = "arXiv",
    primaryClass = "hep-th",
    doi = "10.1090/pspum/107/01954",
    journal = "Proc. Symp. Pure Math.",
    volume = "107",
    pages = "247--276",
    year = "2024"
}

@article{Witten:2023xze,
    author = "Witten, Edward",
    title = "{A background-independent algebra in quantum gravity}",
    eprint = "2308.03663",
    archivePrefix = "arXiv",
    primaryClass = "hep-th",
    doi = "10.1007/JHEP03(2024)077",
    journal = "JHEP",
    volume = "03",
    pages = "077",
    year = "2024"
}

@article{Akers:2024bel,
    author = "Akers, Chris and Sorce, Jonathan",
    title = "{Relative State Counting for Semiclassical Black Holes}",
    eprint = "2404.16098",
    archivePrefix = "arXiv",
    primaryClass = "hep-th",
    reportNumber = "MIT-CTP/5712",
    doi = "10.1103/PhysRevLett.133.201601",
    journal = "Phys. Rev. Lett.",
    volume = "133",
    number = "20",
    pages = "201601",
    year = "2024"
}

@article{Kudler-Flam:2024psh,
    author = "Kudler-Flam, Jonah and Leutheusser, Samuel and Satishchandran, Gautam",
    title = "{Algebraic Observational Cosmology}",
    eprint = "2406.01669",
    archivePrefix = "arXiv",
    primaryClass = "hep-th",
    month = "6",
    year = "2024"
}

@article{Chen:2024rpx,
    author = "Chen, Chang-Han and Penington, Geoff",
    title = "{A clock is just a way to tell the time: gravitational algebras in cosmological spacetimes}",
    eprint = "2406.02116",
    archivePrefix = "arXiv",
    primaryClass = "hep-th",
    month = "6",
    year = "2024"
}

@book{Haag:1996hvx,
    author = "Haag, Rudolf",
    title = "{Local Quantum Physics}",
    doi = "10.1007/978-3-642-61458-3",
    isbn = "978-3-540-61049-6, 978-3-642-61458-3",
    publisher = "Springer",
    address = "Berlin",
    series = "Theoretical and Mathematical Physics",
    year = "1996"
}

@article{Fan:2021wsb,
    author = "Fan, Yale and Mertens, Thomas G.",
    title = "{Supergroup structure of Jackiw-Teitelboim supergravity}",
    eprint = "2106.09353",
    archivePrefix = "arXiv",
    primaryClass = "hep-th",
    reportNumber = "UTTG-05-21",
    doi = "10.1007/JHEP08(2022)002",
    journal = "JHEP",
    volume = "08",
    pages = "002",
    year = "2022"
}

@article{Jafferis:2022wez,
    author = "Jafferis, Daniel Louis and Kolchmeyer, David K. and Mukhametzhanov, Baur and Sonner, Julian",
    title = "{Jackiw-Teitelboim gravity with matter, generalized eigenstate thermalization hypothesis, and random matrices}",
    eprint = "2209.02131",
    archivePrefix = "arXiv",
    primaryClass = "hep-th",
    doi = "10.1103/PhysRevD.108.066015",
    journal = "Phys. Rev. D",
    volume = "108",
    number = "6",
    pages = "066015",
    year = "2023"
}

@article{Emparan:2025qqf,
    author = "Emparan, Roberto and Trezzi, Stefano",
    title = "{Quantum Transparency of Near-extremal Black Holes}",
    eprint = "2507.03398",
    archivePrefix = "arXiv",
    primaryClass = "hep-th",
    month = "7",
    year = "2025"
}

@article{Cotler:2016fpe,
    author = "Cotler, Jordan S. and Gur-Ari, Guy and Hanada, Masanori and Polchinski, Joseph and Saad, Phil and Shenker, Stephen H. and Stanford, Douglas and Streicher, Alexandre and Tezuka, Masaki",
    title = "{Black Holes and Random Matrices}",
    eprint = "1611.04650",
    archivePrefix = "arXiv",
    primaryClass = "hep-th",
    reportNumber = "SU-ITP-16-19, SU-ITP-16/19, YITP-16-124",
    doi = "10.1007/JHEP05(2017)118",
    journal = "JHEP",
    volume = "05",
    pages = "118",
    year = "2017",
    note = "[Erratum: JHEP 09, 002 (2018)]"
}

@article{Berkooz:2018jqr,
    author = "Berkooz, Micha and Isachenkov, Mikhail and Narovlansky, Vladimir and Torrents, Genis",
    title = "{Towards a full solution of the large N double-scaled SYK model}",
    eprint = "1811.02584",
    archivePrefix = "arXiv",
    primaryClass = "hep-th",
    doi = "10.1007/JHEP03(2019)079",
    journal = "JHEP",
    volume = "03",
    pages = "079",
    year = "2019"
}

@article{Mertens:2017mtv,
    author = "Mertens, Thomas G. and Turiaci, Gustavo J. and Verlinde, Herman L.",
    title = "{Solving the Schwarzian via the Conformal Bootstrap}",
    eprint = "1705.08408",
    archivePrefix = "arXiv",
    primaryClass = "hep-th",
    doi = "10.1007/JHEP08(2017)136",
    journal = "JHEP",
    volume = "08",
    pages = "136",
    year = "2017"
}

@article{Berkooz:2022mfk,
    author = "Berkooz, Micha and Isachenkov, Mikhail and Narayan, Prithvi and Narovlansky, Vladimir",
    title = "{Quantum groups, non-commutative AdS$_{2}$, and chords in the double-scaled SYK model}",
    eprint = "2212.13668",
    archivePrefix = "arXiv",
    primaryClass = "hep-th",
    doi = "10.1007/JHEP08(2023)076",
    journal = "JHEP",
    volume = "08",
    pages = "076",
    year = "2023"
}

@article{Lin:2022rbf,
    author = "Lin, Henry W.",
    title = "{The bulk Hilbert space of double scaled SYK}",
    eprint = "2208.07032",
    archivePrefix = "arXiv",
    primaryClass = "hep-th",
    doi = "10.1007/JHEP11(2022)060",
    journal = "JHEP",
    volume = "11",
    pages = "060",
    year = "2022"
}

@article{Belaey:2025ijg,
    author = "Belaey, Andreas and Mertens, Thomas G. and Tappeiner, Thomas",
    title = "{Quantum group origins of edge states in double-scaled SYK}",
    eprint = "2503.20691",
    archivePrefix = "arXiv",
    primaryClass = "hep-th",
    month = "3",
    year = "2025"
}

@article{Blommaert:2023opb,
    author = "Blommaert, Andreas and Mertens, Thomas G. and Yao, Shunyu",
    title = "{Dynamical actions and q-representation theory for double-scaled SYK}",
    eprint = "2306.00941",
    archivePrefix = "arXiv",
    primaryClass = "hep-th",
    doi = "10.1007/JHEP02(2024)067",
    journal = "JHEP",
    volume = "02",
    pages = "067",
    year = "2024"
}

@article{Mertens:2020hbs,
    author = "Mertens, Thomas G. and Turiaci, Gustavo J.",
    title = "{Liouville quantum gravity -- holography, JT and matrices}",
    eprint = "2006.07072",
    archivePrefix = "arXiv",
    primaryClass = "hep-th",
    doi = "10.1007/JHEP01(2021)073",
    journal = "JHEP",
    volume = "01",
    pages = "073",
    year = "2021"
}

@article{Fan:2021bwt,
    author = "Fan, Yale and Mertens, Thomas G.",
    title = "{From quantum groups to Liouville and dilaton quantum gravity}",
    eprint = "2109.07770",
    archivePrefix = "arXiv",
    primaryClass = "hep-th",
    doi = "10.1007/JHEP05(2022)092",
    journal = "JHEP",
    volume = "05",
    pages = "092",
    year = "2022"
}

@article{Lin:2022zxd,
    author = "Lin, Henry W. and Maldacena, Juan and Rozenberg, Liza and Shan, Jieru",
    title = "{Looking at supersymmetric black holes for a very long time}",
    eprint = "2207.00408",
    archivePrefix = "arXiv",
    primaryClass = "hep-th",
    doi = "10.21468/SciPostPhys.14.5.128",
    journal = "SciPost Phys.",
    volume = "14",
    number = "5",
    pages = "128",
    year = "2023"
}

@article{Stanford:2019vob,
    author = "Stanford, Douglas and Witten, Edward",
    title = "{JT gravity and the ensembles of random matrix theory}",
    eprint = "1907.03363",
    archivePrefix = "arXiv",
    primaryClass = "hep-th",
    doi = "10.4310/ATMP.2020.v24.n6.a4",
    journal = "Adv. Theor. Math. Phys.",
    volume = "24",
    number = "6",
    pages = "1475--1680",
    year = "2020"
}

@article{Engelsoy:2016xyb,
      author         = "Engelsoy, Julius and Mertens, Thomas G. and Verlinde,
                        Herman",
      title          = "{An investigation of AdS$_{2}$ backreaction and
                        holography}",
      journal        = "JHEP",
      volume         = "07",
      year           = "2016",
      pages          = "139",
      doi            = "10.1007/JHEP07(2016)139",
      eprint         = "1606.03438",
      archivePrefix  = "arXiv",
      primaryClass   = "hep-th",
      SLACcitation   = "%%CITATION = ARXIV:1606.03438;%%"
}

@article{osti_4665531,
  author       = {Araki, H},
  title        = {A GENERALIZATION OF BORCHERS THEOREM},
  url          = {https://www.osti.gov/biblio/4665531},
  journal      = {Helvetica Physica Acta  (Switzerland)},
  volume       = {Vol: 36},
  place        = {Country unknown/Code not available},
  year         = {1963},
  month        = {01}}

@article{Strohmaier:2023hhy,
    author = "Strohmaier, Alexander and Witten, Edward",
    title = "{Analytic States in Quantum Field Theory on Curved Spacetimes}",
    eprint = "2302.02709",
    archivePrefix = "arXiv",
    primaryClass = "math-ph",
    doi = "10.1007/s00023-024-01419-0",
    journal = "Annales Henri Poincare",
    volume = "25",
    number = "10",
    pages = "4543--4590",
    year = "2024"
}

@article{Strohmaier:2023opz,
    author = "Strohmaier, Alexander and Witten, Edward",
    title = "{The Timelike Tube Theorem in Curved Spacetime}",
    eprint = "2303.16380",
    archivePrefix = "arXiv",
    primaryClass = "hep-th",
    doi = "10.1007/s00220-024-05009-3",
    journal = "Commun. Math. Phys.",
    volume = "405",
    number = "7",
    pages = "153",
    year = "2024"
}

@Article{Borchers1961,
author={Borchers, H. J.},
title={{\"U}ber die Vollst{\"a}ndigkeit lorentzinvarianter Felder in einer zeitartigen R{\"o}hre},
journal={Il Nuovo Cimento (1955-1965)},
year={1961},
month={Feb},
day={01},
volume={19},
number={4},
pages={787-793},
issn={1827-6121},
doi={10.1007/BF02733373},
url={https://doi.org/10.1007/BF02733373}
}

@article{Rovelli:1990ph,
    author = "Rovelli, Carlo",
    title = "{What Is Observable in Classical and Quantum Gravity?}",
    reportNumber = "PITT-90-10",
    doi = "10.1088/0264-9381/8/2/011",
    journal = "Class. Quant. Grav.",
    volume = "8",
    pages = "297--316",
    year = "1991"
}

@article{Giddings:2005id,
    author = "Giddings, Steven B. and Marolf, Donald and Hartle, James B.",
    title = "{Observables in effective gravity}",
    eprint = "hep-th/0512200",
    archivePrefix = "arXiv",
    doi = "10.1103/PhysRevD.74.064018",
    journal = "Phys. Rev. D",
    volume = "74",
    pages = "064018",
    year = "2006"
}

@article{Donnelly:2016rvo,
    author = "Donnelly, William and Giddings, Steven B.",
    title = "{Observables, gravitational dressing, and obstructions to locality and subsystems}",
    eprint = "1607.01025",
    archivePrefix = "arXiv",
    primaryClass = "hep-th",
    doi = "10.1103/PhysRevD.94.104038",
    journal = "Phys. Rev. D",
    volume = "94",
    number = "10",
    pages = "104038",
    year = "2016"
}

@article{Caminiti:2025hjq,
    author = "Caminiti, Jacqueline and Capeccia, Federico and Ciambelli, Luca and Myers, Robert C.",
    title = "{Geometric modular flows in 2d CFT and beyond}",
    eprint = "2502.02633",
    archivePrefix = "arXiv",
    primaryClass = "hep-th",
    doi = "10.1007/JHEP08(2025)166",
    journal = "JHEP",
    volume = "08",
    pages = "166",
    year = "2025"
}

@article{Klinger:2023auu,
    author = "Klinger, Marc S. and Leigh, Robert G.",
    title = "{Crossed products, conditional expectations and constraint quantization}",
    eprint = "2312.16678",
    archivePrefix = "arXiv",
    primaryClass = "hep-th",
    doi = "10.1016/j.nuclphysb.2024.116622",
    journal = "Nucl. Phys. B",
    volume = "1006",
    pages = "116622",
    year = "2024"
}

@book{Daele_1978, place={Cambridge}, series={London Mathematical Society Lecture Note Series}, title={Continuous Crossed Products and Type III Von Neumann Algebras}, publisher={Cambridge University Press}, author={Daele, A. van}, year={1978}, collection={London Mathematical Society Lecture Note Series}}

@article{SORCE2024110420,
title = {A short proof of {T}omita's theorem},
journal = {J. Funct. Anal.},
volume = {286},
number = {12},
pages = {110420},
year = {2024},
issn = {0022-1236},
doi = {https://doi.org/10.1016/j.jfa.2024.110420},
url = {https://www.sciencedirect.com/science/article/pii/S0022123624001083},
author = {Jonathan Sorce}
}

@article{Holzhey:1994we,
	Archiveprefix = {arXiv},
	Author = {Christoph Holzhey and Finn Larsen and Frank Wilczek},
	Doi = {10.1016/0550-3213(94)90402-2},
	Eprint = {hep-th/9403108},
	Journal = {Nucl.Phys.},
	Pages = {443--467},
	Primaryclass = {hep-th},
	Reportnumber = {PUPT-1454, IASSNS-HEP-93-88},
	Slaccitation = {%%CITATION = HEP-TH/9403108;%%},
	Title = {{Geometric and renormalized entropy in conformal field theory}},
	Volume = {B424},
	Year = {1994}}

@incollection{JT,
	Address = {Bristol},
	Author = {Jackiw, R. and Teitelboim, C.},
	Booktitle = {Quantum Theory Of Gravity},
	Editor = {Christensen, S.},
	Publisher = {Adam Hilger},
	Year = 1984}

@Article{ Solodukhin:2011gn,
	Archiveprefix = "arXiv",
	Author = "Sergey N. Solodukhin",
	Eprint = "1104.3712",
	Journal = "Living Rev.Rel.",
	Pages = "8",
	Primaryclass = "hep-th",
	Slaccitation = "%%CITATION = ARXIV:1104.3712;%%",
	Title = "{Entanglement entropy of black holes}",
	Volume = "14",
	Year = "2011"
}

@article{Dabholkar:1994ai,
    author = "Dabholkar, Atish",
    editor = "Das, Sumit R. and Mandal, Gautam and Mukhi, S. and Wadia, S. R.",
    title = "{Strings on a cone and black hole entropy}",
    eprint = "hep-th/9408098",
    archivePrefix = "arXiv",
    reportNumber = "HUTP-94-A019",
    doi = "10.1016/0550-3213(95)00050-3",
    journal = "Nucl. Phys. B",
    volume = "439",
    pages = "650--664",
    year = "1995"
}

@article{Lowe:1994ah,
    author = "Lowe, David A. and Strominger, Andrew",
    title = "{Strings near a Rindler or black hole horizon}",
    eprint = "hep-th/9410215",
    archivePrefix = "arXiv",
    reportNumber = "UCSBTH-94-42",
    doi = "10.1103/PhysRevD.51.1793",
    journal = "Phys. Rev. D",
    volume = "51",
    pages = "1793--1799",
    year = "1995"
}

@article{He:2014gva,
    author = "He, Song and Numasawa, Tokiro and Takayanagi, Tadashi and Watanabe, Kento",
    title = "{Notes on Entanglement Entropy in String Theory}",
    eprint = "1412.5606",
    archivePrefix = "arXiv",
    primaryClass = "hep-th",
    reportNumber = "YITP-14-105, IPMU14-0358",
    doi = "10.1007/JHEP05(2015)106",
    journal = "JHEP",
    volume = "05",
    pages = "106",
    year = "2015"
}

@article{Mertens:2016tqv,
    author = "Mertens, Thomas G. and Verschelde, Henri and Zakharov, Valentin I.",
    title = "{String Theory in Polar Coordinates and the Vanishing of the One-Loop Rindler Entropy}",
    eprint = "1606.06632",
    archivePrefix = "arXiv",
    primaryClass = "hep-th",
    doi = "10.1007/JHEP08(2016)113",
    journal = "JHEP",
    volume = "08",
    pages = "113",
    year = "2016"
}

@article{Balasubramanian:2018axm,
    author = "Balasubramanian, Vijay and Parrikar, Onkar",
    title = "{Remarks on entanglement entropy in string theory}",
    eprint = "1801.03517",
    archivePrefix = "arXiv",
    primaryClass = "hep-th",
    doi = "10.1103/PhysRevD.97.066025",
    journal = "Phys. Rev. D",
    volume = "97",
    number = "6",
    pages = "066025",
    year = "2018"
}

@article{Witten:2018xfj,
    author = "Witten, Edward",
    title = "{Open Strings On The Rindler Horizon}",
    eprint = "1810.11912",
    archivePrefix = "arXiv",
    primaryClass = "hep-th",
    doi = "10.1007/JHEP01(2019)126",
    journal = "JHEP",
    volume = "01",
    pages = "126",
    year = "2019"
}

@article{Mazenc:2019ety,
    author = "Mazenc, Edward A. and Ranard, Daniel",
    title = "{Target space entanglement entropy}",
    eprint = "1910.07449",
    archivePrefix = "arXiv",
    primaryClass = "hep-th",
    doi = "10.1007/JHEP03(2023)111",
    journal = "JHEP",
    volume = "03",
    pages = "111",
    year = "2023"
}

@article{Dabholkar:2022mxo,
    author = "Dabholkar, Atish",
    title = "{Quantum Entanglement in String Theory}",
    eprint = "2207.03624",
    archivePrefix = "arXiv",
    primaryClass = "hep-th",
    month = "7",
    year = "2022"
}

@article{Dabholkar:2023ows,
    author = "Dabholkar, Atish and Moitra, Upamanyu",
    title = "{Finite entanglement entropy in string theory}",
    eprint = "2306.00990",
    archivePrefix = "arXiv",
    primaryClass = "hep-th",
    doi = "10.1103/PhysRevD.109.L121901",
    journal = "Phys. Rev. D",
    volume = "109",
    number = "12",
    pages = "L121901",
    year = "2024"
}

@article{Susskind:1993ws,
    author = "Susskind, Leonard",
    editor = "Teitelboim, C. and Zanelli, J.",
    title = "{Some speculations about black hole entropy in string theory}",
    eprint = "hep-th/9309145",
    archivePrefix = "arXiv",
    reportNumber = "RU-93-44",
    pages = "118--131",
    month = "10",
    year = "1993"
}

@article{Horowitz:1996nw,
    author = "Horowitz, Gary T. and Polchinski, Joseph",
    title = "{A Correspondence principle for black holes and strings}",
    eprint = "hep-th/9612146",
    archivePrefix = "arXiv",
    reportNumber = "NSF-ITP-96-144",
    doi = "10.1103/PhysRevD.55.6189",
    journal = "Phys. Rev. D",
    volume = "55",
    pages = "6189--6197",
    year = "1997"
}

@article{Chen:2021emg,
    author = "Chen, Yiming and Maldacena, Juan",
    title = "{String scale black holes at large D}",
    eprint = "2106.02169",
    archivePrefix = "arXiv",
    primaryClass = "hep-th",
    doi = "10.1007/JHEP01(2022)095",
    journal = "JHEP",
    volume = "01",
    pages = "095",
    year = "2022"
}

@article{Engelhardt:2020qpv,
    author = "Engelhardt, Netta and Fischetti, Sebastian and Maloney, Alexander",
    title = "{Free energy from replica wormholes}",
    eprint = "2007.07444",
    archivePrefix = "arXiv",
    primaryClass = "hep-th",
    doi = "10.1103/PhysRevD.103.046021",
    journal = "Phys. Rev. D",
    volume = "103",
    number = "4",
    pages = "046021",
    year = "2021"
}

@article{Johnson:2021rsh,
    author = "Johnson, Clifford V.",
    title = "{On the Quenched Free Energy of JT Gravity and Supergravity}",
    eprint = "2104.02733",
    archivePrefix = "arXiv",
    primaryClass = "hep-th",
    month = "4",
    year = "2021"
}

@article{Iliesiu:2024cnh,
    author = "Iliesiu, Luca V. and Levine, Adam and Lin, Henry W. and Maxfield, Henry and Mezei, M\'ark",
    title = "{On the non-perturbative bulk Hilbert space of JT gravity}",
    eprint = "2403.08696",
    archivePrefix = "arXiv",
    primaryClass = "hep-th",
    doi = "10.1007/JHEP10(2024)220",
    journal = "JHEP",
    volume = "10",
    pages = "220",
    year = "2024"
}

@article{Goeller:2022rsx,
    author = "Goeller, Christophe and Hoehn, Philipp A. and Kirklin, Josh",
    title = "{Diffeomorphism-invariant observables and dynamical frames in gravity: reconciling bulk locality with general covariance}",
    eprint = "2206.01193",
    archivePrefix = "arXiv",
    primaryClass = "hep-th",
    month = "6",
    year = "2022"
}

@article{Tambornino:2011vg,
    author = "Tambornino, Johannes",
    title = "{Relational Observables in Gravity: a Review}",
    eprint = "1109.0740",
    archivePrefix = "arXiv",
    primaryClass = "gr-qc",
    doi = "10.3842/SIGMA.2012.017",
    journal = "SIGMA",
    volume = "8",
    pages = "017",
    year = "2012"
}

@article{Chen:2021dsw,
    author = "Chen, Yiming and Maldacena, Juan and Witten, Edward",
    title = "{On the black hole/string transition}",
    eprint = "2109.08563",
    archivePrefix = "arXiv",
    primaryClass = "hep-th",
    doi = "10.1007/JHEP01(2023)103",
    journal = "JHEP",
    volume = "01",
    pages = "103",
    year = "2023"
}

@article{Urbach:2022xzw,
    author = "Urbach, Erez Y.",
    title = "{String stars in anti de Sitter space}",
    eprint = "2202.06966",
    archivePrefix = "arXiv",
    primaryClass = "hep-th",
    doi = "10.1007/JHEP04(2022)072",
    journal = "JHEP",
    volume = "04",
    pages = "072",
    year = "2022"
}

@article{Agia:2023skp,
    author = "Agia, Nicholas and Jafferis, Daniel L.",
    title = "{AdS$_3$ String Stars at Pure NSNS Flux}",
    eprint = "2311.04956",
    archivePrefix = "arXiv",
    primaryClass = "hep-th",
    month = "11",
    year = "2023"
}

@article{Kutasov:2005rr,
    author = "Kutasov, David",
    title = "{Accelerating branes and the string/black hole transition}",
    eprint = "hep-th/0509170",
    archivePrefix = "arXiv",
    month = "9",
    year = "2005"
}

@article{Giveon:2005mi,
    author = "Giveon, A. and Kutasov, D. and Rabinovici, E. and Sever, A.",
    title = "{Phases of quantum gravity in AdS(3) and linear dilaton backgrounds}",
    eprint = "hep-th/0503121",
    archivePrefix = "arXiv",
    doi = "10.1016/j.nuclphysb.2005.04.015",
    journal = "Nucl. Phys. B",
    volume = "719",
    pages = "3--34",
    year = "2005"
}
\end{document}